\documentclass[a4paper]{JHEP3}

\usepackage[english]{babel}
\usepackage{amsmath,amssymb}
\usepackage[dvips]{graphicx}
\usepackage{ifthen,array}
\usepackage[comma,square,sort&compress]{natbib}
\usepackage{amsfonts}
\usepackage{bbm}

\allowdisplaybreaks

\title{Interplay of type I and type II seesaw contributions to neutrino mass}

\author{ E. Kh. Akhmedov\thanks{On leave from the National Research
    Centre Kurchatov Institute, Moscow, Russia} \\

Department of Theoretical Physics,
Royal Institute of Technology \\
AlbaNova University Center,
SE-106 91 Stockholm, Sweden \\
E-mail:  \email{akhmedov@ictp.trieste.it}}

\author{M. Frigerio \\

Service de Physique Th\'eorique, CEA-Saclay\\
91191 Gif-sur-Yvette Cedex, France \\
E-mail:  \email{frigerio@spht.saclay.cea.fr}}

\newcommand{\beq}{\begin{equation}}
\newcommand{\eeq}{\end{equation}}
\newcommand{\bea}{\begin{array}}
\newcommand{\eea}{\end{array}}
\newcommand{\bec}{\begin{center}}
\newcommand{\eec}{\end{center}}
\newcommand{\bei}{\begin{itemize}}
\newcommand{\eei}{\end{itemize}}

\abstract{
Type I and type II seesaw contributions to the mass matrix of light neutrinos
are inherently related if left-right symmetry is realized at high energy 
scales. We investigate implications of such a relation for the interpretation
of neutrino data. We proved recently that the left-right symmetric seesaw
equation has eight solutions, related by a duality property, for the mass 
matrix of right-handed neutrinos $M_R$. In this paper the eight allowed 
structures of $M_R$ are reconstructed analytically and analyzed numerically
in a bottom-up approach. We study the dependence of right-handed  
neutrino masses on the mass spectrum of light neutrinos, mixing angle 
$\theta_{13}$, leptonic CP violation, scale of left-right symmetry breaking
and on the hierarchy in neutrino Yukawa couplings. The structure of the seesaw 
formula in several specific $SO(10)$ models is explored in the light of 
the duality. The outcome of leptogenesis may depend crucially on the choice 
among the allowed structures of $M_R$ and on the level crossing between 
right-handed neutrino masses. 
}


\preprint{SACLAY-T06-105}

\begin{document}

\section{Introduction}

Experiments with solar, atmospheric, accelerator and reactor neutrinos 
have given an unambiguous evidence for neutrino oscillations and therefore 
for neutrino mass. This implies a physics beyond the standard model (SM) of 
particle physics, since neutrinos are strictly massless in that model. 
At the same time, neutrino mass is different from zero in almost every 
extension of the SM. Moreover, many models 
contain more than one source of neutrino mass.
A crucial issue is therefore the ability to discriminate, within 
a given theoretical framework, among these sources using neutrino data 
as well as other neutrino-related experimental input.

If lepton number is not conserved, neutrinos are Majorana particles and 
their mass originates from the dimension five operator $LL\phi\phi$ 
via the electroweak symmetry breaking ($L\equiv (\nu_L~l_L)^T$ and 
$\phi\equiv (\phi^+~\phi^0)^T$ are the SM lepton and
Higgs isodoublets, respectively). New physics can contribute to this
operator either at tree or at loop level. We consider here the first
possibility; the super-heavy particles exchanged at tree level can then 
be either isosinglet fermions $N_R$, that is, right-handed (RH) neutrinos
\cite{mink}, 
or isotriplet scalars $\Delta_L$
\cite{MW}.\footnote{We will  
not consider the unique remaining possibility, that is, isotriplet 
fermions \cite{FLHJ}, 
which are more difficult to embed in minimal 
extensions of the SM.} The two cases are known under the names of type I and 
type II seesaw mechanism, respectively. Besides providing a very natural 
explanation of the smallness of the neutrino mass, the seesaw 
mechanism has a simple and elegant built-in mechanism for generating the 
observed baryon asymmetry of the universe -- baryogenesis via leptogenesis 
\cite{FY}. If the masses of $N_R$ and $\Delta_L$ are much larger than the 
electroweak scale, then neutrino experiments can only probe the couplings of 
the effective operator.  Therefore in general the data cannot tell type I from 
type II seesaw contribution to the masses of light neutrinos. Nonetheless, in 
several (possibly unified) models of fermion masses and mixing, both type I 
and type II terms are naturally present, and the knowledge of their relative 
size is crucial for achieving predictivity.\footnote{Even in very constrained 
scenarios, it is far from trivial to control the relative size of the 
contributions of the two seesaw types. This problem in the minimal SUSY 
$SO(10)$ model has been extensively discussed in \cite{DMM2, 
BMS} (for a different model see, e.g., \cite{fuku}).}

In this paper we study in detail the interplay of type I and type II 
seesaw contributions to neutrino mass.  We consider a theoretically well 
motivated connection between these two contributions, which stems 
from the proportionality of the Majorana mass terms of left-handed and
RH neutrinos. 
Let the Lagrangian of the theory contain the following term: 
\begin{equation} 
{\cal L}_M=
- \frac{f}{2} ( \nu_L^T C \nu_L \Delta_L^0 + N_L^{c T} C N^c_L
\Delta_R^0) + {\rm h.c.}\,, 
\label{LRc} 
\end{equation} 
where $C=i\gamma_2\gamma_0$ is the charge conjugation matrix, 
$N^c_L\equiv (N_R)^c$ and $\Delta_{L,R}^0$ are neutral scalar fields. 
In the case of more than one generation, $\nu_L$ and $N_L^c$ carry a 
flavor index and $f$ is a symmetric matrix of couplings. 
When $\Delta_{L,R}^0$ develop non-zero vacuum expectation values 
(VEVs) $v_{L,R}$, two Majorana mass terms, proportional to each other, are 
generated, one for left-handed and one for RH neutrinos. 
In general, a Dirac-type Yukawa coupling term is also present in the theory: 
\beq 
{\cal L}_D=- y \nu_L^T C N^c_L \phi^0 + {\rm h.c.}\,, 
\label{yuk}
\eeq 
where the neutral scalar $\phi^0$ has a VEV $v$. When $v_R\gg v$, the 
effective Majorana mass matrix for light left-handed neutrinos takes 
the form 
\beq 
m_\nu \simeq v_L f - \frac{v^2}{v_R} y
f^{-1} y^T \,. 
\label{master} 
\eeq 
The first (second) term on the r.h.s. 
of this equation is known as type II (type I) contribution to the light 
neutrino mass.\footnote{
Sometimes in the literature the mechanism leading to  
eq.~(\ref{master}) is called type II seesaw. We prefer the terminology 
where it is called type I+II seesaw, whereas the term ``type II seesaw''
is reserved for the situation when the Higgs triplet $\Delta_L$ is the 
sole source of neutrino mass.} 
The same coupling $f$ enters in both contributions because of the
assumption made in eq. (\ref{LRc}), which follows from a discrete
left-right symmetry of the underlying theory.

In fact, the seesaw formula takes the form given in eq. (\ref{master}) in 
models with left-right (LR) symmetry above some energy scale larger than 
the electroweak scale \cite{PS,MoPa}. 
By this we mean that the gauge group contains (or coincides with) $SU(3)_c
\times
SU(2)_L \times SU(2)_R\times U(1)_{B-L}$ {\it and} a discrete symmetry
guarantees that the $SU(2)_L$ and $SU(2)_R$ gauge couplings are
asymptotically equal.\footnote{Classic extensions of this minimal
LR symmetric setting are provided by the Pati-Salam model, based on $SU(4)_c 
\times SU(2)_L\times SU(2)_R \equiv SU_{422}$ \cite{PS}, and by the unified 
models based on $SO(10)$ \cite{G10}. 
} Such models incorporate 
naturally RH neutrinos (as well as isotriplet scalars) and 
explain the maximal parity violation of low-energy weak interactions 
as a spontaneous symmetry breaking phenomenon.  
Note also that LR models can be easily made supersymmetric, and the exact 
R-parity at low energies can be obtained through the spontaneous breaking 
of $U(1)_{B-L}$ \cite{exactR}.

In this paper we perform a thorough phenomenological analysis of the seesaw 
formula (\ref{master}). A duality property of this formula, which will be at 
the basis of our analysis, was identified in a recent paper by the authors 
\cite{AF}. The implications of such duality in a 
specific $SO(10)$ model are studied in \cite{HLS}.

The paper is organized as follows. In section 2 we discuss the duality 
property of the left-right symmetric seesaw. We also give the motivation for 
the reconstruction of the matrix $f$ (and therefore of the mass matrix of 
heavy RH neutrinos) from the seesaw formula (\ref{master}) 
and study the multiplicity of the solutions. 
In section 3 we develop a method of exact analytic reconstruction of $f$ 
in the cases of one, two and  three lepton generations. 
In section 4 
a number of numerical examples, illustrating our analytic results, are 
given  and the conditions for the 
existence of light RH 
neutrinos are discussed.
In section 5 we discuss the seesaw duality in specific models with 
LR symmetry, including several $SO(10)$ models.
In section 6 we briefly discuss some further issues pertaining to our analysis 
-- stability of the results with respect to the renormalization group 
evolution effects and baryogenesis through leptogenesis.  
Discussion and  summary are given in section 7. 
The appendices contain further analytic treatment of relevant topics:
the reconstruction of $f$ in the case of antisymmetric $y$ (Appendix A), 
the estimate of the relative size of type I and II seesaw contributions to
$m_\nu$ (Appendix B), and the generalization of the duality and the 
reconstruction of $f$ in the case of a different realization of the discrete 
LR symmetry (Appendix C).

\section{Bottom-up approach to left-right symmetric seesaw \label{bu}}

The LR symmetric seesaw equation (\ref{master}) relates three vacuum 
expectation values ($v$, $v_L$ and $v_R$) and three $n\times n$ matrices 
($m_\nu$, $y$ and $f$), where $n$ is the number of lepton generations.
The matrix $f$ is not directly constrained by any experiment. Moreover, 
being a Majorana-type Yukawa coupling matrix, $f$ has no analogy with the 
only Yukawa coupling matrices that can be presently accessed experimentally 
($y_u,~y_d,~y_e$), which are all of Dirac type. 

It is therefore natural to employ the seesaw formula for reconstructing  the 
matrix $f$, taking the quantities $m_\nu$, $y$, $v$, $v_{L,R}$ as input 
parameters. The purpose is to provide an insight into the underlying theory 
at the seesaw scale, which is deeply characterized by the structure of $f$ 
and which is not accessible to direct experimental studies. Notice that 
eq.~(\ref{master}) is a non-linear $n\times n$ matrix equation for $f$ and 
solving it in the case $n > 1$ is highly non-trivial.

In this section we first review the experimental and theoretical constraints 
on the input parameters. Then we discuss the duality property 
of the LR symmetric  seesaw formula (\ref{master}) and the multiplicity 
of its solutions for the matrix $f$.

\subsection{Parameter space \label{PS}}

Let us examine the possible values of the various quantities involved in
eq. (\ref{master}).

The neutrino mass matrix $m_\nu$ may be completely determined, at least in 
in principle, from low energy experiments. In terms of the mass 
eigenvalues $m_i$ ($i=1,2,3$) and leptonic mixing matrix $U$, one has 
\beq
m_\nu=U^* diag (m_1,\,m_2,\,m_3)\, U^\dag\,,
\label{mnu}
\eeq
where $U$ depends on three mixing angles $\theta_{12}$, 
$\theta_{23}$ and $\theta_{13}$, one Dirac-type CP-violating phase 
$\delta$ and two Majorana-type CP-violating phases $\rho$ and 
$\sigma$. From global fits of low energy experiments 
\cite{nuglobal1,StVi} 
one finds 
\beq
\Delta m_{sol}^2\equiv m_2^2-m_1^2 = (7.9^{+1.0}_{-0.8})
\times 10^{-5}~{\rm eV}^2\,,\qquad 
\Delta m_{atm}^2 \equiv |m_3^2-m_1^2| = (2.6\pm0.6)
\times 10^{-3}~{\rm eV}^2\,,
\label{dm2}
\eeq
\beq
\theta_{12}= (33^{+6}_{-4})^\circ\,,\qquad 
\theta_{23}= (45\pm10)^\circ\,,\qquad 
\theta_{13}\lesssim 12^\circ\,, 
\label{angles}
\eeq
where the best fit values and $3\sigma$ intervals are taken from the
last update of \cite{nuglobal1} (v5).
Cosmology yields the most stringent upper limit on the sum of 
the neutrino masses, $m_1+m_2+m_3\lesssim 0.4(0.7)$ eV 
at $99.9\%$ C.L. with(without) including Lyman-$\alpha$ data
in the fit \cite{w3GHMT}. 
The values of 
the CP-violating phases are completely unknown at present.

As far as the vacuum expectation values are concerned, we choose the 
convention of real and positive $v$, $v_L$ and $v_R$ (this can always be
achieved by redefining the phases of $\nu_L$, $N^c_L$ and $y$ in eqs. 
(\ref{LRc}) and (\ref{yuk})). The electroweak symmetry breaking 
parameter $v\approx 174$ GeV is accurately known, while only weak constraints 
are available for the triplet VEVs $v_{L,R}$. No lower bound on $v_L$ exists,  
and actually conditions are known under which $\Delta_L$ does not 
acquire any induced VEV. In contrast to this, $v_L$ is bounded from above by 
its contribution to the $\rho$-parameter ($\Delta\rho \approx-2 v_L^2/v^2$), 
so that precision electroweak measurements imply $v_L\lesssim$ GeV) \cite
{PDG}. Since we do not know the scale for the onset of LR symmetry, 
there is essentially no upper bound on $v_R$; we will be assuming in the 
following $v_R< M_{\rm Pl}$. The value of $v_R$ is bounded from below by the 
non-observation of RH weak currents ($v_R\gtrsim$ TeV) \cite{PDG}. It is 
useful to translate these bounds into bounds on the parameter $x\equiv 
v_L v_R/v^2$. 
If all dimensionless couplings in the scalar potential are of order one, its 
natural value is $x\sim 1$. 
However, the scalar sector may well be very complicated and the potential may 
depend on various mass scales which are not constrained {\em a priori}. 
Therefore, 
$x$  should be considered a free parameter in a model independent analysis.  
Using the above experimental constraints, one finds
$0\le x \lesssim 10^{14}$.

The Dirac-type neutrino Yukawa coupling matrix $y$ is not directly measurable, 
at least at present.  Nonetheless, in LR-symmetric models and in their 
embedding in models with partial or grand unification, $y$ is usually related 
to known quark and/or charged lepton Yukawa couplings. A detailed discussion 
of the structure of $y$ in some of these models will be given in section 
\ref{models}. Here we constrain ourselves to a few examples.
In the minimal LR symmetric model, in the supersymmetric case one has 
$y= \tan\beta y_e$, where $y_e$ is the Yukawa coupling matrix of charged 
leptons and $\tan\beta$ is the usual ratio of VEVs.  
In the minimal Pati-Salam model, one obtains instead
$y =y_{u}$, 
where $y_u$ is the Yukawa coupling matrix of the up-type quarks.  This 
relation also holds in the $SO(10)$ model where the only Higgs multiplet 
contributing to the fermion masses is in the fundamental representation 
${\bf 10}_H$. More complicated (and realistic) examples will be discussed in 
section \ref{models}.

One should bear in mind, however, that the nature of the high energy theory 
is essentially unknown and therefore the Yukawa coupling 
matrix of neutrinos may in fact be very different from those of charged 
fermions. In particular, the neutrino flavor sector may have radically 
different symmetries. 
As an interesting example, one may consider $y=diag(a,b,b)$ \cite{GLabb}, 
which is motivated by the maximal or nearly maximal 2-3 mixing observed  
in the atmospheric neutrino oscillations. 
It should be also noted that, in the presence of low energy supersymmetry,  
lepton flavor violating processes can be directly sensitive 
to the neutrino Yukawa couplings. For example, in constrained mSUGRA seesaw 
models one may need $y_{ij}\ll 1$ for some $i,j$, in order to suppress 
processes like $\tau\rightarrow \mu\gamma$ \cite{LMS}.

This brief survey illustrates that although certain choices of $y$ may be 
well motivated, this matrix is in fact model dependent. In the 
following we will keep the form of $y$ as general as possible, so as to  
maintain the generality of our results within the chosen LR symmetric 
framework.

\subsection{Seesaw duality \label{duel}}

Before discussing the general case, it will be useful to find the matrix 
$f$ approximately in the limits when either of the two contributions 
to $m_\nu$ in eq. (\ref{master}) dominates (these limits are  well 
defined as long as all the matrix elements in $m_\nu$ are dominated by the 
same type of seesaw).
In the case of dominant type I seesaw, one obtains  
\beq
\bea{l} f_I = -\dfrac 1x y^T m^{-1} y + \dfrac{1}{x^2}
(y m^{-1} y)^T m^{-1} (y m^{-1} y) \\ \\ - \dfrac{1}{x^3} (m^{-1} y
m^{-1} y)^T (y^T m^{-1} y + y m^{-1} y^T) (m^{-1} y m^{-1} y) + \dots\,,
\label{exp1}
\eea
\eeq 
where we used the notation
\beq
m \equiv \frac{m_\nu }{ v_L}\,,~~~~~~x\equiv \frac{v_L v_R}{v^2} \,.
\eeq
In the case of dominant type II seesaw, one finds 
\beq
f_{II} = m  + \frac 1x y m^{-1} y^T  - \frac{1}{x^2} (y m^{-1} y) m^{-1} 
(y m^{-1} y)^T + \dots\,.
\label{exp2}
\eeq
In this latter case $m_\nu\approx v_L f_{II}$, so that low energy neutrino
data allow one to reconstruct directly the mass matrix of RH neutrinos 
$M_R\approx v_R f_{II}$. In particular, 
the spectrum of the heavy neutrinos coincides (up to an overall factor) 
with that of light neutrinos \cite{JPR}.

In general, there is no {\em a priori} reason to expect that one type 
of seesaw dominates. If both contributions to $m_\nu$ are comparable, 
the solution of eq. (\ref{master}) cannot be obtained expanding around 
the purely type I or II solution and a different
approach is needed.

The LR symmetric type I+II seesaw formula in eq. (\ref{master}) has the 
following intriguing duality property. Suppose that a matrix $f$ solves 
this equation. Then one can verify that 
\beq 
\hat{f}\equiv m  - f = -\frac 1x y f^{-1} y^T 
\label{ft}
\eeq 
is also a solution, provided that the matrix $y$ is invertible and symmetric 
(or antisymmetric). These conditions on $y$ turn out to be both necessary and 
sufficient for $\hat{f}$ to solve eq. (\ref{master}). In particular, they 
guarantee that the matrix $\hat{f}$ is invertible, as it must be in order to 
satisfy eq. (\ref{master}). It is easy to see that the duality operation is 
closed, i.e. the dual of $\hat{f}$ coincides with $f$.

An important example of duality is provided by the solutions for $f$ in
the case of one seesaw type dominance, given in eqs. (\ref{exp1}) and
(\ref{exp2}): if $y=\pm y^T$, one immediately sees that $f_I$ and
$f_{II}$ are dual to each other. This allows us to establish an important 
result: if there is a solution of the seesaw equation with dominant type I 
seesaw, then there is also an alternative one, with dominant type II 
seesaw, and vice versa. More generally, 
we will show that the LR symmetric seesaw equation always has a pair of 
dual solutions which reduce to 
$f_I$ and $f_{II}$ when the input parameters satisfy certain conditions.

In the following we will focus on the case where $y$ is symmetric. This 
is true in models with the discrete LR symmetry $\nu_L\leftrightarrow 
N_L^c$ (see eq. (\ref{yuk})), including the minimal LR symmetric model 
and $SO(10)$ models where only the couplings to ${\bf 10}_H$ and 
${\overline{\bf 126}}_H$ Higgs multiplets contribute to $y$ (for 
further details see section \ref{models}). The case of antisymmetric $y$ 
will be discussed in Appendix \ref{antiY}. When $y$ is symmetric, one can 
write $y=U y_d U^T$, where $y_d$ is diagonal, real and positive and $U$ is 
a unitary matrix. (We consider $m_\nu$ in the basis where the mass matrix 
of charged leptons is diagonal, so that $U$ describes the mismatch between 
the left-handed rotations that diagonalize the Yukawa couplings of 
charged leptons and neutrinos.) As a consequence, eq. (\ref{master}) 
can be rewritten as 
\beq 
U^{\dag} m_\nu
U^* = v_L (U^\dag f U^*) - \frac{v^2}{v_R} y_d (U^\dag f U^*)^{-1} y_d
\,. 
\label{madia}\eeq 
Therefore, one can always work (and we will) in the basis where $y$ is 
diagonal, by redefining $(U^\dag m_\nu U^*) \rightarrow m_\nu$ and 
$(U^\dag f U^*) \rightarrow f$. 
If $U\approx \mathbbm{1}$, the input from the low-energy neutrino data will 
still (approximately) determine the left-hand side of eq. (\ref{madia}). 
Note that $U$ is the leptonic analogue of the CKM mixing matrix in the quark 
sector, where mixing is known to be small. Therefore the condition 
$U\approx \mathbbm{1}$ may be motivated by a quark-lepton symmetry.

\subsection{Multiplicity of solutions \label{1n}}

We have found that, if a matrix $f$ solves the seesaw equation 
(\ref{master}), so does $\hat{f}\equiv m-f$, i.e. this equation does 
not have a unique solution. This is hardly surprising, as eq. (\ref{master}) 
is a non-linear matrix equation for $f$, or, equivalently, a system of 
non-linear coupled equations for its elements $f_{ij}$. From the duality 
property of eq. (\ref{master}) it immediately follows that the number of 
its solutions must be even. We shall now show that for $n$ lepton generations 
the multiplicity of solutions is $2^n$. \footnote{
This result was first obtained in \cite{AF}, though in a less straightforward 
way. An alternative derivation can be found in \cite{HLS}.}

Let us introduce the matrices $\tilde{m}$ and $\tilde{f}$ through the 
relations 
\beq
m=\frac{1}{\sqrt{x}}y^{1/2}\,\tilde{m}\,y^{1/2}\,,\qquad\qquad
f=\frac{1}{\sqrt{x}}\,y^{1/2}\tilde{f}\,y^{1/2}\,,
\label{relat}
\eeq
where $y^{1/2}$ satisfies $(y^{1/2})^2=y$. From eq. (\ref{master}) with 
symmetric and invertible $y$ one then obtains
\beq 
\tilde{m}=\tilde{f}-\tilde{f}^{-1}\,.
\label{tildess}
\eeq
The duality we discussed above is especially clearly seen in 
this equation, where it corresponds to the invariance with respect to 
$\tilde{f}\leftrightarrow -\tilde{f}^{-1} \;(=\tilde{m}-\tilde{f}$).
Multiplying eq.~(\ref{tildess}) by $\tilde{f}$ on the left or on the 
right, we find that $\tilde{f}$ satisfies 
\beq
\tilde{f}^2-\tilde{m}\tilde{f}-\mathbbm{1}=0 
\label{tildef1}
\eeq
and that $[\tilde{m},\tilde{f}]=0$. The commutativity of $\tilde{m}$ and 
$\tilde{f}$ allows one to find a formal solution of eq. (\ref{tildef1}). 
Indeed, let us write $\tilde{f}$ in the form $\tilde{f}=\tilde{m}/2+R$, 
where the matrix $R$ is yet to be determined. Substituting this into 
eq.~(\ref{tildef1}) in which the term $\tilde{m}\tilde{f}$ is rewritten as  
$(1/2)\{\tilde{m}, \tilde{f}\}$, one finds that the matrix $R$ satisfies 
$R^2=\frac{\tilde{m}^2}{4}+\mathbbm{1}$, so that finally one obtains
\beq
\tilde{f}=\frac{\tilde{m}}{2}+\sqrt{\frac{\tilde{m}^2}{4}+\mathbbm{1}}\,.
\label{tildef2e}
\eeq
Since any non-singular $n\times n$ matrix with non-degenerate eigenvalues 
has $2^n$ square roots \cite{Gantm}, eq.~(\ref{tildess}) (and so eq. 
(\ref{master})) has $2^n$ solutions. 
Obviously, if $R_0$ is a square root of $R^2$, so is $-R_0$; 
therefore the solutions in eq. (\ref{tildef2e}) form $2^{n-1}$ 
dual pairs with $\tilde{f}+\hat{\tilde{f}}=\tilde{m}$.

\section{Analytic reconstruction of the coupling matrix $f$ \label{analytic}}

In this section we will solve analytically the seesaw equation for $f$ 
in the cases of one, two and three lepton generations. For one generation, 
the analysis is straightforward; for more than one lepton flavor, the 
effects of mixing complicate the reconstruction of $f$ considerably and, 
for this purpose, we shall need to develop new algebraic techniques. We 
will provide general solutions for the case of symmetric $y$, where the  
seesaw duality occurs (the generalization to antisymmetric $y$ is given 
in Appendix \ref{antiY}).

We will also identify the criteria to quantify the dominance of one or 
the other type of seesaw. In the presence of mixing, this identification 
will turn out to be a subtle issue, whose details are given in Appendix 
\ref{dominance}.

A short account of the main results of this section was published in
\cite{AF}. An alternative approach for the analytic reconstruction of $f$
was developed in \cite{HLS}, for the case of symmetric $y$.

\subsection{The case of one lepton generation  \label{1g}}

In this case $m_\nu$, $y$ and $f$ are merely complex numbers.  
Eq. (\ref{master}), being quadratic in $f$, has two solutions, which we 
denote $f_\pm$:
\beq
v_L f_\pm=\frac 12 \left[m_\nu \pm \left( m^2_\nu + \frac{4 v^2 y^2 v_L}
{v_R}\right)^{1/2}\right]
= \frac{m_\nu}{2} \left[1\pm \left(1+d\right)^{1/2}\right]\,,
\label{exact}\eeq
where
\beq 
d\equiv \frac{v_L}{v_R} \frac{4 v^2 y^2}{m^2_\nu} \,.
\label{domina}\eeq
Obviously, the solutions $f_{\pm}$ are dual to each other: $m=f_+ + 
f_-$.

At this point it is useful to state our convention for the assignment of
the complex phases. The freedom to rephase the fields in eqs. (\ref{LRc}) 
and (\ref{yuk}) allows us to take $v$, $v_L$, $v_R$ as well as $m_\nu$ 
real and positive. Next, we define $y^2 \equiv |y|^2 e^{i \chi}$, so 
that $\arg d=\chi$ and the phase of $f$ is then determined by eq. 
(\ref{exact}). 

Type I and II contributions to $m_\nu $ are defined as 
\beq
m_\nu^{I}\equiv -\frac{v^2 y^2}{v_R f}\,,~~~~~~~~~
m_\nu^{II}\equiv v_L f = m_\nu - m_\nu^{I} \,.
\label{III}\eeq
Their relative size is determined by the value of $d$:
\beq
r_\pm^{II/I}\equiv \frac{m_{\nu\pm}^{II}}{m_{\nu\pm}^I}= 
-\frac{[1\pm(1+d)^{1/2}]^2}{d} \,.
\eeq
We identify three possible physical regimes:\\
\begin{tabular}{llll}\\
(i)~~ & $|d|\ll 1$ 
~~~~~ & $|m_{\nu+}^I| \ll |m_{\nu+}^{II}|$ and $|m_{\nu-}^{II}| \ll 
|m_{\nu-}^{I}| $~~~~~ & single type dominance \\
(ii)~~ & $|d|\sim 1$  
& $|m_\nu^I|\sim |m_\nu^{II}|$ & hybrid seesaw \\
(iii)~~ & $|d|\gg 1$  
& $m_\nu^I \approx - m_\nu^{II}$ & cancellation regime\\ &&&
\end{tabular}\\
Let us discuss these cases in turn.

(i) The dominant seesaw type is I (II) 
in the case of the $f_-$ ($f_+$) solution. 
For $|d|\ll 1$, eq. (\ref{exact}) becomes
\beq
f_- \approx -\frac {v^2 y^2}{v_R m_\nu}\,,~~~~~~~~~~
f_+ \approx \frac{m_\nu}{v_L} + \frac {v^2 y^2}{v_R m_\nu}  \,.
\label{expD}\eeq  
Therefore  
a value $|d|\ll 1$ implies that one type of seesaw is dominant, but it 
does not determine which one. Thus, {\it both type I and type II dominance 
limits correspond to the same condition $|d|\ll 1$}. This may look 
counter-intuitive, as these two limits correspond to apparently opposite  
conditions $|m_\nu^I|\gg |m_\nu^{II}|$ and  $|m_\nu^I|\ll |m_\nu^{II}|$. 
However, it is easy to see that, when expressed in terms of the ``input'' 
parameters only, i.e. when the corresponding solutions for $f$ are 
substituted, both conditions reduce to $|d|\ll 1$.

A solution $f$ is meaningful (perturbative) only if $|f| \lesssim 1$: 
for $v_L \ll m_ \nu$, the solution $f_+$ violates perturbative unitarity 
and should be discarded, so that only $f_-$ is viable (dominant type I). 
A necessary condition for both solutions to be perturbative is $v_R\gtrsim
v^2|y|^2/m_\nu$, so that either type I or type II contribution to 
$m_\nu$ can dominate. Notice that for $|d|\ll1$ the expansions (\ref{exp1}) 
and (\ref{exp2}) apply for $f_-$ and $f_+$, respectively; eq. (\ref{expD})
just gives a simplified version of such expansions. 

(ii) For $|d|\sim 1$ both seesaw types give sizable contributions to $m_\nu$. 
Notice that when $d=-1$ (i.e., $4v_Lv^2|y|^2=v_R m_\nu^2$ and $\chi\equiv\arg 
y^2=\pi$), the equality $m_\nu^I=m_\nu^{II}=m_\nu/2$ is realized. This 
degeneracy point corresponds to the absence of CP violation. \footnote{
Note that in general CP violation is present in the one-generation case due 
to the presence of the Higgs triplets. Although it does not manifest itself 
at low energies, it can lead to a successful leptogenesis (more on this 
in section \ref{lept}). 
}

(iii) The case $|d|\gg 1$ corresponds to a cancellation between the two 
seesaw contributions to $m_\nu$. To bar a too strong (``unnatural") 
cancellation, one may demand $|m_\nu^{I,II}| \lesssim 1 $ eV. 
Eq. (\ref{III}) then implies a lower bound on the scale $v_R$ of 
$SU(2)_R$ symmetry breaking: $v_R \gtrsim v^2 |y|^2 / (|f|\cdot 1$ eV$) 
\approx |y^2/f| 3 \cdot 10^{13}$ GeV.  If one makes the additional 
(``natural'') assumption $x\equiv v_R v_L /v^2 \sim 1$, eq. (\ref{III})
will also imply $v_R\gtrsim v^2 |f|/$eV$ \approx |f| 3 \cdot 10^{13}$ GeV.

\FIGURE[t]{
\includegraphics[width=10cm]{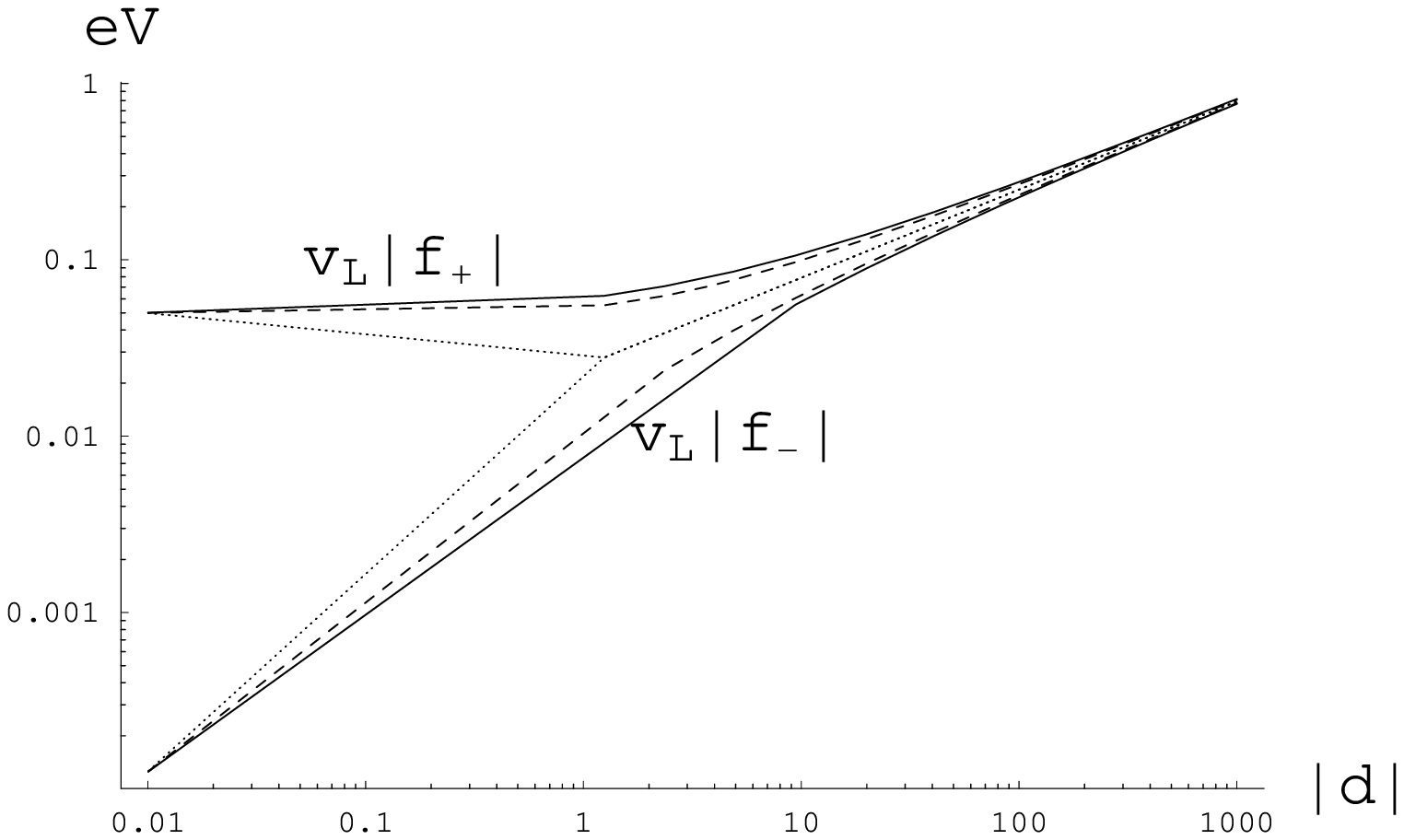}
\caption{Moduli of the solutions $v_L f_\pm$ of the seesaw formula, as 
functions of $|d|$. We chose $m_\nu=\sqrt{\Delta m^2_ {atm}}=0.05$ eV. 
The solid, dashed and dotted curves correspond to $\arg d = 0,~\pi/2$ and 
$\pi$, respectively. Recall that $m_{\nu\pm}^{II}=v_L f_\pm$ and 
$m_{\nu\pm}^I = m_\nu - m_{\nu\pm}^{II} = v_L f_\mp$. 
\label{decom}}
}

\FIGURE[t]{
\includegraphics[width=10cm]{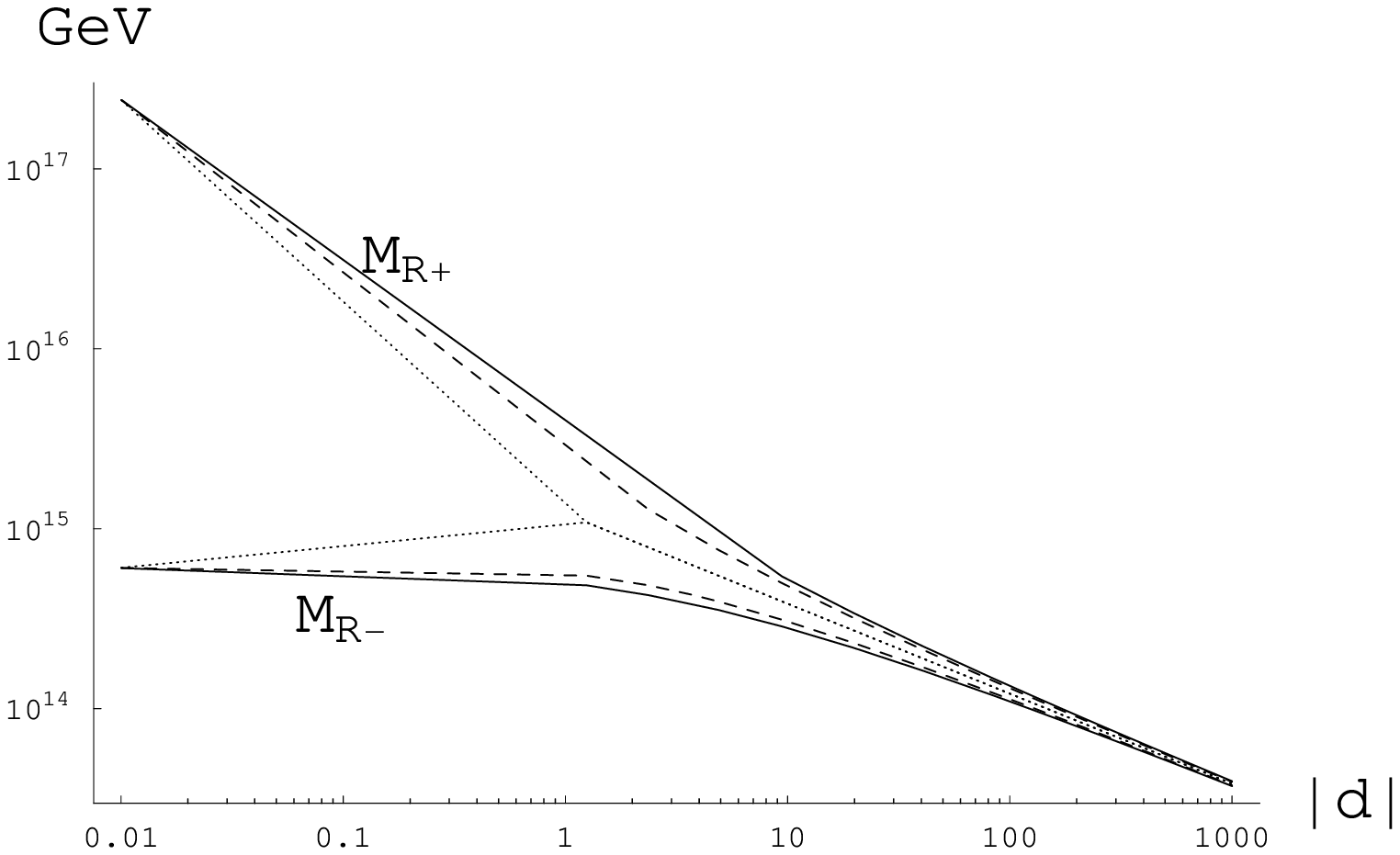}
\caption{The mass $M_{R\pm} = v_R |f_\pm|$ of the right-handed neutrino 
for the two solutions $f_\pm$ of the seesaw formula. We chose $m_\nu=
\sqrt{\Delta m^2_{atm}}=0.05$ eV and $|y|=1$. For fixed $|d|$, $M_R$ 
scales as $|y|^2$. The solid, dashed and dotted curves correspond to 
$\arg d = 0,~\pi/2$ and $\pi$, respectively. 
\label{Heavymasses}}
}

For illustration, let us assume that the light neutrino mass is given by
the atmospheric scale, $m_\nu = 0.05$ eV. 
The moduli $v_L|f_\pm|$ of the solutions of eq. (\ref{exact}) as functions 
of $|d|$ are then shown in fig. \ref{decom}. 
The three regimes (i), (ii) and (iii) correspond to the left, central
and right regions of the plot, respectively. The value of $|f_+|$
($|f_-|$) is a slightly decreasing (increasing) function of $\arg d
= \chi$ between $0$ and $\pi$, as shown in the figure. For $\chi=\pi$
and $|d|\ge 1$ the moduli of the two solutions coincide.

The mass of the RH neutrino is given by 
\beq M_{R\pm} \equiv v_R |f_\pm|= \frac{v^2}{m_\nu} 
\left|\frac{2y^2 [1\pm (1+d)^{1/2}]}{d}\right| \,. 
\label{MR1}
\eeq 
In fig. \ref{Heavymasses} we plot $M_R$ as a function of $|d|$ for 
$m_\nu = 0.05$ eV and $|y|=1$. 
The mass of the scalar triplet $M_\Delta$ is not directly related to the 
parameters in the seesaw formula. One expects $M_\Delta \sim v^2/v_L \sim 
v_R$, but very different values are also naturally possible, depending 
on the details of the mechanism which induces $v_L$.

Consider now the case of $n$ lepton generations with no mixing (which means 
that $f$ and $y$ are diagonal in the same basis). This case can be 
described as $n$ replicas of the one-generation case, each with two 
solutions $f_{\pm}^{(i)}$ ($i=1,...,n$).
For each generation, the relative size of type I and
II contribution to the light neutrino mass $m_{\nu i}$ depends on the
dominance parameter 
\beq d_{i}\equiv \frac{v_L}{v_R} \frac{4 v^2
y_i^2}{m^2_{\nu i}}\,,~~~ i=1,\dots,n\,. 
\eeq 
For a given value of $v_L/v_R$, one can have $|d_i|\ll 1$ (single type 
dominance) for an \mbox{$i$-th} generation, while at the same time the 
hybrid 
seesaw or cancellation regimes may be realized for $j\ne i$. We will 
see that in the presence of flavor mixing it is much less trivial to 
assess the relative size of the two seesaw types.

\subsection{The case of two lepton generations \label{2g}}

In this case in the flavor basis, where the mass matrix of charged 
leptons is diagonal (say, $m_l=diag(m_\mu,m_\tau$)), 
the seesaw formula (\ref{master}) can be explicitly written as 
\beq 
\left(\bea{cc}
m_{\mu\mu} & m_{\mu\tau} \\ m_{\mu\tau} & m_{\tau\tau} \eea\right) =
\left(\bea{cc} f_{22} & f_{23} \\ f_{23} & f_{33} \eea\right) - \frac
{1} {x F} \left(\bea{cc} y_{\mu 2} & y_{\mu 3} \\ y_{\tau 2} & y_{\tau
3} \eea\right) \left(\bea{cc} f_{33} & -f_{23} \\ -f_{23} & f_{22}
\eea\right) \left(\bea{cc} y_{\mu2} & y_{\tau2} \\ y_{\mu 3} & y_{\tau
3} \eea\right)\,, 
\label{2eq}
\eeq 
where $F\equiv \det f = f_{22}f_{33} - f_{23}^2$. The matrix equation 
(\ref{2eq}) is equivalent to the system of three coupled non-linear 
equations for $f_{22}$, $f_ {33}$ and $f_{23}$: 
\beq
\bea{l} xF(f_{22}-m_{\mu\mu}) = f_{33} y_{\mu 2}^2 -
2 f_{23}y_{\mu 2}y_{\mu 3} + f_{22} y_{\mu 3}^2 \,, \\
xF(f_{23}-m_{\mu\tau}) = f_{33} y_{\mu 2}y_{\tau 2} -
f_{23}(y_{\mu 3}y_{\tau 2}+y_{\mu 2}y_{\tau 3}) + f_{22} y_{\mu 3}
y_{\tau 3} \,, \\ xF(f_{33}-m_{\tau\tau}) = f_{33} y_{\tau2}^2 - 
2f_{23}y_{\tau2}y_{\tau3} + f_{22} y_{\tau3}^2 \,. 
\eea 
\label{system}
\eeq
To solve the system (\ref{system}), we use the following procedure. 
Let us define $f'= f/\sqrt{\lambda}$, $m'=m/\sqrt{\lambda}$ and $y'=
y/\sqrt{\lambda}$, where $\lambda$ is an as yet arbitrary complex number. 
The scaling law was chosen in such a way that in terms of the primed 
variables the system of equations for $f_{ij}'$ has the same form as 
eq.~(\ref{system}). Next, we fix the value of $\lambda$ by requiring 
$F'\equiv det f' = 1$. The system of equations for $f_{ij}'$ then becomes 
linear and can be readily solved. Expressing the primed variables back 
through the unprimed ones and substituting them into the condition 
$F'(\lambda)=1$, one obtains a 4th order polynomial  equation for $\lambda$. 
In general, it has four complex solutions $\lambda_i$ ($i=1,\dots,4$), 
leading to four allowed matrix structures $f_i$. Notice that $F_i\equiv 
\det f_i=\lambda_i$.
This procedure proves that eq.~(\ref{2eq}) has four solutions for a 
generic structure of $y$ and, therefore, it generalizes (for the case
$n=2$) the proof given in section \ref{1n} for symmetric $y$.

We now present the general analytic solution in the case where the 
matrix $y$ is invertible and symmetric, so 
that the duality holds.
As follows from eq. (\ref{madia}), when $y=y^T$ we can choose the basis
where $y$ is diagonal: $y_{\mu 3}=y_{\tau 2}=0,~y_{\mu 2}\equiv
y_2,~y_{\tau 3}\equiv y_3$. 
With a little abuse of notation, we will still denote the matrix elements 
of $m$ and $f$ in the new basis as $m_{\alpha\beta}$ and $f_{ij}$.  
The system (\ref{system}) can be linearized as described above and is 
easily solved: 
\beq 
f=\dfrac{x\lambda}{(x\lambda)^2-y_2^2 y_3^2}
\left(\bea{cc} x\lambda m_{\mu\mu} + y_2^2 m_{\tau\tau} & m_{\mu\tau}
(x\lambda -y_2 y_3) \\ 
\dots & x\lambda m_{\tau\tau} + y_3^2 m_{\mu\mu}\eea\right) \,, 
\label{sol2}
\eeq 
where $\lambda$ is a solution the following quartic equation: 
\beq 
\left[(x\lambda)^2 - y_2^2 y_3^2\right]^2 - 
x\left[\det m (x\lambda - y_2 y_3)^2 x\lambda + (m_{\mu\mu} y_3 + m_
{\tau\tau} 
y_2)^2 (x\lambda)^2 \right] = 0 \,. 
\label{lam2}
\eeq 
Taking the determinant of the equality $\hat{f}\equiv m -f = - y f^{-1} 
y/x$, one obtains 
\beq x^2 \lambda \hat{\lambda}\equiv x^2 F \cdot \hat{F} = y_2^2 y_3^2\,, 
\label{detdu}
\eeq 
where $\hat{F}\equiv \det \hat{f}$.  With the help of this relation, it 
is straightforward to check that the four solutions $f_i$ defined by 
eqs. (\ref{sol2}) and (\ref{lam2}) form two dual pairs. 
Actually, the duality makes it easy to express the four solutions of
eq. (\ref{lam2}) in the closed form as follows: 
\beq x\lambda_i = \frac 14
\left[x \det m + r_\pm \pm \sqrt{2(x \det m)^2 + 4 k x + 2 r_\pm x\det
m}\right]\,, 
\label{solu}
\eeq
where
\beq k\equiv m_{\mu\mu}^2
y_3^2+2m_{\mu\tau}^2y_2y_3+m_{\tau\tau}^2y_2^2\,,~~~~~ r_\pm = \pm
\sqrt{(x\det m)^2+4kx + 16 y_2^2 y_3^2 }  \,.
\label{KR}\eeq 
One pair of dual solutions corresponds to $r_+$ and the other one to $r_-$;
within each pair, a solution is distinguished from its dual by the sign in 
front of the radical in eq. (\ref {solu}).
Summarizing, eqs. (\ref{sol2}) and (\ref{solu}) define explicitly the four 
solutions $f_i$ as functions of the input parameters $m\equiv m_\nu/v_L$, 
$x\equiv v_L v_R/v^2$ and $y$.

We can now study the structure of $f_i$ in any regions of parameters 
of physical interest. To start with, let us consider the limit 
$y_2\rightarrow 0$. 
This can be justified if neutrino Yukawa couplings are related to 
those of charged fermions, so that $y_2\ll y_3\lesssim 1$. 
Then eq.~(\ref{solu}) can be expanded as
\beq
x \lambda_{1,3}=\frac{x\det m+ r_\pm^0
}{2} + \frac{2 x  y_3  
m_{\mu\tau}^2}{r_\pm^0 
}y_2+{\cal O}(y_2^2)\,,~~~~~~~~x\lambda_{2,4}
\equiv 
x\hat{\lambda}_{1,3}= \frac{y_2^2 y_3^2}{x\lambda_{1,3}}\,,
\eeq
where $r_\pm^0 =\pm \sqrt{(x\det m)^2+4 x y_3^2 m_{\mu\mu}^2 }$. The 
solutions for $f$ take the form
\beq
f_{1,3}=m-f_{2,4}
\,,~~~~~~~~
f_{2,4}=\left(\bea{cc}
0 & \dfrac{y_2 y_3 m_{\mu\tau}}{x\lambda_{1,3}} \\
\dots & - \dfrac{y_3^2 m_{\mu\mu}}{x\lambda_{1,3}}
\eea\right) + {\cal O}(y_2^2) \,.
\eeq
Since $\lambda_{2,4}=\det f_{2,4}$ are proportional to $y_2^2$, in the case 
of the solutions $f_{2,4}$ one RH neutrino mass becomes much 
smaller than $v_R$ for very small $y_2$. 
Notice that, even though $\lambda_{2,4}$ go to zero for $y_2\rightarrow
0$, the matrices $f_{2,4}$ are finite and invertible (and therefore
acceptable solutions) for any $y_2\ne 0$.

Let us define, in analogy with eq. (\ref{domina}), a dominance parameter 
that controls the relative size of type I and type II seesaw contributions:
\beq
d \equiv \frac{v_L}{v_R} \frac{4 v^2 y_3^2}{(m_\nu)_{\tau\tau}^2}
=\frac{4y_3^2}{xm_{\tau\tau}^2} \,.
\label{dom2}\eeq
Obviously, other analogous parameters can be defined by replacing 
$m_{\tau\tau}$ and/or $y_3$ with the other entries of $m$ and $y$. 
A detailed discussion of this issue is postponed to Appendix~\ref{dominance}; 
here we just study the main features of the dependence of the solutions on 
$d$. Fig.~\ref{2gXL} shows the dependence of the values of $|x\lambda_i|$ on 
$|d|$ for a specific realistic set of the input parameters.
In the region $|d|\ll 1$, there is a solution 
$|x\lambda_1|\gg|y_iy_j|$, which leads to $f_1\simeq m$ (dominant type II 
seesaw), while for the dual solution, one has $|x\lambda_2|\ll |y_iy_j|$ 
(dominant type I seesaw). In general, the other pair of dual solutions 
corresponds to hybrid seesaw. Only if the value of $\det m$ is strongly 
suppressed, one finds $f_3\simeq f_1$ (dominant type II seesaw) and 
$f_4\simeq f_2$ (dominant type I seesaw). In the region $|d|\sim 1$, 
all four solutions are of hybrid type. Finally, for $|d|\gg 1$ 
a cancellation between type I and II seesaw contributions to $m_\nu$ occurs.

\FIGURE[t]{
\includegraphics[width=10cm]{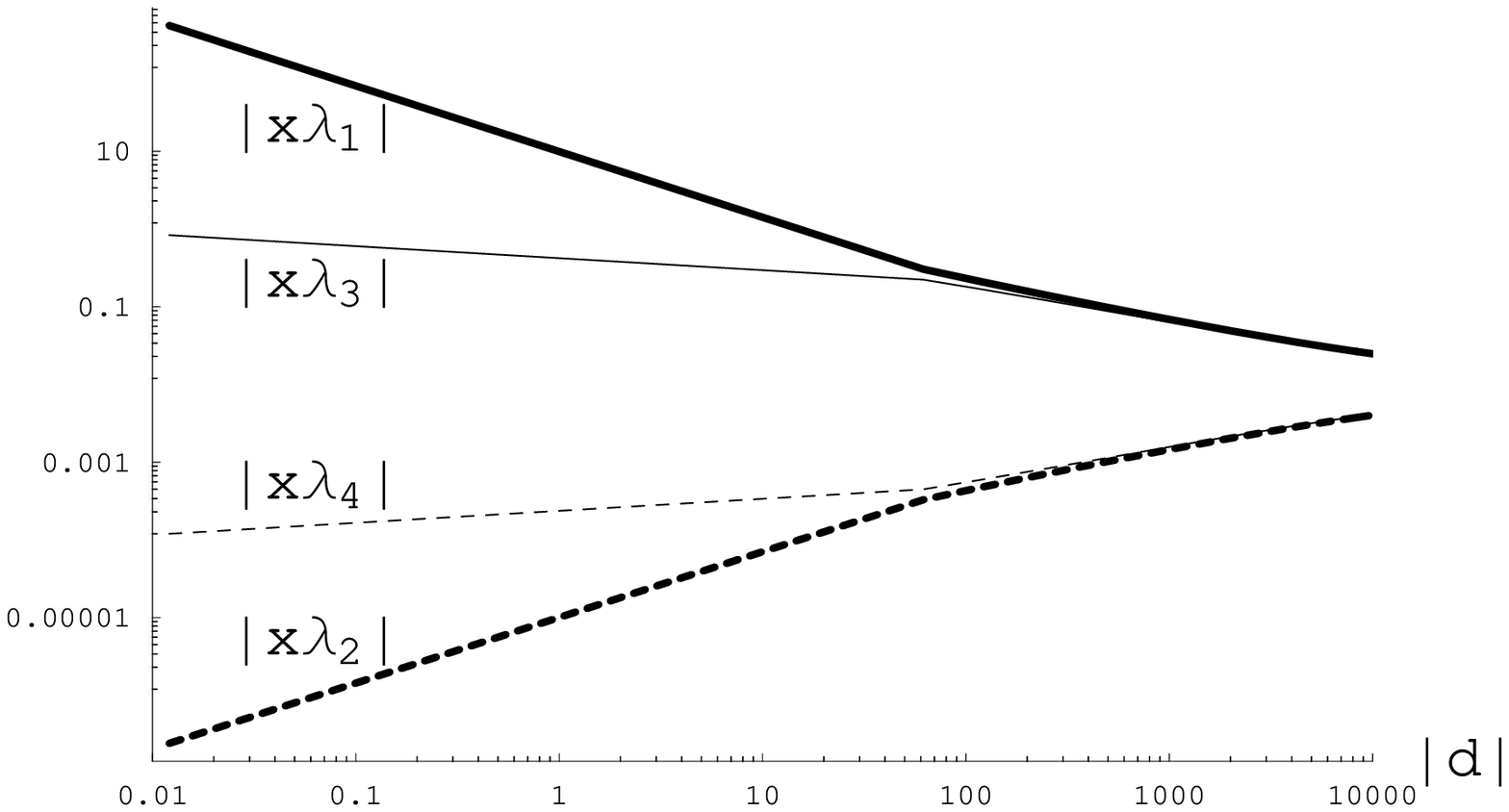} 
\caption{The four solutions $x \lambda_i$  of eq. (\ref{solu}) as functions 
of $|d|\equiv 4y_3^2/(x m_{\tau\tau}^2)$. We chose $m_{\mu \mu}=m_{\tau\tau}
=(0.02$ eV$)/v_L$ and $m_{\mu\tau}=(0.03$ eV$)/v_L$, so that $\theta_{23}=
\pi/4$ and $\Delta m^2_{23}=2.4 \cdot 10^{-3}$ eV. We also took $y_3=1$ and 
$y_2= 10^{-2}$.
\label{2gXL}}
}

\FIGURE[t]{
\includegraphics[width=15.1cm]{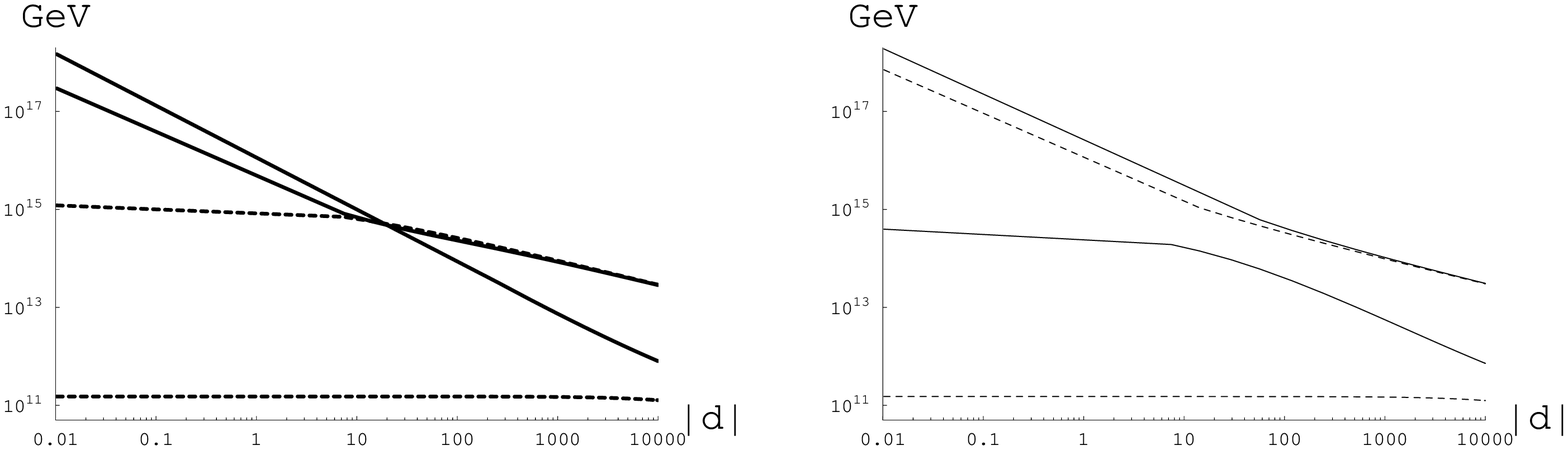} 
\caption{The masses $M_{2,3}$ of the RH neutrinos as functions of $|d|$ for 
the same set of input parameters as in fig. \ref{2gXL}. Thick solid (dashed) 
curves in the left panel correspond to the solution $f_1$ ($f_2$), thin solid 
(dashed) curves in the right panel, to the solution $f_3$ ($f_4$). 
\label{2gMR}}
}

In fig. \ref{2gMR} we plot the values $M_{2,3}$ of the two RH neutrino masses
(that is, the eigenvalues of $M_R=v_R f$) versus $|d|$ for the same set of 
the input parameters as in fig. \ref{2gXL}. In order to assess the size of 
$M_{2,3}$, it is useful 
to define a ``seesaw scale'' $M_s \equiv v^2/m_3$, which for $m_3=0.05$ 
eV is equal to $6\cdot 10^{14}$ GeV. 
Since for $|d|\ll 1$ the solution $f_1$ ($f_2$) corresponds to type II (I) 
seesaw dominance, one has $M_{2,3} \gg M_s$ for $f_1$ and $M_{2,3}\sim 
y^2_{2,3} M_s$ for $f_2$, as shown in the left panel of fig. \ref{2gMR}.
For the solutions  $f_{3,4}$, the situation is intermediate
and one finds $M_{3} \gg M_s$, $M_2 \sim y^2_{2,3} M_s$, as
shown in the right panel of fig. \ref{2gMR}.
For $|d|\gg 1$, the asymptotic values of $M_{2}$ are the same for all four 
solutions; the same is also true for $M_3$, with $M_2/M_3\sim y_2/y_3$. 
For the solution $f_1$, one observes a level-crossing of the two RH 
neutrino masses for $|d|\sim 20$. One could further characterize the sector 
of RH neutrinos by studying the dependence of their mixing on $d$.

\subsection{The case of three lepton generations \label{3g}}

In this case, the symmetric matrix $f$ contains 6 independent elements 
$f_{ij}$. The LR symmetric seesaw relation (\ref{master}) can be written 
as a system of 6 nonlinear coupled equations for these elements, analogous  
to that in eq. (\ref{system}). Here we solve this system analytically for 
the case of a symmetric neutrino Yukawa coupling matrix $y$, for which the 
duality holds. Using the freedom to work in the basis where $y$ is diagonal, 
$y=diag(y_1,y_2,y_3)$, 
one can write this system as 
\beq
xF(f_{ij}-m_{ij}) = y_i y_j F_{ij}\,,
\label{eqsf}
\eeq
where $F\equiv \det{f}$ and
\beq
F_{ij}\equiv \frac 12 \epsilon_{ikl}\epsilon_{jmn} f_{km}f_{ln} \,.
\eeq
The six corresponding equations for the dual matrix $\hat{f}\equiv m-f$ are
\beq
x\hat{F}(\hat{f}_{ij}-m_{ij}) = -x\hat{F}f_{ij}=y_i y_j \hat{F}_{ij}\,,
\label{eqsft}
\eeq
where $\hat{F}\equiv\det{\hat{f}}$ and
\beq\bea{c}
\hat{F}_{ij}\equiv \dfrac 12 \epsilon_{ikl}\epsilon_{jmn} \hat{f}_{km}
\hat{f}_{ln} =F_{ij}-T_{ij}+M_{ij}\,,~~ \\
T_{ij}\equiv \epsilon_{ikl}\epsilon_{jmn} f_{km} m_{ln}\,,~~
M_{ij}\equiv \dfrac 12 \epsilon_{ikl}\epsilon_{jmn} m_{km}m_{ln} \,.
\eea
\label{tensor}
\eeq
Taking the determinant of the equality $\hat{f} = -yf^{-1}y/x$, one proves 
immediately that
\beq
x^3 F \hat{F} = - y_1^2 y_2^2 y_3^2\,, 
\label{du3}\eeq
which is an analogue of eq. (\ref{detdu}).

To solve the system (\ref{eqsf}) for $f$, we make use of a procedure 
similar to that employed in the 2-generation case.
Let us define $f'= f/\lambda^{1/3}$, $m'=m/\lambda^{1/3}$ and $y'= y/ 
\lambda^{1/3}$ where $\lambda$ is determined from the condition  $F'(\lambda)
\equiv \det f'(\lambda) =1 $.  At this point the left-hand (right-hand) side 
of equations (\ref{eqsf}), written in terms of the primed variables, becomes 
linear (quadratic) in $f'_{ij}$. Using the duality (eqs. 
(\ref{eqsft})-(\ref{du3})), the quadratic terms, $F'_{ij}$, can be rewritten as
\beq
F'_{ij}=\dfrac{(y'_1 y'_2 y'_3)^2}{x^2 y'_i y'_j} f'_{ij}-M'_{ij}+T'_{ij}\,.
\eeq
This allows one to linearize the system for $f'_{ij}$:
\beq
[x^3-(y'_1 y'_2 y'_3)^2] f'_{ij} - x^3m'_{ij}= x^2 y'_i y'_j 
(T'_{ij}-M'_{ij})\,.
\label{full}\eeq
After some 
algebra, the solution for $f=\lambda^{1/3}f'$ can be written in a rather   
compact form (to be compared with eq. (\ref{sol2})):
\beq
f_{ij}  = \dfrac{  
\lambda^2 \left[ (\lambda^2-Y^2)^2 -  Y^2 \lambda \det m + Y^4 S \right] 
m_{ij} 
+  \lambda \left(\lambda^4 - Y^4\right) A_{ij} 
-  Y^2 \lambda^2(\lambda^2+Y^2) S_{ij} }{
(\lambda^2-Y^2)^3 - Y^2 \lambda^2 (\lambda^2-Y^2)  S - 2  Y^2 
\lambda^3 \det m }\,,
\label{solone}
\eeq
where 
\beq
Y^2\equiv \frac{(y_1y_2y_3)^2}{x^3}\,,~~~~ S \equiv 
{\displaystyle \sum_{k,l=1}^3} \left(\dfrac{m^2_{kl} x}{y_k y_l}\right)\,, 
~~~~A_{ij} \equiv \dfrac{y_i y_j M_{ij}}{x}\,,~~~~
S_{ij}\equiv {\displaystyle \sum_{k,l=1}^3} \left(m_{ik}m_{jl}
\dfrac{m_{kl} x}{y_k y_l}\right)\,.
\label{sums}\eeq

The  value of $\lambda$ is determined from the equation 
\beq
F(\lambda)\equiv \det f(\lambda)=\lambda\,,
\label{eq18}
\eeq 
where $f(\lambda)$ was defined in eq. (\ref{solone}). Eq. (\ref{eq18}) turns 
out to be an 8th order polynomial equation for $\lambda$, \footnote{
To recognize that, one has to use duality relations to identify and
cancel a common polynomial factor in the numerator and denominator
of $F(\lambda)$. 
} 
which has in general eight complex solutions. Defining 
\beq
A\equiv 
\sum_{k,l=1}^3 \left(\frac{y_k y_l M_{kl}^2}{x}\right)\,,
\eeq
this equation reads
\beq\bea{c}
\left[(\lambda^2 - Y^2)^2 - Y^2 \lambda^2 S \right]^2
- \lambda^2 (\lambda^2 + Y^2)^2 A -  Y^2 \lambda^4 (\det m)^2 \\
- \lambda \left[\lambda^6 + Y^2 \lambda^2 (\lambda^2- Y^2) 
\left(5+ S \right) - Y^6 \right] \det m\,=\,0\,, 
\eea
\label{final}
\eeq
to be compared with eq. (\ref{lam2}). The eight solutions of eq. (\ref{final})
form four dual pairs $\lambda_i,~\hat{\lambda_i}$ ($i=1,\dots,4)$. The duality 
is described by the relation $\lambda_i \hat{\lambda_i}= - Y^2$, which follows 
from eq. (\ref{du3}).
As a consequence, eq. (\ref{final}) can be rewritten as
\beq
0=\prod_{i=1}^4 (\lambda-\lambda_i)\left(\lambda+\frac{Y^2}{\lambda_i}
\right)=
\prod_{i=1}^4 (\lambda^2-z_i \lambda - Y^2)\,,~~~~
z_i\equiv \lambda_i-\frac{Y^2}{\lambda_i}\,.
\label{simpl}\eeq
By comparing eqs. (\ref{final}) and (\ref{simpl}), it is easy to verify 
that $z_i$ are the roots of the following quartic equation:
\beq
z^4 - \det m~ z^3 -(2Y^2 S+A)z^2 - Y^2(8+S)\det m~ z + Y^2
[Y^2S^2-4A-(\det m)^2]=0\,.
\label{zzzz}
\eeq
Thus, although the general order 8 algebraic equation does not 
admit analytic solutions, the equation for $\lambda$ can be solved in 
radicals since, due to duality, it reduces to a quartic equation in $z$. 
For the latter a general solution in radicals is known, though complicated. 
The 8 solutions for $f$ are obtained by plugging into eq. (\ref{solone}) 
the values of $\lambda$ given by
\beq
\lambda_i(\hat{\lambda}_i) = 
\frac{z_i\pm\sqrt{z_i^2+4Y^2}}{2}\,,~~~~(i=1,\dots,4)\,,
\label{laz}\eeq
where the sign in front of the radical distinguishes between $\lambda_i$ and 
$\hat{\lambda}_i$. 

Given the complicated algebraic form of the solutions for $f$, it may seem 
a hopeless task to quantify the dominance of one or the other seesaw type in 
the eight cases. However, it turns out that the techniques developed for  
analyzing this issue in the one- and two-generation cases can be generalized 
here. The full details of the analysis will be given in Appendix 
\ref{dominance}. Here we only describe the generic case, in which the relative 
size of type I and type II contributions does not change much from one 
element of the matrix $m$ to another and, in addition, no special 
cancellations among the entries of $m$, such as leading to $|\det m| \ll 
|m_{\alpha\beta}|^3$, occur.  
Then, in terms of the dominance parameter $|d|$ defined in 
eq. (\ref{dom2}), the classification goes as follows: for $|d|\ll 1$ 
there is one pair of dual solution with one seesaw type dominance, while
the other three pairs 
correspond to hybrid seesaw. All three RH neutrino masses are generically 
larger than $v^2/m_i$ in the solution with type II dominance, only two 
of them are larger than $v^2/m_i$ in three of the hybrid solutions, only one 
mass satisfies this condition in the corresponding three duals, and finally 
all RH neutrino masses are of the order of $v^2/m_i$ in the case of dominant 
type I seesaw. When $|d|\gtrsim1$, all 8 solutions for $f$ lead to hybrid 
seesaw.

In the next two subsections we will specialize our general analytic solution
to some physically interesting limits. The reader more interested in numerical 
examples may proceed directly to section \ref{numex}.

\subsubsection{Limit of hierarchical  neutrino Yukawa couplings \label{y0}}

A considerable simplification of the general solution for $f$ occurs when 
type I contributions to the elements of $m_\nu$, proportional to $y_1$,  
are negligible, i.e. in the limit $y_1\ll 1$. In fact, this limit is 
physically well motivated, in view of the tininess of the Yukawa couplings of 
the charged fermions of first generation. Strictly speaking, for $y_1=0$ the 
matrix $y$ is not invertible, so that the duality of solutions for $f$ does 
not hold. Nonetheless, eq. (\ref{solone}) is valid for any small but non-zero 
value of $y_1$ and therefore can be expanded in powers of $y_1$. Assuming 
that $\lambda$ is finite in the limit $y_1\rightarrow 0$, one finds
\beq
f = \left(\bea{ccc}
m_{ee} & m_{e\mu} & m_{e\tau} \\
\dots &
\dfrac {m_{\mu\mu} + \dfrac{y_2^2}{x \lambda} \left(M_{22}-\dfrac{y_3^2 
m_{ee}m_{e
\mu}^2}{x\lambda}\right)} 
{1 - \dfrac{y_2^2 y_3^2 m_{ee}^2}{(x \lambda)^2}} &  
\dfrac{m_{\mu\tau}+\dfrac{y_2y_3 m_{e\mu} m_{e\tau}}{x\lambda}}
{1 + \dfrac{y_2y_3m_{ee}}{x\lambda}} \\
\dots & \dots &
\dfrac {m_{\tau\tau} + \dfrac{y_3^2}{x\lambda} \left(M_{33}-
\dfrac{y_2^2 m_{ee}m_{e\tau}^2}{x\lambda}\right)} 
{1 -\dfrac{y_2^2 y_3^2 m_{ee}^2}{(x\lambda)^2}}
\eea\right) + {\cal O}(y_1) \,.
\label{big}
\eeq
The $e$-row of the mass matrix of light neutrinos $m$ directly determines
the first row of $f$ (pure type II seesaw), whereas the $2-3$ sector of $f$ 
depends on the interplay of the elements of the $e$-row and $\mu\tau$-block 
of $m$, as well as on the values of $y_2$, $y_3$ and $x\lambda$.

Consider now the behavior of $\lambda$ in the limit $y_1 \rightarrow 0$.
For $y_1=0$, the 8th order polynomial equation (\ref{final}) reduces to 
\beq
\lambda^4 \left\{[(x\lambda)^2 -  m_{ee}^2y_2^2 y_3^2]^2 
- x \left[\det m (x\lambda - m_{ee}y_2 y_3 )^2 x\lambda  
+ (M_{22} y_2 + M_{33} y_3)^2 (x\lambda)^2\right]\right\} = 0 \,.
\label{lam32}
\eeq
The zeros of the term in curly brackets in eq. (\ref {lam32}) are (to leading 
order) the four solutions $\lambda$ which are finite for $y_1\rightarrow 0$.
The four duals $\hat {\lambda}=-y_1^2 y_2^2 y_3^2 /(x^3\lambda)$ 
vanish in this limit, so that for them eq. (\ref{big}) does not hold.
However, the corresponding solutions can be obtained by duality: 
$\hat{f} = m - f$ with $f$ given in eq. (\ref{big}). By expanding directly 
eq.~(\ref{solone}), it is easy to check that $\hat{f}_{11}\sim y_1^2$, 
$\hat{f}_{12,13}\sim y_1$, while the other entries are finite.
As a consequence, one RH neutrino mass is much smaller than 
$v_R$ since it is proportional to $y_1^2$.

When $y_1=0$, one has only four (instead of eight) solutions, since their 
duals become singular. From the comparison of the term in curly brackets 
in eq. (\ref{lam32}) with eq. (\ref{lam2}), a strong analogy with the pure 
two-generation case becomes evident. In fact, a {\it different duality} 
among the four remaining solutions is present: if $\lambda\ne 0$
satisfies eq. (\ref{lam32}), also $\tilde{\lambda}\equiv y_2^2 y_3^2
m_{ee}^2 /(x^2 \lambda)$ does, and it corresponds to $\tilde{f}\equiv
\tilde{m}-f$, where $\tilde{m}_{\alpha\beta}=m_{\alpha\beta}+
m_{e\alpha}m_{e\beta}/m_{ee}$. There are two pairs of such solutions.

A very simple yet non-trivial scenario corresponds to the case when  type I 
contributions to $m_\nu$ proportional to $y_2$ are also negligible
($y_2\rightarrow 0$). In this limit the relation 
$f_{\alpha\beta}=m_{\alpha\beta}$ (pure type II seesaw) holds for all the 
entries of $f$ but $f_{33}$ \cite{JPR4,JPR}. There are two solutions for 
$f_{33}$, defined by
\beq
(f_{33}-m_{\tau\tau})^2 + \frac{\det m}{M_{33}} (f_{33}-m_{\tau\tau})-\frac
{y_3^2}{x} = 0 \,.
\eeq
In particular, this equation admits $|f_{33}|\gg |m_{\tau\tau}| \approx 
|m_{\mu\mu}|$, that is, $f$ with hierarchical structure can lead to $m_\nu$ 
with large $2-3$ mixing.

\subsubsection{Limit of hierarchical light neutrino masses \label{det0}}

When the mass spectrum of light neutrinos has normal (inverted)
ordering, the lightest neutrino mass $m_1$ ($m_3$) can be negligibly 
small. In this case the determinant of $m\equiv m_\nu / v_L$ vanishes.
For $\det m =0$ (more generically, for $|\det m|\ll |Y|$), 
eq. (\ref{final}) reduces to a pair of quartic equations:
\beq
(\lambda^2-Y^2)^2 - [\pm \sqrt{A} \lambda (\lambda^2+Y^2)+Y^2 S 
\lambda^2] = 0 \,.
\label{lar}\eeq
If $\lambda_i$ is a solution of the ``sign +'' equation, then also
$Y^2/\lambda_i$ is, while $-\lambda_i$ and $-Y^2/\lambda_i$ are 
solutions of the ``sign $-$'' equation. 
Therefore, two pairs of dual solutions for $\lambda$ are equal in absolute 
value and opposite in sign to the other two pairs. Moreover, eq. (\ref{lar}) 
has the same form as eq. (\ref{lam2}), if one makes the identifications
\beq
Y^2 \leftrightarrow \frac{y_2^2 y_3^2}{x^2}\,,~~~~
\pm\sqrt{A} \leftrightarrow \det m \,,~~~~
S \leftrightarrow \frac{xk}{y_2^2 y_3^2}\,,
\label{ridef}
\eeq
where the quantities on the right hand sides refer to the two-generation
case. 
Therefore, the quartic equations (\ref{lar}) can be explicitly 
solved and analyzed as in section \ref{2g}.

Since the solar mass squared difference is much smaller than the atmospheric 
one, in the case of the normal hierarchy one can neglect, in first 
approximation, both $m_1$ and $m_2$. In this limit
$m_\nu$ becomes a rank-1 matrix, and one can write
\beq
m = \left(\bea{ccc} a^2 & ab & ac \\ ab & b^2 & bc \\ ac & bc & c^2 
\eea\right) \,.
\label{fact3}
\eeq
Almost maximal atmospheric mixing implies $b\sim c$ as well as $\theta_{13}
\sim a/b$. The solar mass scale is zero in the limit of eq. (\ref{fact3}), so 
that the 1-2 mixing is undefined. 
Notice that plugging eq. (\ref{fact3}) in eq. (\ref{sums}), one finds 
$A_{ij}=0$ and $S_{ij}=m_{ij}S$.  
As a consequence, the expression for $f$ given in eq. (\ref{solone}) becomes 
singular in this limit, since $f\propto m$ and thus is not invertible.
One therefore has to resort to a different approach in order to find $f$. It 
turns out that in this limit the seesaw equation (\ref{eqsf}) has 
an infinite number of solutions. To illustrate this phenomenon, consider 
for simplicity the case $y_1=0$. Eqs.~(\ref{big}) and (\ref{lam32}) 
cannot be used since they would yield $f=m$, an unacceptable result 
when the matrix $m$ is not invertible.
A direct solution of eq. (\ref{eqsf}) gives in this case
\beq
f=\left(\bea{ccc} a^2 & ab & ac \\ ab & b^2 + y_2 \cos\alpha & 
bc+\sqrt{y_2y_3} \sin\alpha \\ ac & bc+\sqrt{y_2y_3} \sin\alpha & c^2 - 
y_3 \cos\alpha \eea\right) \,,
\eeq
where $\alpha$ is an arbitrary complex number different from zero. 
(For $\alpha=0$ a different ambiguity appears: 
\beq
f=\left(\bea{ccc} a^2 & ab & ac \\ ab & b^2 \pm y_2 & bc \\ ac & bc & c^2 
\pm y_3 
\eea\right) \,,
\eeq
where the two ``$\pm$'' signs are uncorrelated so that, for this value of 
$\alpha$, there are four solutions.)
Summarizing, when the matrix $m$ is of rank 1
there is an infinite number of solutions for $f$.

\section{Numerical examples \label{numex}}

We have shown that, for arbitrary values of $y_{1,2,3}$, $v_{L,R}$ and 
of the elements of $m_\nu$, one generically finds eight solutions 
for the matrix $f$, defined by eqs. (\ref{solone}), (\ref{laz}) and 
(\ref{zzzz}). The 
allowed values of the input parameters were discussed in section \ref{PS}. 
Here we present several numerical examples which are realistic, in the sense 
that they reproduce the observed neutrino oscillation parameters and respect 
all the other experimental constraints. This will enable us to identify the 
most interesting features of the allowed structures of $f$.


{\bf 1.~} As a first realistic numerical example that we will use as a 
benchmark point in the space of the input parameters, let us take
\beq
m\equiv \frac{m_\nu}{v_L}= \left(\bea{ccc} 0 & 0.1 & -0.1 \\ 0.1 & 0.55 & 0.45 
\\ -0.1 & 0.45 & 0.55 \eea\right) \,,
\label{real}\eeq
which corresponds to the tri-bi-maximal mixing ($\tan^2\theta_{23}=1$,
$\tan^2\theta_{12}=1/2$, $\tan^2\theta_{13}=0$), no CP violation and 
$\Delta m^2_{sol}/\Delta m^2_{atm} \approx 0.031$, in agreement with all the 
current data. The eigenvalues of $m$ are ($-0.1,~0.2,~1$), so that the 
spectrum has the normal hierarchy and the lightest neutrino has CP parity 
opposite to that of the other two. The above choice of $m$ fixes also the 
value $v_L \approx \sqrt{\Delta m^2_{atm}}\approx 0.05$ eV (see eq. 
(\ref{dm2})). We also take $x\equiv v_Lv_R/v^2=1$, which is a natural value 
if the dimensionless parameters in the scalar potential are all of order one. 
This determines $v_R\approx 6\cdot 10^{14}$ GeV. Finally, we take the 
hierarchy among the eigenvalues of $y$ to be slightly weaker than that for 
Yukawa couplings of the charged fermions: $y_1=10^{-2}$, $y_2=10^{-1}$, 
$y_3=1$. We neglect possible (CKM-like) rotations between the flavor basis 
and the basis where $y$ is diagonal.

With these choices, the solutions of the LR seesaw formula are
\beq\bea{ll}
f_1\approx\left(\bea{ccc} -0.001 & 0.10 & -0.14 \\ \dots & 0.56 & 0.49 
\\ \dots & \dots & 0.88
\eea\right) \,,~~~~ &
\hat{f}_1\approx\left(\bea{ccc} 0.001 & -0.005 & 0.04 \\ 
\dots & -0.006 & -0.04 \\ \dots & 
\dots & -0.33
\eea\right) \,,
\\
f_2\approx\left(\bea{ccc} -0.01 & 0.11 & -0.04 \\ \dots & 0.55 & 0.44 \\ 
\dots & \dots &  -0.88 
\eea\right) \,,~~~~ &
\hat{f}_2\approx\left(\bea{ccc} 0.006 & -0.008 & -0.06 \\ \dots & -0.004 & 
0.01 \\ \dots & \dots &  1.44 
\eea\right) \,,
\\
f_3\approx \left(\bea{ccc} 0.02 & 0.07 & -0.02 \\ \dots & 0.61 & 0.30 \\ 
\dots & \dots &  1.58 \eea\right) \,,~~~~ &
\hat{f}_3\approx \left(\bea{ccc} -0.02 & 0.03 & -0.08 \\ \dots & -0.06 & 
0.15 \\ \dots & \dots &  -1.03 
\eea\right) \,,
\\
f_4\approx \left(\bea{ccc} 0.01 & 0.08 & 0.08 \\ \dots & 0.60 & 0.25 \\ 
\dots & \dots &  -0.19
\eea\right) \,,~~~~ &
\hat{f}_4\approx \left(\bea{ccc} -0.01 & 0.02 & -0.18 \\ 
\dots & -0.05 & 0.20 \\ 
\dots & \dots &  0.74
\eea\right) \,.
\eea\label{numer}
\eeq
The rounding off in the numerical values of $f_{ij}$ is chosen so as to 
clearly illustrate the matrix structure of the solutions.
We have checked that for the high precision solutions the duality
relation $f_i+\hat{f}_i=m$ is satisfied very accurately for each dual pair. 
The structure (\ref{real}) of $m$ is recovered by plugging $f_i$ 
or $\hat{f}_i$ back into eq. (\ref {master}), but very accurate values of 
their entries (not shown in eq.~(\ref{numer})) should be used. 
This requirement of high accuracy is a consequence of the strong hierarchy 
among $y_i$ and exactly vanishing $ee$-entry of $m$.

For the solutions $f_i$ (the first column in eq. (\ref{numer})) the dominant 
$\mu\tau$-block of $m$ is reflected in a dominant $23$-block of $f$. The  
solutions $\hat{f}_i$ (the second column in eq. (\ref{numer})) exhibit a  
hierarchical structure, with a dominant $33$-entry; the maximal 2-3 mixing 
angle in $m$ is generated from a small mixing angle in $\hat{f}_i$. All the 
solutions $f_i$ and $\hat{f}_i$ 
have one order one eigenvalue, so that for the heaviest RH 
neutrino mass one finds $M_3\sim v_R$. The other two eigenvalues can 
be as small as $\sim 10^{-3}$. 
Type II (I) seesaw contributions dominate the entries $m_{e\mu}$, $m_{\mu\mu}$ 
and $m_{\mu\tau} $ for the solutions $f_{1,2}$ ($\hat{f}_{1,2}$). 
The other entries receive significant contributions from both seesaw types
(hybrid seesaw). This is also true for all the entries of the dual pairs of 
solutions $f_3$, $\hat{f}_3$ and $f_4$, $\hat{f}_4$.


{\bf 2.~} The largest uncertainty in the structure of $m_\nu$ is due to the
unknown absolute mass scale of the light neutrinos. Therefore, it is 
interesting to study the dependence of the structure of $f$ on this scale. 
Let us consider the same set of input parameters as above, but change the 
eigenvalues of $m_\nu$ as follows: $v_L(-0.1,0.2,1)\rightarrow 
(-m_1,\sqrt{m_1^2+\Delta m^2_{sol}}, \sqrt{m_1^2+\Delta m^2_{atm}})$. 
The solutions for $f$ will now depend on $m_1$, which may vary between 
zero and $\sim 0.23$ eV (due to the cosmological upper bound 
$\sum_i m_i\lesssim 0.7$ eV). The structures of $f$ given in eq. (\ref{numer}) 
correspond to $m_1=0.1v_L\approx 0.005$ eV. In fig.~\ref{NHm1}, we plot the 
masses $M_{1,2,3}$ of the RH neutrinos, given by the eigenvalues of $M_R\equiv 
v_R f$, as functions of $m_1$. We chose for illustration the first and the  
third pairs of dual solutions.\footnote{
Here and below in this section we adopt the numbering of solutions 
according to which $f_i$ go to the corresponding solutions given 
in eq.~(\ref{numer}), when the input parameters approach those chosen 
for eq.~(\ref{numer}). 
}

\FIGURE[t]{
\includegraphics[width=10cm]{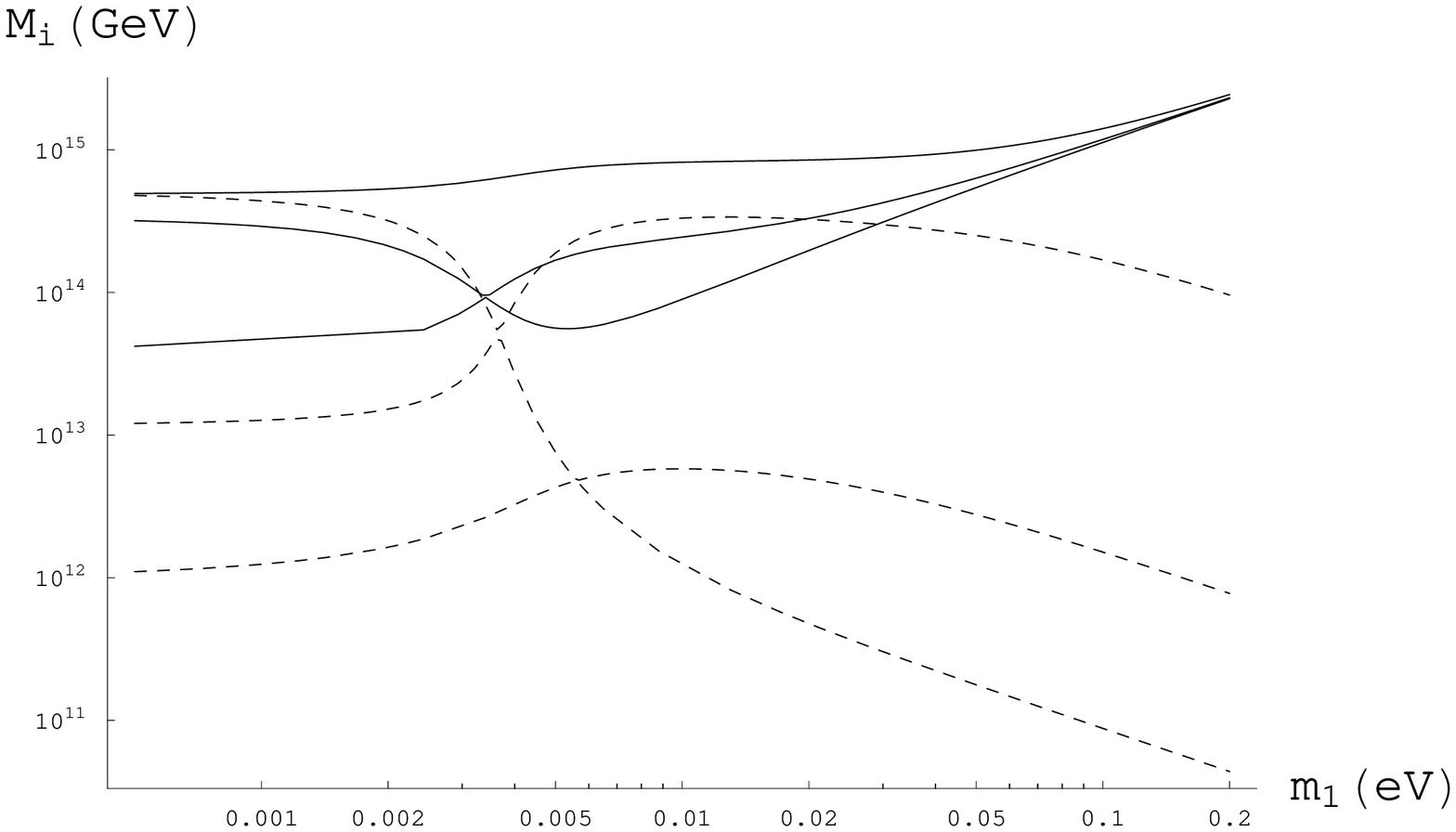} 
\includegraphics[width=10cm]{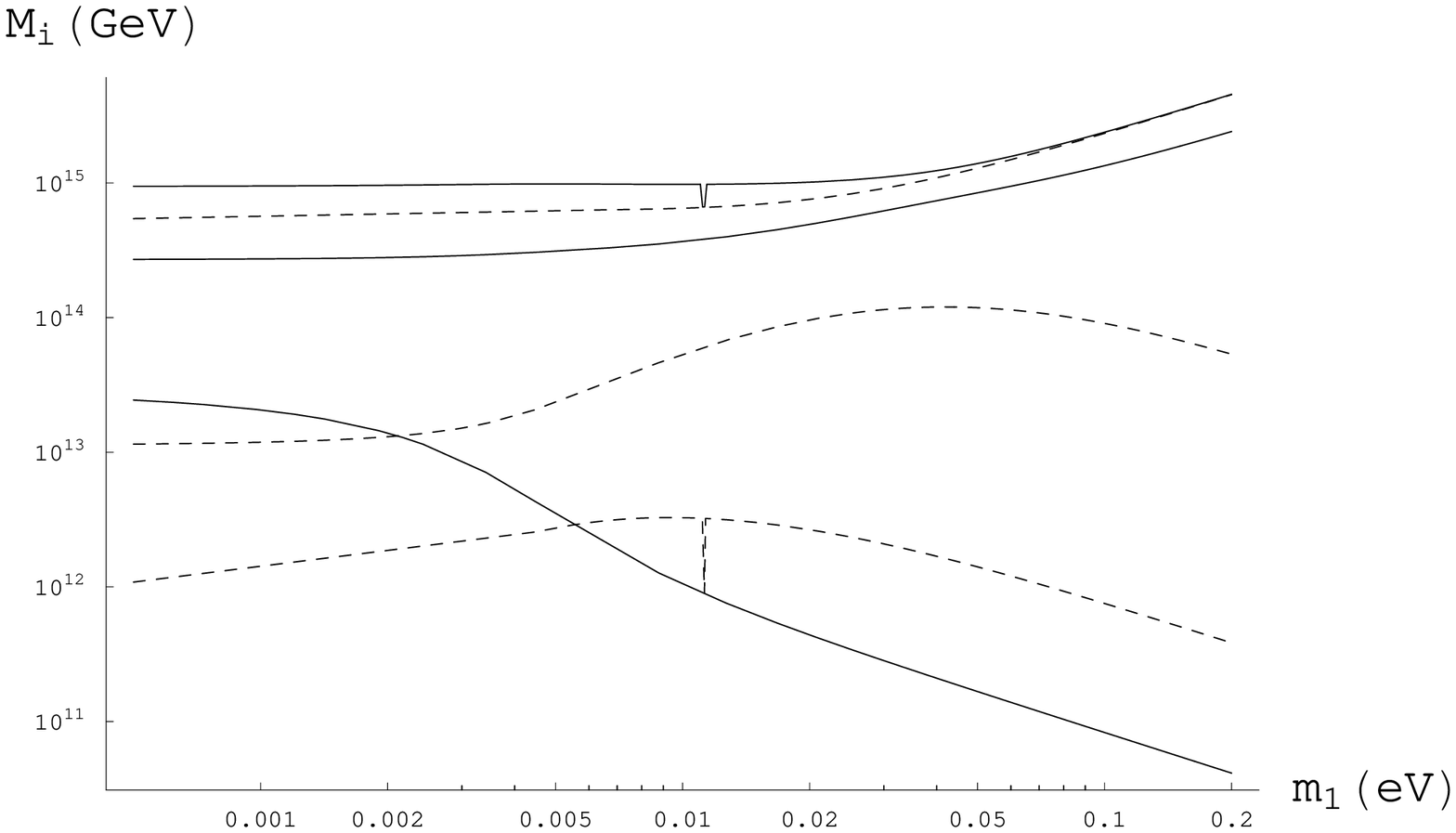}
\caption{The masses $M_{1,2,3}$ of the three RH neutrinos versus the 
light neutrino mass scale $m_1$. In the upper (lower) panel solid curves 
correspond to the solution $f_1$ ($f_3$) and dashed curves to the dual 
solution $\hat{f}_1$ ($\hat{f}_3$). We chose the tri-bi-maximal mixing, the 
light neutrino mass spectrum 
$(-m_1,\sqrt{m_1^2+\Delta m^2_{sol}},\sqrt{m_1^2+\Delta m^2_{atm}})$, 
$v_L\approx 0.051$ eV, $v_R\approx 5.9\cdot 10^{14}$ GeV and 
$y_{1,2,3}=10^{-2,-1,0}$. The irregularities around $m_1=0.011$ eV in the 
lower panel are a numerical artifact.
\label{NHm1}}
}

For the solution  $f_1$, the mass spectrum of RH neutrinos spans at most 
one order of magnitude, 
all three masses being rather close to $v_R$. 
The two lightest RH neutrinos undergo a level crossing at 
$m_1 \approx 0.035$ eV. 
Such crossing points were already identified in \cite{M4} for the case of 
pure type I seesaw. In these points the lepton asymmetry generated in the 
decays of the RH neutrinos can be resonantly enhanced, which may be crucial 
for reproducing the observed value of the baryon asymmetry of the universe 
\cite{M4}. For $m_1\gtrsim 0.1$ eV (quasi-degenerate light neutrinos), the 
three RH neutrino masses also become quasi-degenerate, since $f_1$ becomes a 
solution with dominant type II seesaw, so that $f_1\approx m$.

The dual solution $\hat{f}_1$ has a substantially different spectrum.
The RH neutrino masses span about three orders of magnitude, and two of them 
can be much smaller than $v_R$. Level crossings occur between the two 
heaviest and between the two lightest RH neutrinos. For $m_1\gtrsim 0.1$ eV, 
by duality $\hat{f}_1$ corresponds to the dominant type I seesaw. The masses 
$M_i$ of RH neutrinos then scale approximately quadratically with the 
Dirac-type Yukawa couplings $y_i$.

For the solution $f_3$, two RH neutrino masses are close to $v_R$
for all values of $m_1$, while the lightest RH 
neutrino mass $M_1$ decreases by almost three orders of magnitude 
when $m_1$ increases in its allowed range. For the dual solution, the RH 
neutrino masses lie between $10^{15}$ and $10^{12}$ GeV. This dual pair 
corresponds to hybrid seesaw for all values of $m_1$.


{\bf 3.~} The ordering (normal or inverted) is another important information 
about the light neutrino mass spectrum that is presently missing. Let us 
consider the input parameters chosen in example ${\bf 1}$, but replace the 
normally ordered spectrum of $m$ with an inverted one: $(-0.1,0.2,1)
\rightarrow (-1+\epsilon,1+\epsilon,\epsilon)$, where $\epsilon \equiv 
(1/4) \Delta m^2_{sol}/\Delta m^2_{atm} \lesssim 0.01$. This amounts to 
replace eq. (\ref{real}) by
\beq
m = \frac 16 \left(\bea{ccc} -2 & 4 & -4 \\ 4 & 1 & -1 
\\ -4 & -1 & 1 \eea\right) + \epsilon \left(\bea{ccc} 1 & 0 & 0 \\ 0 & 1 & 0 \\
0 & 0 & 1 \eea\right)\,.
\label{realIH}\eeq
Then, one pair of dual solutions is given by
\beq\bea{ll}
f_1\approx\left(\bea{ccc} -0.33 & 0.67 & -0.67 \\ \dots & 0.19 & -0.08 \\ 
\dots & \dots &  1.08
\eea\right) \,,~~~~ &
\hat{f}_1\approx\left(\bea{ccc} 4\cdot 10^{-5} & -0.001 & 0.001 \\ \dots 
& -0.01 & -0.08 \\ \dots & \dots &  -0.91 
\eea\right) \,,
\eea\label{numIH1}
\eeq
and the pair $f_2,~\hat{f}_2$ exhibits a qualitatively similar structure. 
Type II (I) seesaw dominates the entries of the first row of $f_1$ 
($\hat{f}_1$), while hybrid seesaw determines the entries of the $23$-block. 
A third pair of dual solutions is 
\beq\bea{ll}
f_3\approx\left(\bea{ccc} -0.35 & 0.51 & -0.52 \\ \dots & -0.77 & 0.68 \\ 
\dots & \dots &  -1.67
\eea\right) \,,~~~~ &
\hat{f}_3\approx\left(\bea{ccc} 0.02 & 0.15 & -0.15 \\ \dots & 0.95 & 
-0.84 \\ \dots & \dots &  1.85 
\eea\right)\,,
\eea\label{numIH2}\eeq
and the pair $f_4,~\hat{f}_4$ exhibits a qualitatively similar structure. 
All entries of these matrices receive significant contribution from both 
type I and type II seesaw.

\FIGURE[t]{
\includegraphics[width=7.4cm]{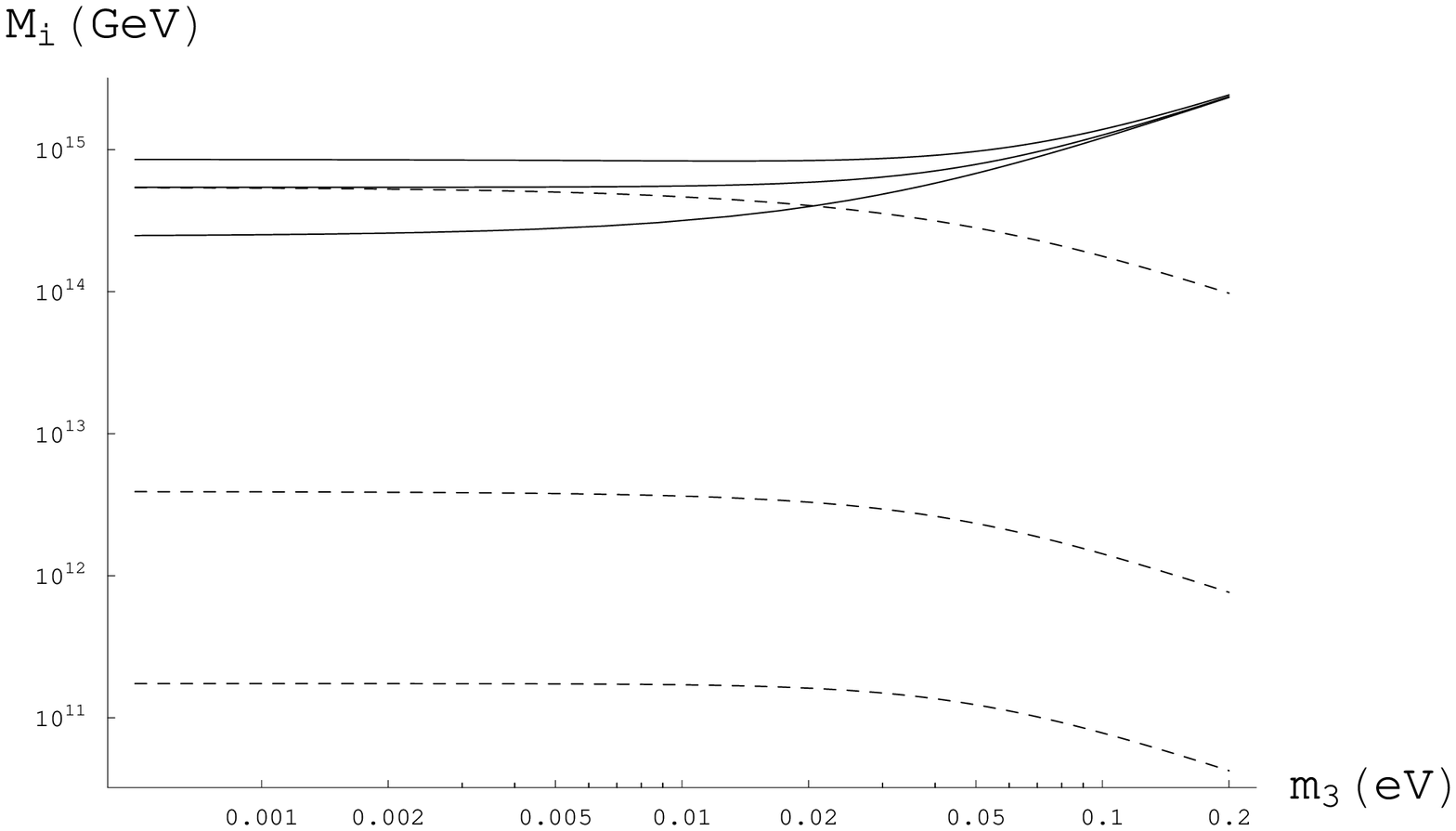}
\includegraphics[width=7.4cm]{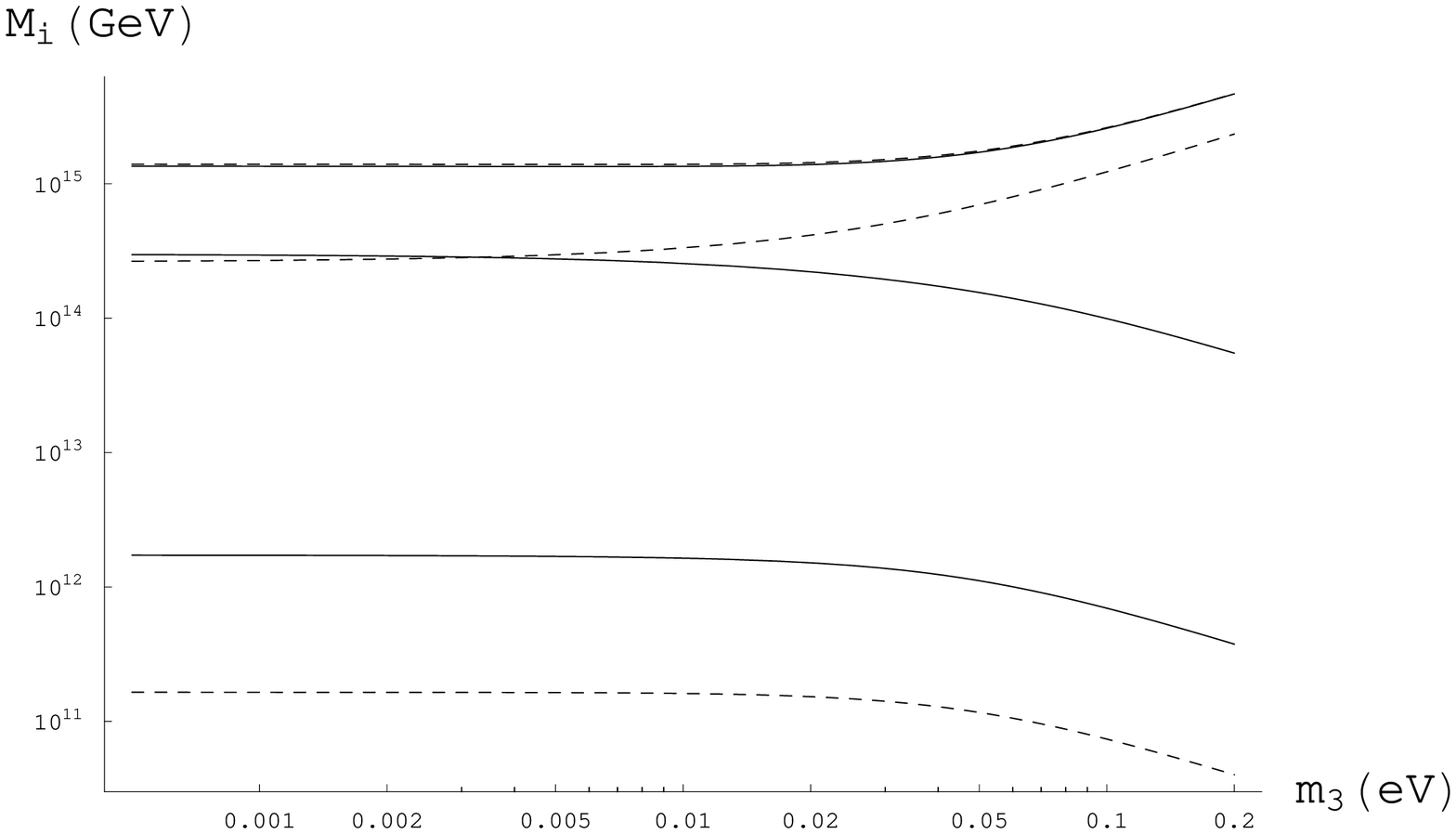} 
\caption{The masses $M_{1,2,3}$ of the three RH neutrinos versus the light 
neutrino mass scale $m_3$. We chose the same input parameters as in
fig. \ref{NHm1}, but the inverted light neutrino mass spectrum 
$(-\sqrt{m_3^2+\Delta m^2_{atm}-\Delta m^2_{sol}/2},~
\sqrt{m_3^2+\Delta m^2_{atm}+\Delta m^2_{sol}/2},~m_3)$.
In the left (right) panel, the solid curves correspond to the solution 
$f_1$ ($f_3$) and the dashed curves to its dual solution $\hat{f}_1$ 
($\hat{f}_3$).
\label{IHm1bis}}
}

In fig. \ref{IHm1bis} the RH neutrino spectrum is shown as a function of 
the absolute mass scale $m_3$ of light neutrinos. 
The eigenvalues of $m_\nu$ are chosen as
$(-\sqrt{m_3^2+\Delta m^2_{atm}-\Delta m^2_{sol}/2},$ 
$\sqrt{m_3^2+\Delta m^2_{atm}+\Delta m^2_{sol}/2},$ $m_3)$, which 
generalizes the choice $v_L(-1+\epsilon,1+\epsilon,\epsilon)$ that leads 
to eqs. (\ref{realIH})-(\ref{numIH2}). For the solution $f_1$, the masses 
of the three RH neutrinos are rather close to each other (within a factor 5) 
and to $v_R$. For the solution $\hat{f}_1$, instead, they are spread over 
$3-4$ orders of magnitude, with the lightest one being around $10^{11}$ GeV. 
These features of the mass spectrum of RH neutrinos are rather insensitive to 
variations of $m_3$ between zero and $0.23$ eV. For both $f_3$ and $\hat{f}_3$ 
and in the whole allowed range of $m_3$, the masses of two RH neutrinos are 
close to $v_R$, while the third mass is significantly smaller.  
Actually, despite the fact that all the entries of $f_3$ and $\hat{f}_3$
in eq. (\ref{numIH2}) are of order one, the determinants $\lambda_3$ and 
$\hat{\lambda}_3$ turn out to be much smaller than one, corresponding to 
one RH neutrino mass being much smaller than $v_R$. Notice also that the value 
of $M_3$ is almost the same for these two dual solutions. Contrary to the 
case of the normal mass ordering, the present set of input parameters does 
not lead to the level  crossing phenomenon.


{\bf 4.~} Up to now we were assuming in this section the exact tri-bi-maximal 
mixing. However, the present uncertainties of the values of the leptonic mixing 
angles are quite sizable. In particular, only an upper bound on the 1-3 mixing 
exists, $\sin\theta_{13} \lesssim 0.2$. Measuring this parameter is one of 
the main goals of the near-future experimental neutrino program. Let us 
discuss the modifications to the structures of $f$ when a non-vanishing 
$\theta_{13}$ is allowed.

\FIGURE[t]{
\includegraphics[width=7.4cm]{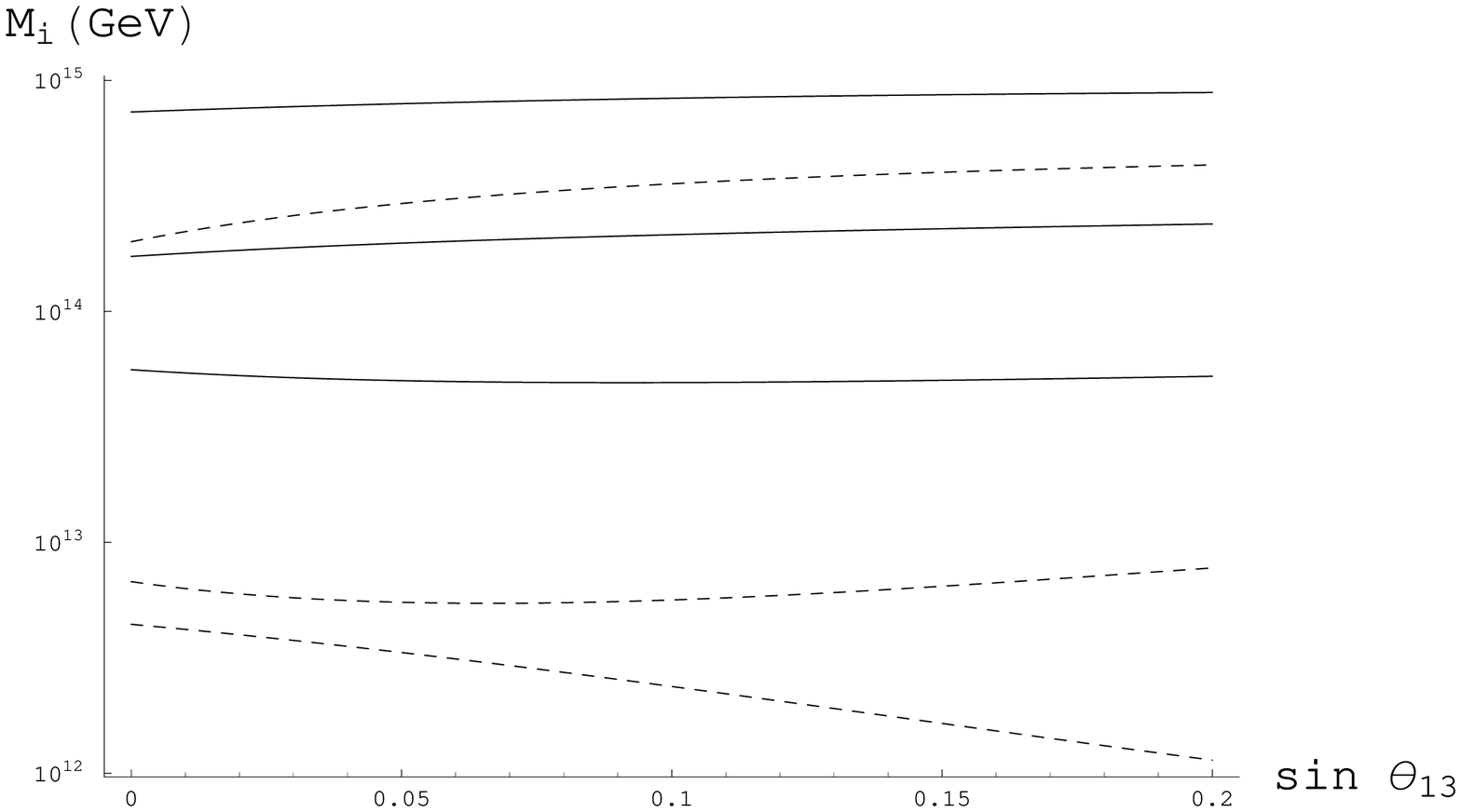} 
\includegraphics[width=7.4cm]{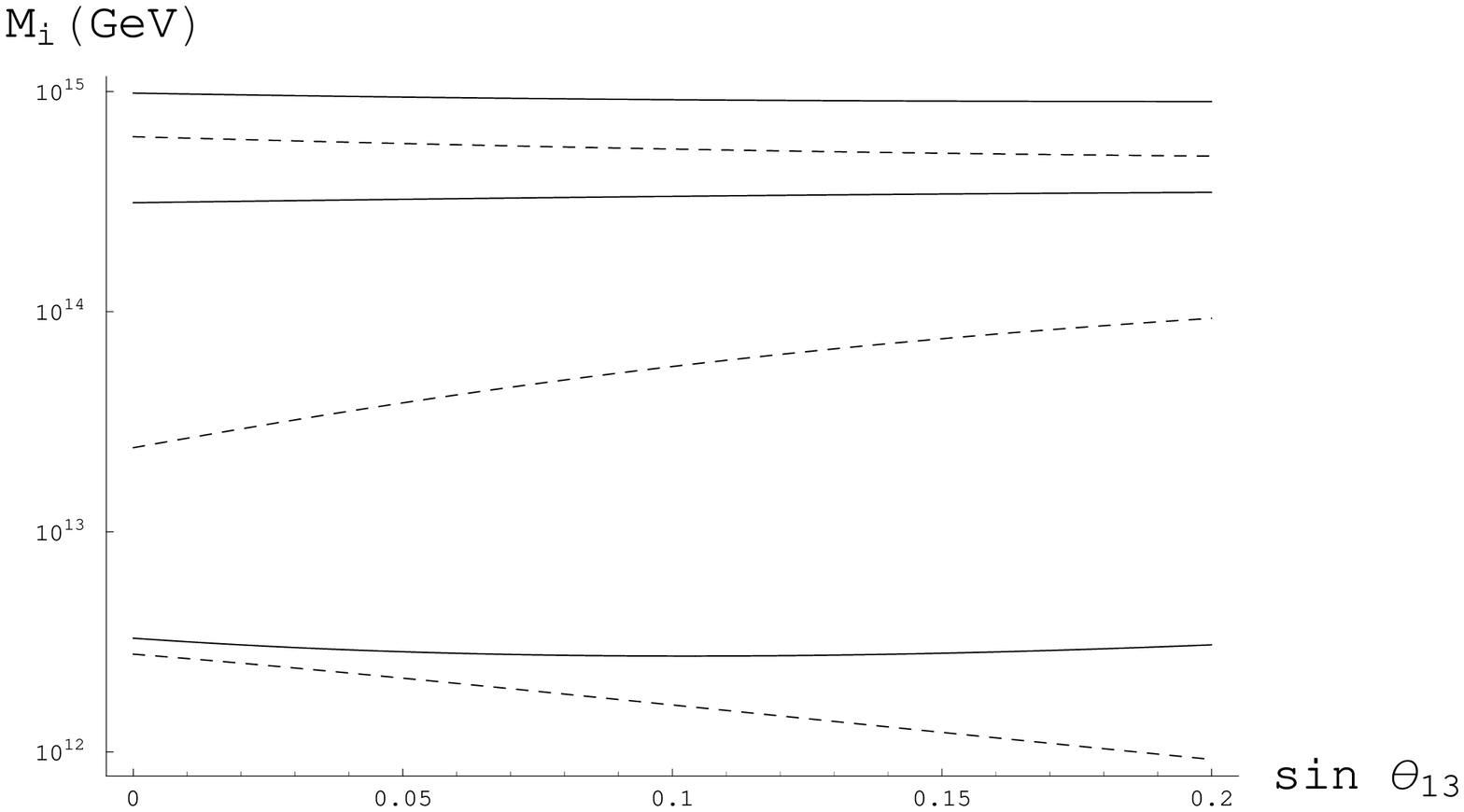} 
\caption{The masses $M_{1,2,3}$ of the three RH neutrinos as functions of 
$\sin\theta_{13}$. We chose the same input parameters as in fig. \ref{NHm1}, 
but fixed $m_1=0.005$ eV and allowed non-zero $\theta_{13}$.
In the left (right) panel, the solid curves correspond to the solution 
$f_1$ ($f_3$) and the dashed curves to its dual solution $\hat{f}_1$ 
($\hat{f}_3$).
\label{s13}}
}

Consider the same input parameters as in example ${\bf 1}$,
but with $\sin\theta_{13}=0.1$. Then eq.~(\ref{real}) is replaced by
\beq
m \approx  \left(\bea{ccc} 0.01 & 0.17 & -0.03 \\ 0.17 & 0.53 & 0.44 
\\ -0.03 & 0.44 & 0.56 \eea\right) \,.
\label{real13}
\eeq
The solutions for $f$ are modified slightly with respect to those in 
eq.~(\ref{numer}), but their qualitative features remain the same. 
However, the relative size of type I and II seesaw contributions to a 
given element of $f$ may change significantly with $\theta_{13}$.
The mass spectrum of RH neutrinos in general depends weakly on $\theta_{13}$, 
as shown in fig. \ref{s13} for the pairs of dual solutions $f_{1,3}$ and 
$\hat{f}_{1,3}$: the masses $M_i$ change at most by about a factor of 
10 for $\sin\theta_{13}$ varying between 0 and 0.2.
This indicates that 
similar underlying theories at the seesaw scale may result in very 
different values of $\theta_{13}$.


{\bf 5.~}  An important experimental and theoretical issue is whether CP 
is violated in the leptonic sector. The Dirac-type CP-violating phase is 
associated to the parameter $\theta_{13}$ and therefore its effect on the
structure of $m_\nu$ is small. On the contrary, the two Majorana-type
CP-violating phases, that is, the relative complex phases of the eigenvalues 
$m_{1,2,3}$ of $m_\nu$, can have large effects on the structure 
of the mass matrix. In particular, the relative phase $\rho$ between $m_1$ 
and $m_2$ affects substantially the parameter $m_{ee}\equiv |(m_\nu)_{ee}|$, 
which determines the decay rate of nuclei undergoing neutrinoless 
$2\beta$-decay. Since next generation $2\beta 0\nu$ decay experiments 
will be mostly sensitive to the quasi-degenerate neutrino mass spectrum, we 
consider here the dependence on $\rho$ assuming three light neutrinos with 
mass $\sim 0.2$ eV.\footnote{ 
In the basis where $y$ is real and diagonal, the matrix $m$ in the
seesaw formula depends on 6 CP-violating  phases, the three low
energy ones plus three phases that are not observable at low energies,
but may be relevant for the reconstruction of $f$. They can be taken into 
account by allowing $y_i$ to be complex. In the examples considered here we 
set them equal to zero.
}

For definiteness,  we once again use the parameters chosen in example 
${\bf 1}$, except that the mass spectrum  of light neutrinos is now taken 
to be  $v_L(e^{-2i\rho},~1+\epsilon,~1+\eta)$, with $v_L=0.2$ eV, $
\epsilon\approx \Delta m^2_{sol}/(2v_L^2)$ and $\eta \approx \Delta 
m^2_{atm}/(2v_L^2)$. The mass matrix of light neutrinos takes the form
\beq
m= \mathbbm{1}_3 +\frac{1-e^{-2i\rho}}{6}  \left(\bea{ccc} -4 & 2 & -2 \\ 2 
& -1 & 1 
\\ -2 & 1 & -1 \eea\right) + \frac{\epsilon}{3} \left(\bea{ccc} 1 & 1 & -1 \\ 1 
& 1 & -1 \\
-1 & -1 & 1 \eea\right) + \frac{\eta}{2} \left(\bea{ccc} 0 & 0 & 0 \\
0 & 1 & 1 \\ 0 & 1 & 1 \eea\right)\,.
\label{realQD}\eeq
In particular, one finds $m_{ee} \approx v_L \sqrt{1-(8/9)\sin^2\rho }$. 
For $\rho=0$ ($\rho=\pi/2$) the CP parity of $\nu_1$ is equal (opposite) to 
that of the other two light neutrinos.  

\FIGURE[t]{
\includegraphics[width=7.4cm]{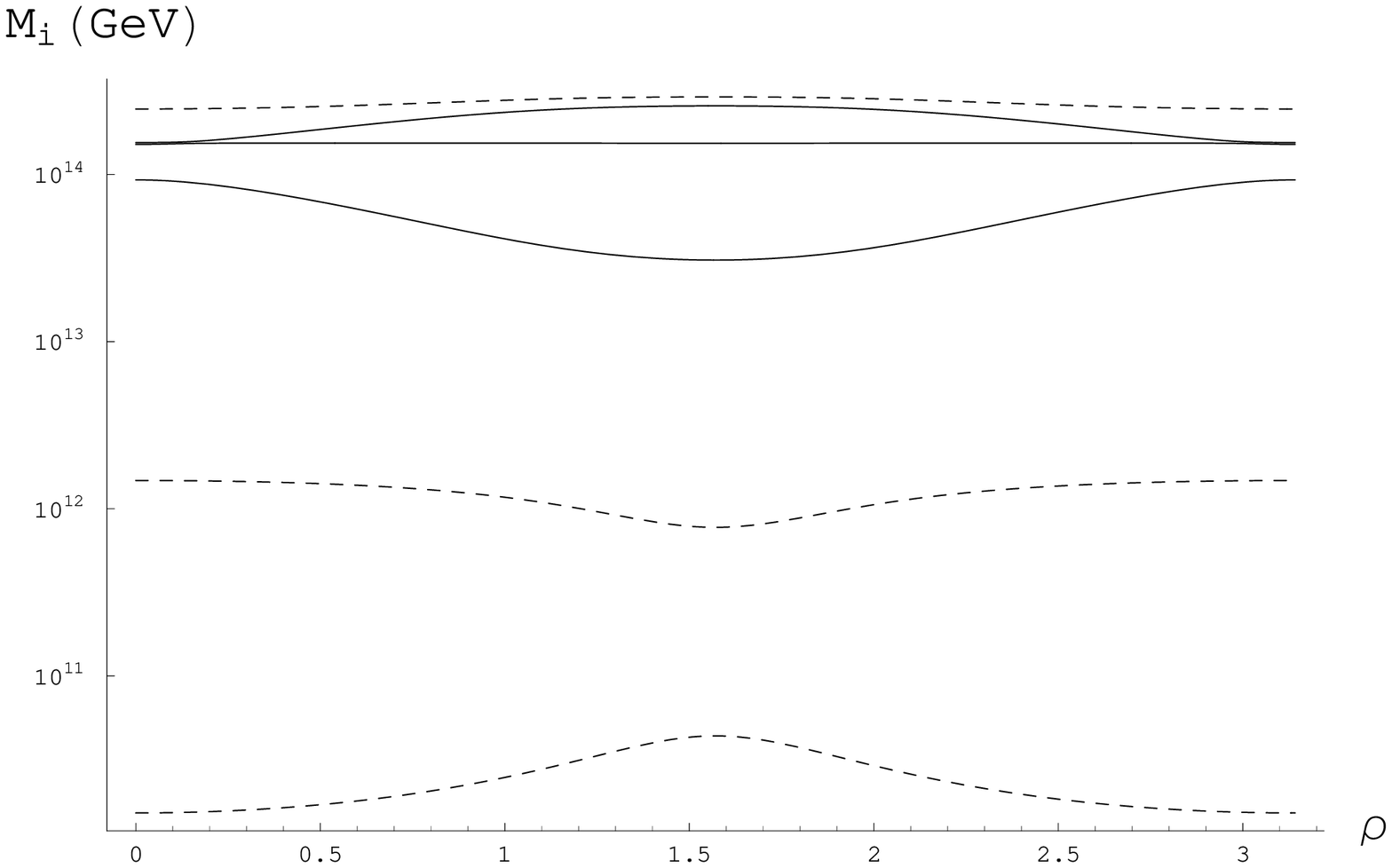} 
\includegraphics[width=7.4cm]{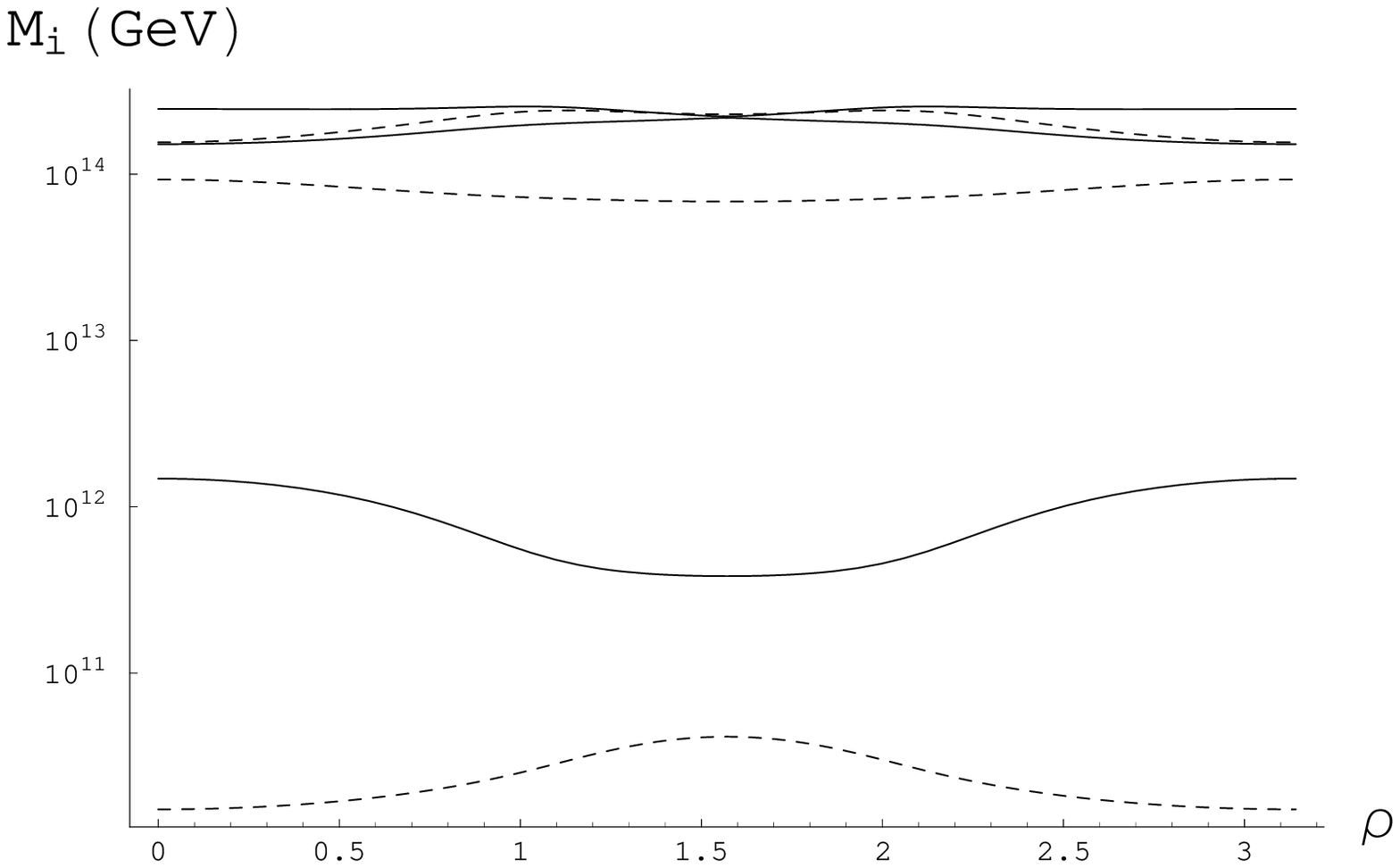} 
\caption{The masses $M_{1,2,3}$ of the three RH neutrinos versus the 
Majorana-type CP violating phase $\rho$. 
In the left (right) panel, the solid curves correspond to the solution 
$f_1$ ($f_2$) and the dashed curves to its dual solution $\hat{f}_1$ 
($\hat{f}_2$).
We chose the same input 
parameters as in fig. \ref{NHm1}, but assumed the quasi-degenerate mass 
spectrum of light neutrinos $v_L[e^{-2i\rho},1+\Delta m^2_{sol}/(2v_L^2),
1+\Delta m^2_{atm}/(2v_L^2)]$  with $v_L=0.2$ eV.
\label{rho2}}
}

The eight solutions for $f$ depend on the value of the phase $\rho$ in the 
interval $[0, \pi]$. The dependence of the masses of RH neutrinos on $\rho$ 
for two pairs of dual solutions
is presented in fig. \ref{rho2}. 
The figure is invariant under the transformation $\rho\leftrightarrow 
\pi-\rho$ because, in the absence of the other complex phases, this 
transformation is equivalent to taking the complex conjugate of $m$ and 
therefore of $f$, so that the masses of RH neutrinos 
remain unchanged. The dependence of $M_i$ on $\rho$ is generically weak 
(they change by less than one order of magnitude). The three RH neutrino 
masses corresponding to the solution $f_1$  are quasi-degenerate for 
$\rho=0$ (type II seesaw dominates: $f_1\approx m \approx \mathbbm{1}_3$), 
whereas for $\rho=\pi/2$ they are split by an order of magnitude. For the 
solution $\hat{f}_{1}$ (dominant type I seesaw) the mass spectrum of RH 
neutrinos is strongly hierarchical independently  of the value of $\rho$.
For the solution $f_2$, level crossing between $M_2$ and $M_3$ occurs at 
$\rho=\pi/2$.


{\bf 6.~} Next, we discuss the dependence of $f$ on
the values of the Dirac-type neutrino Yukawa couplings $y_i$. In the
previous examples we assumed the hierarchical values $y_3=1,~y_2=0.1,
~y_1=0.01$. However, as pointed out in section \ref{PS}, the flavor 
structure of the Dirac-type Yukawa couplings 
of neutrinos may be qualitatively different from that for charged fermions.

Let us keep fixed $y_3 \gg y_1$ and vary $y_2$ in between. In fig. \ref{y2} 
we plot $M_i$ as functions of  $y_2$ for the dual pairs $f_{1,3}$ and 
$\hat{f}_{1,3}$, taking for the other input parameters the  same values as 
in example ${\bf 1}$. The left side of the figure corresponds to the limit 
$y_2=y_1=0.01$, the right side to the limit $y_2=y_3=1$. The central value 
$y_2=0.1$ corresponds to example ${\bf 1}$ (more precisely, to the dual pairs 
$f_{1,3}$ and $\hat{f}_{1,3}$ in eq. (\ref{numer}), that is, to the value 
$m_1=0.005$ eV in fig. \ref{NHm1}). One can see that for the solutions 
$f_{1,3}$ the variations of $M_i$ with $y_2$ are small; this is because 
type I contribution proportional to $y_2$ is subdominant.
In contrast to this, for the solutions $\hat{f}_{1,3}$ the two lightest RH 
neutrino masses decrease significantly with decreasing $y_2$.
For the solution $\hat{f}_1$ a level crossing occurs at $y_2\sim 0.3$.

\FIGURE[t]{
\includegraphics[width=7.4cm]{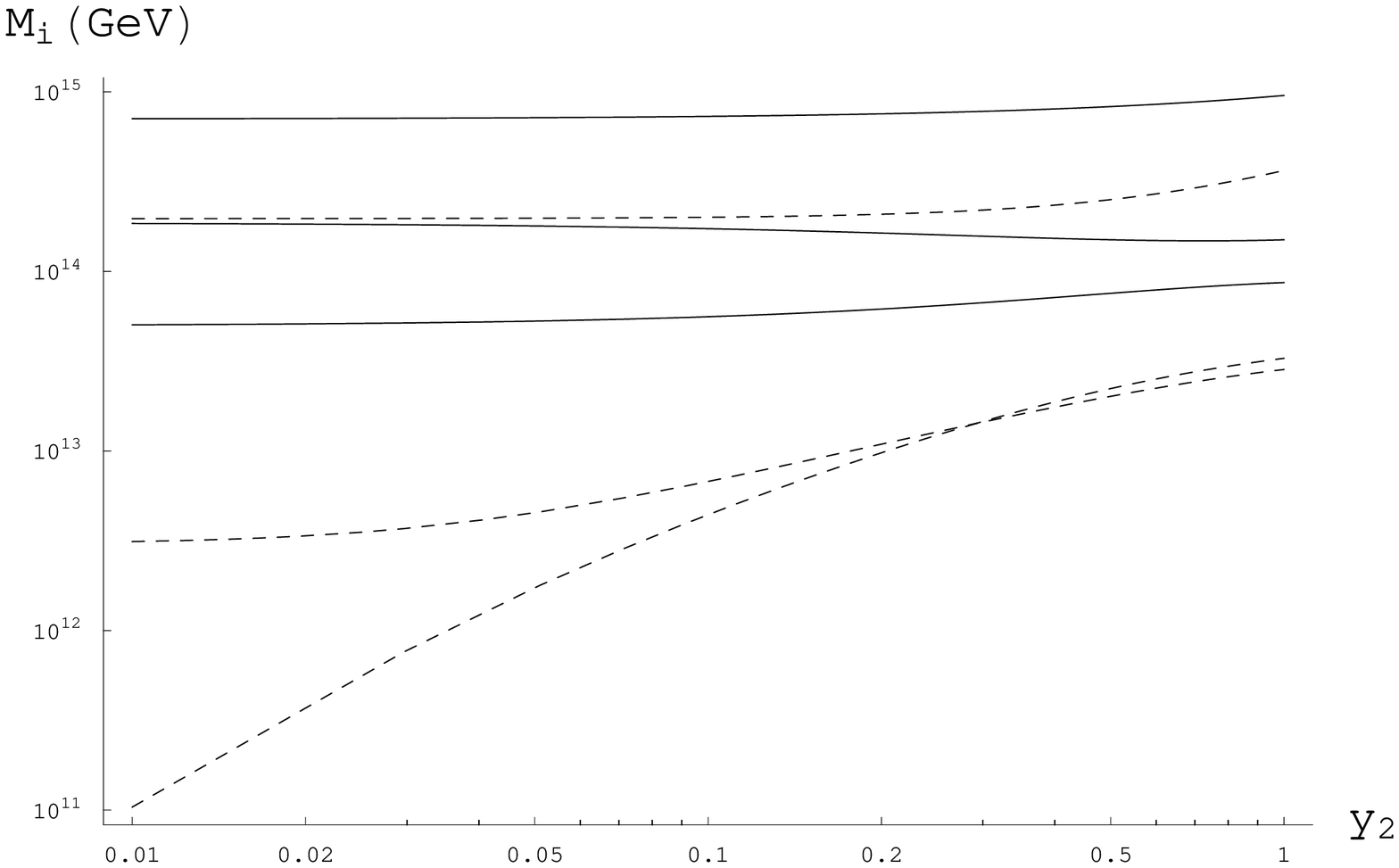} 
\includegraphics[width=7.4cm]{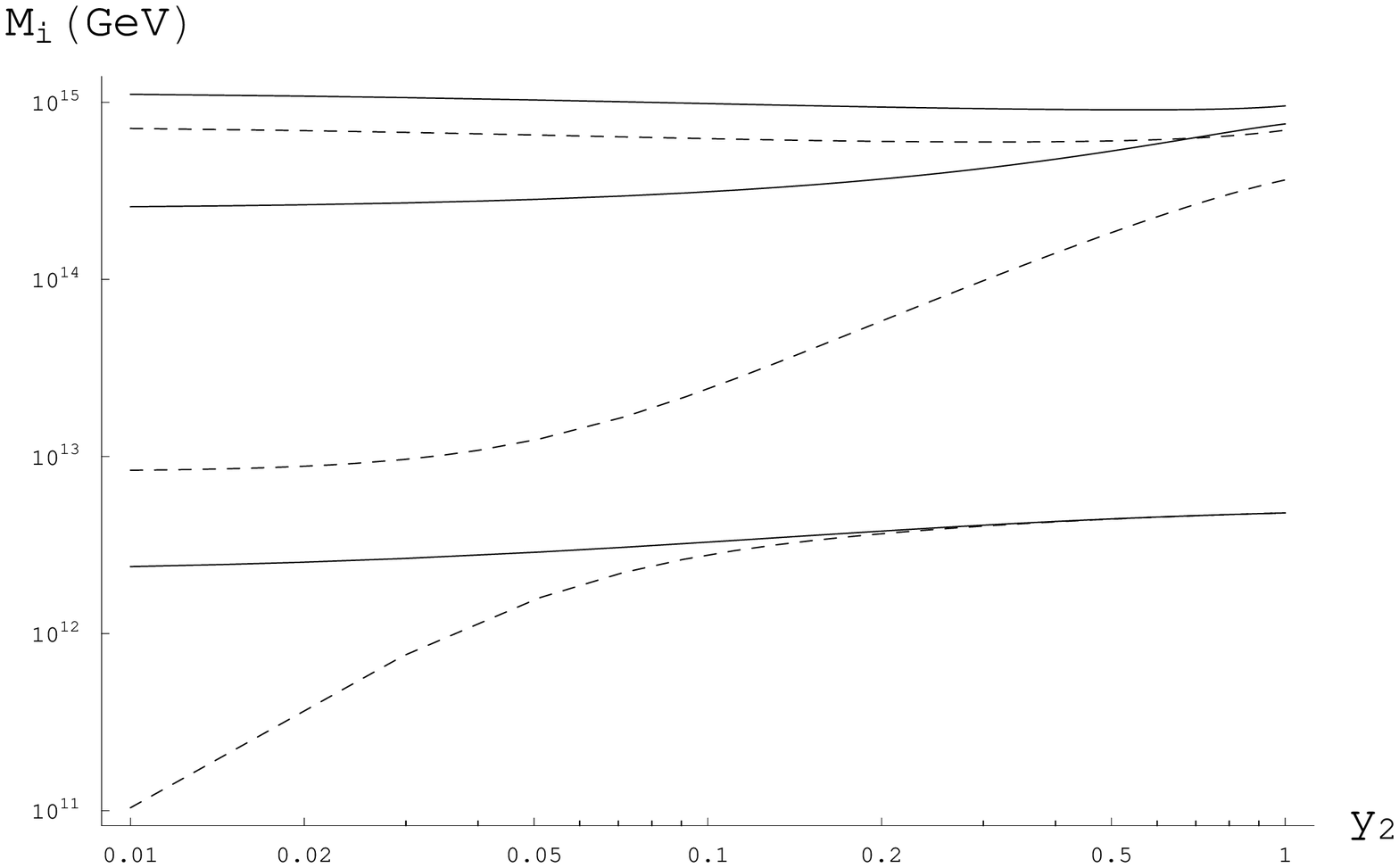} 
\caption{The masses $M_{1,2,3}$ of the three RH neutrinos as functions of 
$y_2$. In the left (right) panel the solid curves correspond to the 
solution $f_1$ ($f_3$) and the dashed curves to its dual solution 
$\hat{f}_1$ ($\hat{f}_3$). We chose the same input parameters as in
fig. \ref{NHm1}, but fixed $m_1=0.005$ eV and varied $y_2$ between $y_1=0.01$ 
and $y_3=1$.
\label{y2}}
}

Let us focus on the special case $y_2=y_3$, which may be motivated by the 
experimental observation $\nu_3 \approx (\nu_\mu+\nu_\tau)/\sqrt{2}$, that 
is, $\theta_{13}\approx 0$ and $\theta_{23}\approx \pi/4$.
For illustration, we show the structures of $f_1$ and $\hat{f}_1$: 
\beq\bea{ll}
f_1\approx\left(\bea{ccc} -0.0003 & 0.14 & -0.14 \\ \dots & 0.86 & 0.76 \\ 
\dots & \dots &  0.86
\eea\right) \,,~~~~ &
\hat{f}_1\approx\left(\bea{ccc} 0.0003 & -0.04 & 0.04 \\ \dots & -0.31 & 
-0.31 \\ \dots & \dots &  -0.31 
\eea\right)\,.
\eea\label{numy}\eeq
In this as well as in the other three dual pairs, one finds $f_{12}=-f_{13}$ 
and $f_{22}=f_{33}$. This is a consequence of the 
choice $m_{e\mu}=-m_{e\tau}$, $m_{\mu\mu}=m_{\tau\tau}$ and $y_2=y_3$, as can 
be directly verified using eq. (\ref{solone}). Moreover, this ``2-3 symmetry'' 
also connects the dual pairs of solutions to each other. This can be best  
seen by considering the seesaw formula after the maximal 2-3 rotation:
\beq\bea{cc}
\left(\bea{ccc} 
m_{ee} & \sqrt{2}m_{e\mu} & 0 \\
\dots & m_{\mu\mu}-m_{\mu\tau} & 0 \\
\dots & \dots & m_{\mu\mu}+m_{\mu\tau} 
\eea\right) = \left(\bea{ccc}
f_{11} & \sqrt{2}f_{12} & 0 \\
\dots & f_{22}-f_{23} & 0 \\
\dots & \dots & f_{22}+f_{23} 
\eea\right) \\ \\
- \dfrac 1x \left(\bea{ccc}
y_1 & 0 & 0 \\ \dots & y_2 & 0 \\ \dots & \dots & y_2 
\eea\right)
\left(\bea{ccc}
f_{11} & \sqrt{2}f_{12} & 0 \\
\dots & f_{22}-f_{23} & 0 \\
\dots & \dots & f_{22}+f_{23} 
\eea\right)^{-1} 
\left(\bea{ccc}
y_1 & 0 & 0 \\ \dots & y_2 & 0 \\ \dots & \dots & y_2 
\eea\right)\,.
\eea\eeq
The equation for the $33$-entry decouples from the rest of the system and is 
quadratic in $f_{22}+f_{23}$. One its root is the third eigenvalue common 
to four solutions for $f$. The other root is the third eigenvalue common 
to the four duals. The equation for the 1-2 block has the structure of the LR 
symmetric seesaw in the case of two generations. As a consequence, there are 
four solutions for $f_{11}$, $f_{12}$ and $f_{22}-f_{23}$. Each of them can be 
combined with one or the other solution for $f_{22}+f_{33}$, giving two 
solutions which have the same first and second eigenvalues.
Note that, despite the presence of this ``2-3  symmetry'' in the neutrino 
sector in the considered example, 
charged leptons break this symmetry badly because $m_\mu\ne m_\tau$.


{\bf 7.} Finally, let us examine the dependence of the masses of the RH 
neutrinos on the dominance parameter $d$ defined in eq. (\ref{dom2}). For the 
one- and two-generation cases we illustrated this dependence in figs. 
\ref{Heavymasses} and \ref{2gMR}, respectively. We now fix all the input 
parameters as in example ${\bf 1}$, but allow $x$ to be different from 1. 
This gives $d\approx 13/x \approx 8\cdot 10^{15}$ GeV$/v_R$. Notice that 
eqs. (\ref{solone}) and (\ref{final}) imply that the dependence of $f$ on 
$x$ amounts to a common rescaling of the three Dirac-type Yukawa couplings: 
$f(x,y_i)=f(1,y_i/\sqrt{x})$.

\FIGURE[t]{
\includegraphics[width=7.4cm]{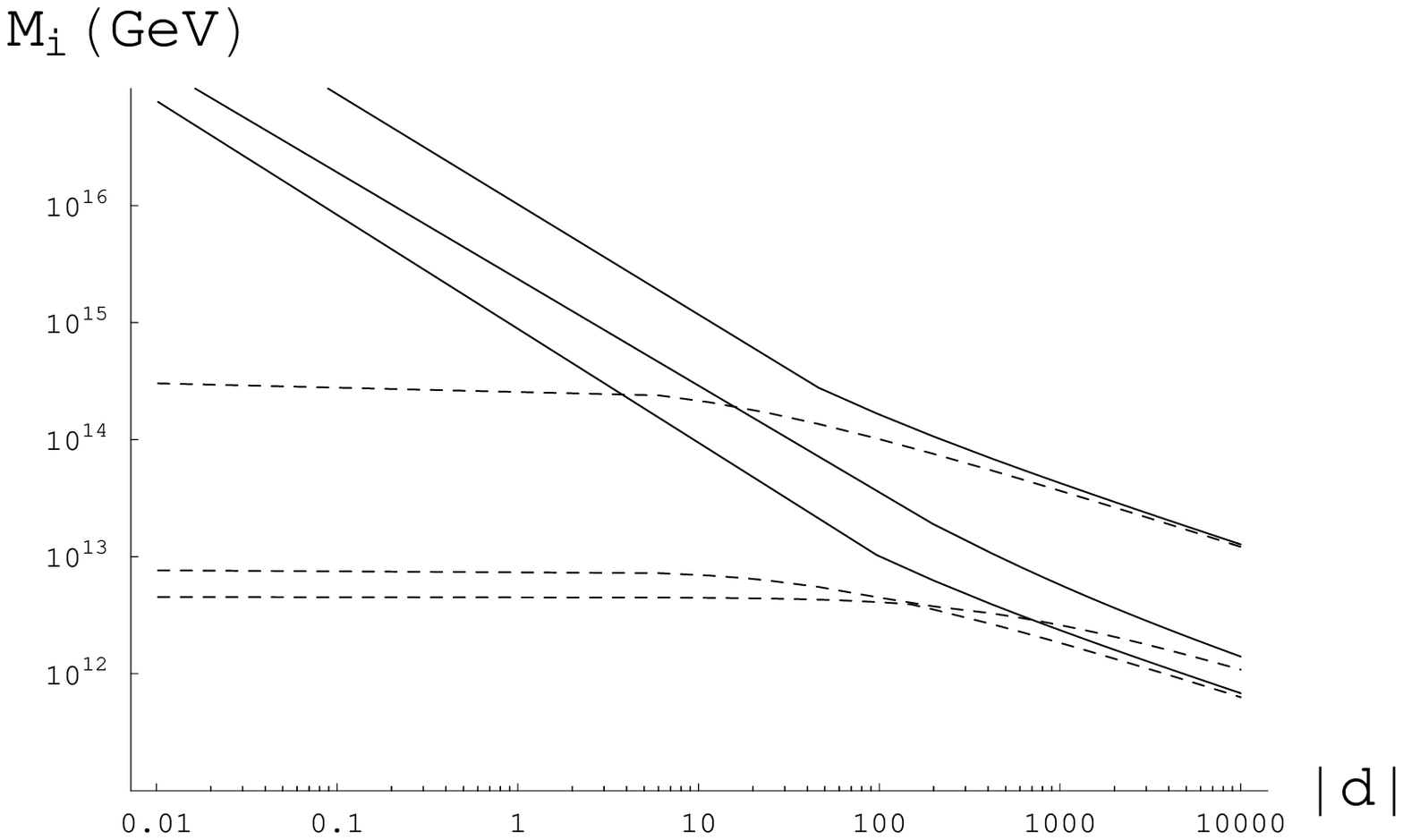} 
\includegraphics[width=7.4cm]{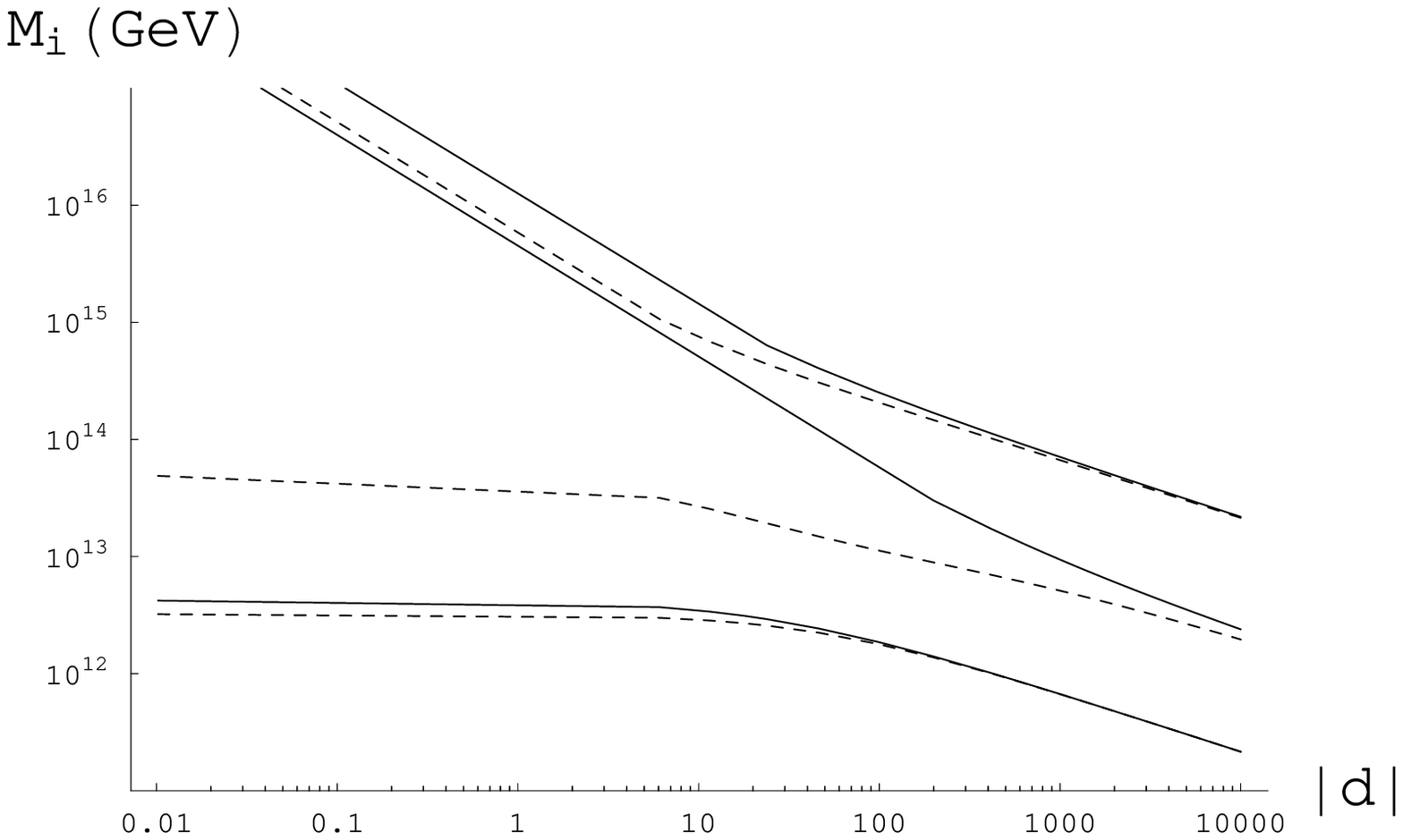} 
\caption{The masses $M_{1,2,3}$ of the three RH neutrinos as functions of 
the dominance parameter $|d|\equiv |4v^2y_3^2/(m_\nu)_{\tau\tau}^2|v_L/v_R$.
We chose the same input parameters as in fig. \ref{NHm1}, but fixed  
$m_1=0.005$ eV and allowed $v_R$ to vary, so that $d = 7.8\cdot 10^{15}$ 
GeV$/v_R$. In the left (right) panel, the solid curves correspond to the 
solution $f_1$ ($f_3$) and the dashed curves to its dual  $\hat{f}_1$ 
($\hat{f}_3$).
\label{d3RH}}
}

In fig. \ref{d3RH} we plot the masses $M_i$ of the three RH neutrinos  
as functions of $d$ for the dual pairs of solutions $f_{1,3}$ and 
$\hat{f}_{1,3}$. Since $M_i$ are given by the eigenvalues of $v_R f$, 
they generally decrease as $v_R\propto1/d$ with increasing $d$.  However, 
this is not the case when type I seesaw dominates, since in this case 
$f\propto 1/v_R$ and $M_i$ tend to constant values, as it was already 
shown in \cite{HLS}. Such type I seesaw dominance 
is realized for $d\ll 1$ for the solution $\hat{f}_{1}$ (left panel in 
fig. \ref{d3RH}). Correspondingly, type II seesaw dominates in $f_1$, so 
that $M_i$ grow linearly with $v_R$ and the mass spectrum of the RH 
neutrinos  is proportional to that of light neutrinos. As discussed at the 
end of section \ref{3g} (see also Appendix \ref{dominance}), the other three 
dual pairs exhibit intermediate features for $d\ll1$, with only $M_1$ ($M_1$ 
and $M_2$) tending to a constant for the solutions $f_{2,3,4}$ 
($\hat{f}_{2,3,4}$), as shown in the right panel of fig. \ref{d3RH}. Large 
$d$ corresponds, instead, to a cancellation between type I and II seesaw 
contributions to $m_\nu$, that is, $f_i\approx - \hat{f}_i$, so that the 
mass spectra of the RH neutrinos for each pair of dual solutions
tend to coincide, as can be seen on the right sides of the plots in
fig. \ref{d3RH}. 
Further discussion of the dependence of the mass spectrum of RH neutrinos on 
$v_R$ can be found in \cite{HLS}.

\subsection{Right-handed neutrinos at TeV scale}

The seesaw mechanism is not accessible to direct experimental tests unless the 
mass scale of the seesaw particles is as low as $\sim$ TeV. Therefore, from the 
phenomenological point of view, it is interesting to investigate what regions 
in the space of the input parameters could lead to new particles with masses at
this scale. In this section we present a number of examples where such a 
scenario is realized.

(1) One vanishing or very small Dirac-type neutrino Yukawa coupling ($y_1
\rightarrow 0$). This limit was studied analytically in section \ref{y0}, 
where we showed that in this case four of the solutions for $f$ have one 
eigenvalue vanishing as $y_1^2$. We illustrate this phenomenon numerically in  
fig. \ref{tiny1}, where $y_3$ and $y_2$ have fixed values (1 and 0.01, 
respectively), whereas $y_1$ is varied. As expected, in each pair of dual 
solutions there is one with $M_1$ decreasing as $y_1^2$. The other RH 
neutrino masses are practically independent of $y_1$ when it becomes smaller 
than $\sim 10^{-4}$. Notice that for $y_1 \sim 10^{-6}$ the mass of the 
lightest RH neutrino $N_1$ is already as small as $\sim 10$ TeV. However, its 
Yukawa couplings to the SM leptons are also suppressed as $y_1$ and therefore 
there is no hope to detect $N_1$ through these couplings. For yet smaller 
values of $y_1$, $N_1$ could be in principle so light as to play a role 
in cosmology or in active-sterile neutrino oscillations.

\FIGURE[t]{ 
\includegraphics[width=7.4cm]{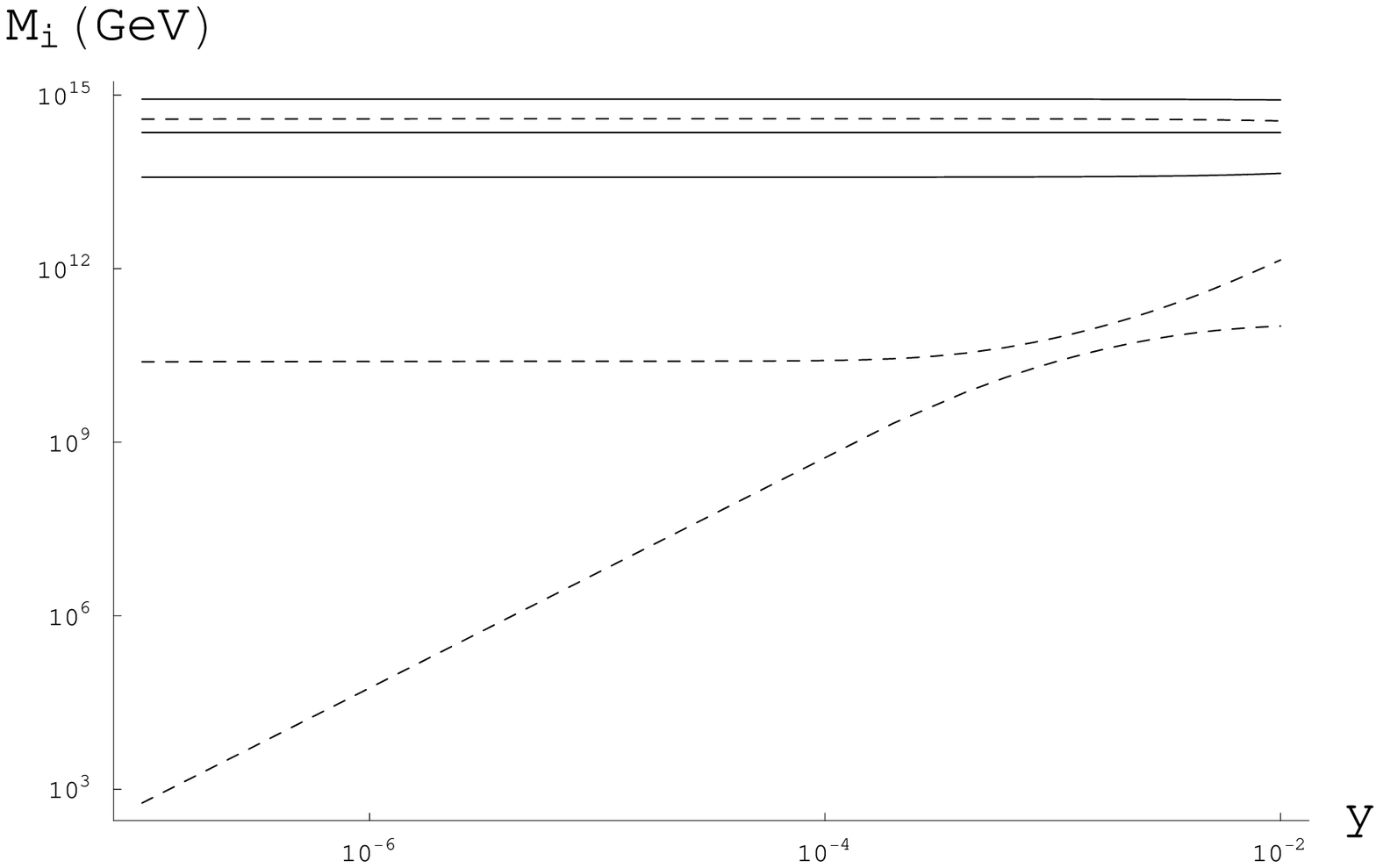} 
\includegraphics[width=7.4cm]{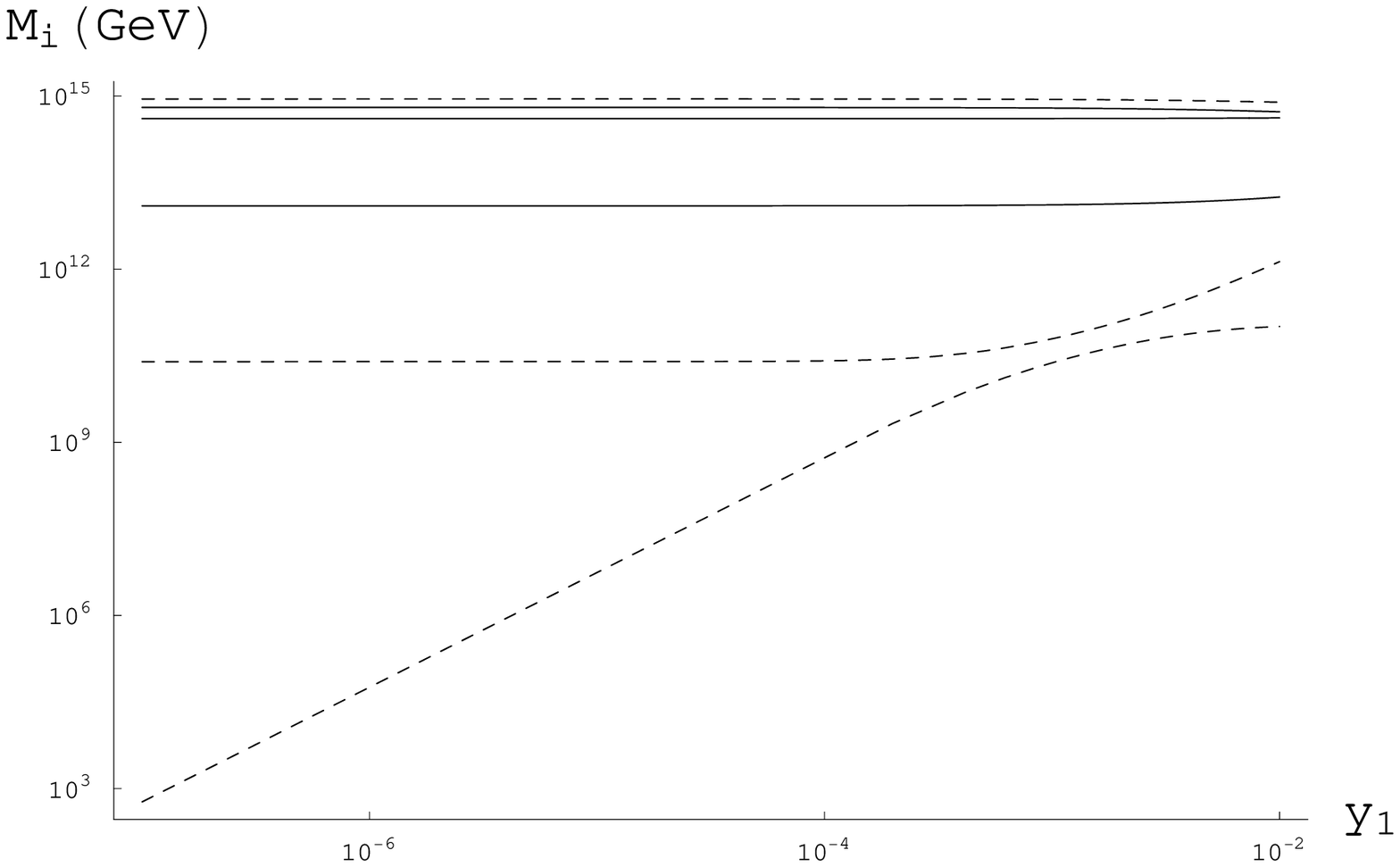} \caption{The masses $M_{1,2,3}$ 
of the three RH neutrinos as functions of $y_1$ for two dual pairs of 
solutions. We chose $y_2=0.01$, $m_1=0.005$ eV, $\sin\theta_{13}=0.1$, 
while the other input parameters are the same as in fig. \ref{NHm1}. In 
both panels dashed curves correspond to the solution with $M_1$ vanishing in 
the limit $y_1\to 0$ and solid curves to the dual solution. \label{tiny1}} 
}

(2) Lowering of the LR symmetry breaking scale $v_R$ down to 
$\sim$ TeV. This is allowed by direct searches of RH currents, as discussed 
in section \ref{PS}. 
Such a scenario is not viable for generic (``natural'') values of the Yukawa 
couplings, since it would result in the masses of light neutrinos which are 
much larger than the observed ones. Nonetheless, in view of the unique 
phenomenological possibility to directly test the breaking of the LR 
symmetry and the seesaw mechanism, it is worth  accepting the necessary 
tuning of Yukawa couplings and investigate its consequences.

The simplest phenomenologically viable scenario with $v_R\sim$TeV 
corresponds to choosing tiny values for all the Dirac-type neutrino Yukawa 
couplings $y_i$. This allows one to avoid conflict with the low-energy 
neutrino data while maintaining the perturbative unitarity of the solutions 
for $f$, $f_{ij}\lesssim 1$. Notice that the seesaw formula 
is invariant 
with respect to a simultaneous decrease of $y_i$ and $x$ provided that 
$y_i^2/x$ remains unchanged. Consider the same choice of the input parameters 
as in example ${\bf 1}$ of section \ref{numex}, except that the values $x=1$, 
$y_3=1$ are replaced with $x=10^{-10}$, $y_3=10^{-5}$, keeping $v_L$ and the 
hierarchy between $y_i$ unchanged. Then the solutions for $f$ are unchanged, 
but since $v_R \approx 6\cdot10^4$ GeV in this case, the masses of RH 
neutrinos are smaller than those in example ${\bf 1}$ of section \ref{numex} 
by a factor $10^{10}$. As one can see in fig. \ref{NHm1}, this means that the 
lightest RH neutrino can have a mass $\sim 100$ GeV. Such light RH neutrinos 
would not be observable through their Dirac-type Yukawa couplings, which are 
tiny; however, because of the LR mixing, they may interact strongly 
with the ordinary matter through the Yukawa couplings $f$ with scalar triplets 
as well as through their coupling to RH gauge bosons. More generally, the RH  
gauge bosons as well as the non-standard Higgs particles related to the LR 
symmetry breaking would provide clear phenomenological signatures if light 
enough. If level crossing occurs between two TeV-scale RH neutrinos, they 
could still lead to a successful leptogenesis via the resonant enhancement of 
the produced lepton asymmetry (see section \ref{lept}).

(3) It should be noted that very small values of $x$, as in scenario 2 above,  
would indicate an approximate symmetry of the scalar potential, since for a 
generic choice of order one couplings one expects $v_L v_R\sim v^2$, that is 
$x\sim 1$. Very interestingly, it is possible to realize a TeV scale LR 
symmetry breaking even preserving this ``naturalness'' relation. Indeed, 
consider a modification of scenario 2 in which $v_L$ is increased by a large 
factor $W$ and, at the same time, all $y_i^2$ are reduced by $W$. Then the 
solutions $f$ of the seesaw formula maintain the same matrix structure, but 
are also reduced by an overall factor $W$. This means, in particular, that the 
RH neutrino masses are much smaller than $v_R$. Using the values of the 
parameters above and taking $W=10^8$, one finds $M_1\sim$ keV. Summarizing, if 
LR symmetry breaking occurs at scales much lower than the Grand Unification 
scale and the natural relation $v_L v_R \sim v^2$ is preserved, this would 
indicate that the Dirac-type and triplet Yukawa couplings $y$ and $f$ are all 
much smaller than one, with the related consequences for phenomenology.

(4) Finally, let us investigate if TeV scale RH neutrinos may be compatible 
with at least some order one Dirac-type Yukawa couplings and $v_R\sim$ 
TeV. This would require a strong fine-tuning in the structure of the Yukawa 
coupling matrices, in order to cancel large contributions to the masses of 
light neutrinos. To provide a simple example, it is convenient to start in 
the basis where the light neutrino mass matrix is diagonal, $m_{diag} \equiv 
(m_\nu)_{diag}/v_L = diag(m_1,m_2,m_3)$. Let us define $r^2 \equiv m_1/m_2$ 
and choose 
\beq 
y=\left(\begin{array}{ccc} r^2 s & \,irs & \,0 \\ irs 
& {}-s & 0 \\ \,0 & \,0 & \,y_3 \end{array}\right). 
\label{y}
\eeq 
In this case the seesaw equation for $f$ has only two solutions, the reason 
being as follows. The matrix $y$ in eq.~(\ref{y}) has one zero eigenvalue, 
which reduces the number of solutions for $f$ from eight to four (section 
\ref{y0}). Furthermore, in the basis where $y$ is diagonal, one has $m_{ee}=0$, 
which makes two of the four remaining solutions singular and thus unphysical 
(see eq.~(\ref{lam32})). The leftover two solutions are 
\beq 
f_{1,2}=diag\left(m_1,~m_2,~\frac{m_3}{2} \left[1 \pm 
\sqrt{1+\frac{4y_3^2}{xm_3^2}} \right]\right). 
\label{2sol}
\eeq 
Notice that the entries of $y$ proportional to $s$ cancel exactly in the 
seesaw formula and therefore do not contribute to the light neutrino masses. 
As a consequence, the coupling $s$ can be as large as one. In contrast to this, 
$y_3^2/x$ has to be smaller than one to guarantee $f_{33} \lesssim 1$. To 
ensure perturbative unitarity of $(f_{1,2})_{ii}$, $v_L$ should be $\,\lesssim$ 
eV; therefore a low LR scale $v_R \sim$ TeV requires $y_3\lesssim 10^{-5}$.

To consider the structure of the seesaw mechanism in this scenario in the 
flavor basis, it is sufficient to rotate on the left and on the right by the 
leptonic mixing matrix $U$:
\beq 
m \rightarrow U^* m_{diag} U^\dag ~,~~~~ y \rightarrow U^* y U^\dag\,,
~~~~ f_{1,2} \rightarrow U^* f_{1,2} U^\dag\,. 
\label{f12}
\eeq 
In this flavor basis, type I contributions to the elements of the light 
neutrino mass matrix $m_{\alpha\beta}$ are proportional to $U^*_{\alpha 3}
U^*_{\beta 3} y_3^2/x$.

The masses of RH neutrinos are given by the eigenvalues of $f_{1,2}$ (i.e. the 
diagonal entries in eq.~(\ref{f12})) multiplied by $v_R \sim$ TeV. The two 
possible solutions for $f$ are distinguished by the mass of the third RH 
neutrino, while $M_{1,2}$ are uniquely determined, with $M^2_2-M^2_1 = 
\Delta m^2_{sol} v_R/v_L$. Two of the RH neutrinos ($N_{1,2}$) have order one 
Yukawa couplings to the light neutrinos and charged leptons.

\section{Structure of seesaw in specific left-right symmetric models
 \label{models}}

In this section, we discuss the basic features of models which incorporate 
type I+II seesaw mechanism. In particular, we analyze their Yukawa sector 
and derive the seesaw formula for the light neutrino mass in the form given 
in eq. (\ref{master}). A crucial issue is the structure of the matrix of 
Dirac-type Yukawa couplings of neutrinos  $y$, which is an input in the 
bottom-up approach we adopted to reconstruct $f$.

\subsection{Minimal LR symmetric model}

We begin by reviewing the structure of the minimal LR model based
on the gauge group $SU(3)_c\times
SU(2)_L \times SU(2)_R\times U(1)_{B-L}$.
It contains the following (color singlet) Higgs multiplets: $\Phi(2,2,0)
$, $\Delta_L (3,1,-2)$ and $\Delta_R(1,3,2)$, where in the brackets 
we indicate the $SU(2)_L\times SU(2)_R \times U(1)_{B-L}$ quantum numbers. 
The leptons are assigned to $L(2,1,-1) = (\nu_L~ l_L)^T$ and $L^c(1,2,1) = 
(N_L^c~l_L^c)^T$, and their Yukawa couplings can be written as  
\beq 
-{\cal L}_Y = \frac {f_L}{2} L^T i\sigma_2 C \Delta_L L +
\frac{f_R}{2} L^{cT} i\sigma_2 C \Delta_R L^c + g L^T C \sigma_2 \Phi
\sigma_2 L^c + h L^T C \Phi^* L^c + {h.c.}\,, 
\label{Lyuk}
\eeq 
where
$\sigma_2$ is the isospin Pauli matrix and the following conventions are 
assumed:
\beq 
\Delta_L=\left(\bea{cc}
\Delta^+_L/\sqrt{2} & \Delta_L^{++} \\ \Delta_L^0 & -\Delta^+_L/\sqrt{2}
\eea\right) \,,~~ \Delta_R=\left(\bea{cc} \Delta^-_R/\sqrt{2} &
\Delta_R^{--} \\ 
\Delta_R^0 & -\Delta^-_R/\sqrt{2} \eea\right) \,,~~
\Phi=\left(\bea{cc} \phi_1^0 & \phi_1^+ \\ \phi_2^- & \phi_2^0
\eea\right) \,. 
\eeq 
Denoting the VEVs of the neutral components of $\Phi$ as $v_{1,2}$, for the 
Dirac-type mass matrices of the charged leptons and neutrinos one finds 
\beq 
m_l \equiv v y_e 
=v_1 g + v_2^* h \,,~~~~~~~ m_{D} 
\equiv v y 
= v_1^* h + v_2 g \,.
\label{dirac}
\eeq 
At low energies the Dirac-type neutrino Yukawa coupling can be rewritten 
in the form of eq.~(\ref{yuk}) by defining $\phi^0 \equiv (v_1\phi_1^{0*} + 
v_2^* \phi_2^0)/v$, 
where $v\equiv \sqrt{|v_1|^2 +|v_2|^2} \approx 174~{\rm GeV}$ is 
the electroweak symmetry breaking parameter.\footnote{
The field $\phi^0$ is thus identified with the neutral component of 
the SM Higgs boson. The orthogonal combination of the neutral scalar fields, 
$\eta^0 \equiv (-v_2\phi_1^{0*} + v_1^* \phi_2^0)/v$, has zero VEV and so does 
not contribute to the fermion masses. Since it mediates flavour changing 
neutral currents, it must be heavy and so decouples from the low-energy 
dynamics.}
Notice that if the theory is supersymmetrized, the coupling $h$ should be 
removed from eq. (\ref{Lyuk}) due to the requirement of the analyticity of 
the superpotential. As a consequence, $y=y_e(v_2/v_1) \equiv y_e\tan\beta$ and 
the two Higgs doublets, $\phi_u\equiv(\phi_1^+~\phi_2^0)^T$ and $\phi_d
\equiv(\phi_1^0~\phi_2^-)^T$, do not mix.\footnote{
Similarly to eq. (\ref{dirac}), for quarks one has $m_d \equiv v y_d = 
v_1 g_q + v_2^* h_q$ and $m_u \equiv v y_u = v_1^* h_q + v_2 g_q$, where 
the matrices $g_q$ and $h_q$ are in general independent of $g$ and $h$,  
and $h_q$ is absent in the SUSY case.}

The Majorana mass matrices of neutrinos are generated by the VEVs $v_
{L,R}$ 
of the neutral components of $\Delta_{L,R}$: 
\beq m_{L} = v_L f_L \,,~~~~~ M_R =v_R f_R \,. 
\eeq 
In the limit where the eigenvalues of of $M_R$ are much larger than 
those 
of $m_{D}$, the mass matrix of light neutrinos  takes the form 
\beq
m_\nu \simeq m_L - m_{D} M_R^{-1} m_{D}^T = v_L f_L -
\dfrac{v^2}{v_R} y f_R^{-1} y^T \,, 
\label{ss1}
\eeq 
which is known as type I+II seesaw formula.

At this point, a discrete LR symmetry should be introduced if one wants to 
guarantee the equality of $SU(2)_L$ and $SU(2)_R$ gauge couplings, 
that is, the asymptotic LR symmetry of the model. There are essentially two 
possibilities: the discrete transformation can act as a charge conjugation 
or as a parity transformation. In the previous sections we were implicitly 
assuming the former possibility (we postpone the discussion of the latter one 
to Appendix \ref{star}). Then, the discrete LR symmetry acts on matter and 
Higgs fields as follows: 
\beq 
L\leftrightarrow L^{c} \,,~~~~~ \Phi \leftrightarrow \Phi^T
\,,~~~~~ \Delta_L\leftrightarrow \Delta_R \,. 
\label{Dp}
\eeq 
This yields 
\beq 
f\equiv f_L = f_R \,,~~~~~ g=g^T \,,~~~~~ h= h^T \,.
\label{symm}
\eeq 
In this case the mass matrix of light neutrinos has the
form given in eq. (\ref{master}). 
Moreover, eq.~(\ref{symm}) implies that $y$ is symmetric, so that the seesaw
duality is realized. (If, instead, the model contained a Higgs bidoublet 
$\Phi'$ with the transformation law $\Phi'\leftrightarrow -\Phi'^T$ under the 
discrete symmetry, the corresponding Dirac-type Yukawa coupling $y'$ would be 
antisymmetric, leading to the seesaw duality as well).

We have seen that in the minimal SUSY LR model the matrix $y$ is proportional 
to the Yukawa coupling matrix of charged leptons $y_e$. Therefore, in the  
basis where $y_e$ is diagonal, also $y$ is, and the ratios of its 
eigenvalues are $m_e/m_\mu$ and $m_\mu/m_\tau$. This scenario is ideal for 
the reconstruction of $f$, since the input related to the structure of $y$ 
is completely determined. However, one should keep in mind that this minimal 
model does not provide a realistic description of the quark sector, as it  
leads to $m_u\propto m_d$. As follows from eq.~(\ref{dirac}), in the non-SUSY 
case, the structure of $y$ can be different from $y_e$. For a recent 
realistic model, see \cite{kiers}. 

Let us briefly consider an upgrade of the minimal LR model to the
Pati-Salam gauge group $SU(2)_L\times SU(2)_R \times SU(4)_c$, with a 
Higgs field in the $(2,2,1)$ representation. 
In this case quark-lepton unification occurs, which for the Yukawa sector 
amounts to identifying the quark and lepton coupling matrices:  
$g_q=g$ and $h_q=h$. This implies, in particular, $y=y_u$ and $y_e=y_d$. 
In this case the eigenvalues of $y$ are determined by the up-type quark masses, 
while the mixing is given by the CKM matrix. In non-minimal Pati-Salam models, 
also Higgs fields transforming as $(2,2,15)$ may couple to fermions. Their 
contribution to the lepton masses differs from those for the quarks by the 
factor $-3$ because of the different respective values of $(B-L)$. If both 
bidoublet types exist and contribute significantly to the fermion masses, 
the direct connection between $y$ and $y_u$ is lost.

\subsection{$SO(10)$ models}

A natural grand unified embedding of LR symmetric models is provided by
$SO(10)$ models. As is well-known, $SO(10)$ accommodates all the standard 
model fermions as well as RH neutrinos in a single  
multiplet ${\bf 16}_F$. 
Fermion bilinears transform under $SO(10)$ as 
\begin{equation}
{\bf 16}_F \times {\bf 16}_F = {\bf 10}_s + {\bf 120}_a + {\bf 126}_s \,,
\end{equation}
where the subscript $s(a)$ indicates symmetry (antisymmetry) in the flavor 
indexes. As a consequence, 
there are three possible types of Higgs multiplets which may contribute 
to fermion masses at the renormalizable level: ${\bf 10}_H$ and 
$\overline{\bf 126}_H$ give symmetric contributions, while ${\bf 120}_H$ 
gives an antisymmetric one. 

The Higgs triplets $\Delta_{L,R}$ are both contained in 
${\overline{\bf 126}}_H $, so that the Majorana mass matrices of 
left-handed and RH neutrinos originate from the coupling
\begin{equation}
f~{\bf 16}_F{\bf 16}_F\overline{\bf 126}_H ~\ni~ f(LL\Delta_L+L^cL^c
\Delta_R) \,.
\label{f10}\end{equation}
One generator of $SO(10)$, known as D-parity, acts as in eq. (\ref{Dp}).
In other words, the  discrete LR symmetry discussed above is promoted to an 
automatic gauge symmetry and this  guarantees the proportionality of the two 
terms on the r.h.s. of eq.~(\ref{f10}). 
The Dirac-type neutrino Yukawa coupling $y$ may receive contributions from 
the Higgs multiplets of all three types, provided that they acquire 
VEVs in the appropriate direction. In fact, both ${\bf 10}_H$ and ${\bf 
120}_H$ contain one bidoublet $\Phi\sim(2,2,1)$ under the Pati-Salam group, 
and both ${\bf 120}_H$ and ${\overline{\bf 126}}_H$ contain one $(2,2,15)$ 
bidoublet. If only ${\bf 10}_H$ and ${\overline{\bf 126}}_H$ contribute 
to $y$, it is symmetric; if only ${\bf 120}_H$ does, it is antisymmetric.

Consider first the minimal possibility where only one $\overline{\bf 126}_H$ 
multiplet develops a VEV in the $\nu_L N^c_L$ direction, that is, $y\propto 
f$. In this case both type I and type II seesaw contributions to $m_\nu$ 
are proportional to $f$. The latter is uniquely determined by $m_\nu$ up to 
an overall factor. 
There is no seesaw duality in that case, since $y$ is not independent 
of $f$ and so it is not invariant under $f\to m-f$. 

If two (or more) $\overline{\bf 126}_{Hi}$ multiplets are introduced, 
the proportionality between 
$m_L \equiv v_{L1} f_1 + v_{L2} f_2$ and $M_R\equiv  v_{R1} f_1 + v_{R2} f_2$ 
is in general lost. The matrix $y$ depends on the same Yukawa 
couplings $f_{1,2}$: $y = -3(v^u_{126,1} f_1 +v^u_{126,2} f_2)$,
where $v^u_{126,i}$ is the VEV of the ``up-type'' Higgs doublet (i.e. of 
the doublet with non-zero VEV in the up-type isospin direction) 
contained in the $(2,2,15)$ component of $\overline{\bf 126}_{Hi}$. 
Nonetheless, if $v_{R2}=0$, an effective seesaw duality is realized, with 
$f_1$ and $f_2$ playing the role of $f$ and $y$, respectively:
\beq
\tilde{m} =  
f_1 - \frac{1}{\tilde{x}} 
\tilde{y} f_1^{-1} \tilde{y} \,,
\label{m126}\eeq
where 
\beq
\tilde{m} \equiv \frac{m_\nu -  
\left(v_{L2}-18 v^u_{126,2}v^u_{126,1}/v_{R1}\right) f_2}
{v_{L1}- (3v^u_{126,1})^2/v_{R1}} \,,~~~~
\tilde{x}\equiv \frac{v_{L1}v_{R1}- (3v^u_{126,1})^2}{v^2}\,,~~~~
v\tilde{y} \equiv -3 v^u_{126,2} f_2 \,.
\eeq
If all the parameters in eq. (\ref{m126}) except $f_1$ were known, one would 
be able to solve (\ref{m126}) for $f_1$ using the techniques developed in 
the previous sections for solving the usual seesaw formula (\ref{master}) 
for $f$. 
For three lepton generation this would yield four dual pairs of solutions.

Consider now the  scenario with one ${\bf 10}_H$ and one $\overline{\bf 
126}_H$ multiplets. Since ${\bf 10}$ is a real representation, ${\bf 10}_H^*$ 
transform as ${\bf 10}_H$ under $SO(10)$, so that in general two new 
Yukawa couplings should be added to eq. (\ref{f10}):
\beq
g~{\bf 16}_F{\bf 16}_F{\bf 10}_H + h~{\bf 16}_F{\bf 16}_F{\bf 10}_H^* \,.
\label{gh}
\eeq 
One then finds
\beq
m_D\equiv v y = v^u_{10} g + v^{d*}_{10} h - 3 v^u_{126} f\,, 
\eeq
where $v^{u,d}_{10}$ are the VEVs of the up- and down-type Higgs doublets
contained in the $(2,2,1)$ component of ${\bf 10}_H$.
Since $y$ depends on $f$, one cannot regard it as an input for solving  
the seesaw formula for $f$. However, one can consider as an input the 
matrix $v \tilde{y}\equiv v^u_{10} g + v^{d*}_{10} h$; note that in this 
class of models the couplings to ${\bf 10}_H$ usually give the dominant 
contribution to the mass matrices of the charged fermions. 

Then, the seesaw formula (\ref{master}) can be 
written as
\begin{equation}
\tilde{m} =  f - \frac{1}{\tilde{x}}\tilde{y}f^{-1}\tilde{y}\,,~~~~~
\tilde{m}\equiv \frac{m_\nu- (6v v^u_{126}/v_R)\tilde{y}}
{v_L- (3v^u_{126})^2/v_R} \,,~~~~~ \tilde{x}\equiv
\frac{v_L v_R - (3v^u_{126})^2}{v^2} \,.
\label{10126}\end{equation}
This equation can be solved for $f$ as usual, taking $\tilde{m}$, $\tilde{y}$ 
and $\tilde{x}$ as input parameters. In some models of this class,  the GUT 
symmetry breaking is such that only the Higgs doublets in ${\bf 10}_H$ develop 
VEVs (a mixing between these doublets and those in ${\bf 126}_H$ is induced 
only in the presence of $\bf{210}_H$ multiplets).  In this case $v^u_{126}=0$, 
and one recovers  $\tilde{m}=m$, $\tilde{y}=y$ and $\tilde{x}= x$.
Moreover, one has $y=y_u$, that is, the matrix $y$ is fully determined if we 
know the Yukawa couplings of up-type quarks. If also $v^d_{126}=0$, one finds 
$y_e=y_d$, which fails to reproduce correctly the masses of the down-type 
fermions of the second and first generations. The corrections coming from the 
$\overline{\bf 126}_H$ (and therefore proportional to $v^d_{126}f$) may cure 
this problem. In this case, however, the form of $f$ is constrained not only 
by the seesaw formula, but also by the values of the masses of charged 
fermions.

A comment on the number of the Yukawa couplings to ${\bf 10}_H$ multiplets is 
in order. If one wants to study minimal models with only one such Yukawa 
coupling matrix, there are two options: (i) if only one ${\bf 10}_H$ is 
introduced and the model is supersymmetric, the coupling $h$ in eq. (\ref{gh}) 
is automatically forbidden by the analyticity of the superpotential; (ii) 
one can forbid $h$ also in the non-supersymmetric case by assuming that 
the multiplet ${\bf 10}_H$ is real. 
Then only the coupling $g$ contributes to the fermion masses, leading to 
$y=y_u\propto y_e=y_d$, which needs substantial corrections in order to 
match the differences among the mass spectra of charged leptons, up- and 
down-type quarks. This turns out to be extremely constraining, if one 
tries 
to fit all the data by adding just one $\overline{\bf 126}_H$ multiplet 
(see \cite{BMS} and references therein). If one is willing to consider models 
with one more Yukawa coupling matrix, in the SUSY case one needs to introduce 
a second ${\bf 10}_H$ multiplet (see e.g. \cite{HLS}), whereas in the 
non-SUSY case one complex ${\bf 10}_H$ is sufficient. In this framework the up 
sector is decoupled from the down sector, but the relation $y_e=y_d$ persists.

Alternatively, instead of adding ${\bf 10}_H$ to the $\overline{\bf 126}_H$ 
multiplet, one can explore the possibility where ${\bf 120}_H$ is added.
The couplings in (\ref{gh}) are then replaced by
\beq
g~{\bf 16}_F{\bf 16}_F{\bf 120}_H + h~{\bf 16}_F{\bf 16}_F{\bf 120}_H^* \,,
\label{gha}\eeq 
where the matrices $g$ and $h$ are antisymmetric. Denoting 
$v^{u(d)}_{120,1}$ the VEVs of the up (down)-type Higgs doublets contained 
in the $(2,2,1)$ component of ${\bf 120}_H$, and similarly $v^{u(d)}_{120,15}$ 
for the $(2,2,15)$ component of ${\bf 120}_H$, one finds 
\beq
m_D\equiv v y = (v^u_{120,1}-3v^u_{120,15}) g + 
(v^{d*}_{120,1}-3v^{d*}_{120,15}) h - 3 v^u_{126} f 
\equiv v\tilde{y} - 3 v^u_{126} f \,,
\eeq
where $\tilde{y}$ is antisymmetric.
Analogously to eq. (\ref{10126}), the seesaw relation takes the form
\begin{equation}
\tilde{m} =  f + \frac{1}{\tilde{x}}\tilde{y}f^{-1}\tilde{y} \,,~~~~~
\tilde{m}=\frac{m_\nu}
{v_L- (3v^u_{126})^2/v_R} \,,~~~~~ \tilde{x}=
\frac{v_L v_R - (3v^u_{126})^2}{v^2} \,.
\label{120126}\end{equation}
This equation can be solved for $f$, taking $\tilde{m}$, $\tilde{y}$ 
and $\tilde{x}$ as input parameters. Using the results of Appendix
\ref{antiY}, which apply to the case of antisymmetric $\tilde{y}$, 
one finds two solutions for $f$.

Finally, let us discuss models with both ${\bf 10}_H$ and ${\bf 120}_H$ 
multiplets added to $\overline{\bf 126}_H$ (for an early analysis of fermion 
masses in this case, see  \cite{fuku}). In the search for the minimal
realistic SUSY $SO(10)$ model, it has been recently shown \cite{aul120} that 
the mass matrices of the charged fermions can be reproduced through 
${\bf 10}_H$ and ${\bf 120}_H$ Yukawa couplings only, at least in the first 
approximation (a three generation fit presents some problems \cite{LKG}). 
In the case when the coupling $f$ to $\overline{\bf 126}_H$ is also 
introduced, satisfactory fits of charged fermion and neutrino parameters have 
been recently obtained \cite{KG}. 
In this scenario there is no seesaw duality (at least in the form 
considered here), since $y$ contains both symmetric and antisymmetric 
contributions. 
Nonetheless, the non-linearity of the seesaw formula still leads to multiple 
solutions for $f$. To take this into account, one should (i) fix
the ${\bf 10}_H$ and ${\bf 120}_H$ Yukawa couplings
by fitting (approximately) the masses of charged fermions;
(ii) use the seesaw formula to derive the different structures of $f$ that 
reproduce a given set of neutrino data; (iii) test which solutions for $f$ 
provide the small corrections needed to achieve a satisfactory fit of the
masses of charged fermions.

\section{Further considerations}

In this section we briefly discuss the following issues pertaining to our 
analysis: stability of our results with respect to the renormalization 
group evolution effects and baryogenesis via leptogenesis.

\subsection{Stability of the seesaw formula}

Up to now we were assuming that the seesaw formula (\ref{master}) describes 
accurately the mass of light neutrinos at low energy scales. However, due to 
the breaking of the discrete LR symmetry at a scale $v_{LR}$, the 
renormalization group (RG) evolution effects below this scale can 
result in a violation of the conditions in eq.~(\ref{symm}), which in turn 
would modify the seesaw formula (\ref{master}). A full study of the RG  
effects in the type I+II LR symmetric seesaw is beyond the scope of this 
paper; 
here we constrain ourselves to a classification of possible effects and
estimate of their size.

Above the scale $v_{LR}$ the discrete LR symmetry ensures $f_L=f_R$ and 
$y=y^T$. Below this scale, the fields related by the LR symmetry acquire in 
general different masses, so that $f_L$ and $f_R$ may evolve differently and 
$y$ may get asymmetric corrections. For example, if $v_{LR}$ is larger 
than $v_R$ (the scale of $SU(2)_R$ breaking), logarithmic in $v_R/v_{LR}$ 
radiative corrections arise.\footnote{
In $SO(10)$ models, the discrete LR symmetry can be actually broken 
already at Grand Unification scale, even  if $SU(2)_R$ is not 
\cite{CMP}.}
The size of this effect depends on the details of the mass spectra of RH 
gauge bosons and Higgs particles responsible for the breaking of $SU(2)_R$, 
which are highly model-dependent. We will therefore not discuss this case 
any further and will just assume that the conditions $f_L=f_R$ and $y^T=y$ 
still hold at the scale $v_R=v_{LR}$.

At this point the fate of the seesaw formula resides with the particle 
spectrum below $v_R$. If the masses of all three RH neutrinos as well as of 
$\Delta_L$ are close to $v_R$, they can be integrated out all together,  
leading to the low-energy effective neutrino mass matrix $m_\nu$ of 
eq.~(\ref{master}). If, on the contrary, one or more RH neutrinos and/or 
$\Delta_L$ have masses that are much smaller than $v_R$, they will contribute 
to the RG evolution of $f_L$, $f_R$ and $y$ down to their mass scale. 
Generically, these corrections induce a splitting between $f_L$ and $f_R$ and 
an asymmetry in $y$. One could envisage two possible approaches to determine 
the effective mass matrix $m_\nu$ at the lightest seesaw scale $M_s$ (the 
smallest among $M_{1}$ and $M_{\Delta_L}$). One option is to evolve 
$f_{L,R}$ and $y$ from $v_R$ down to $M_s$ and then integrate out the RH 
neutrinos and $\Delta_L$ all together.  
This has the advantage of determining $f_{L,R}$ and $y$ as they enter in the 
seesaw formula (\ref{ss1}). Another option is to integrate out $N_{Ri}$ and 
$\Delta_L$ each at its own mass scale, redefining iteratively the effective 
$m_\nu$. Given the $\beta$-functions to some finite order in perturbation 
theory, this second approach should provide a more accurate result.

For type I seesaw models, a detailed analysis of RG effects, including the 
running between the different mass scales of RH neutrinos, has been performed 
in \cite{lindner2}. 
For type I+II seesaw, such an 
analysis is not yet available.  
Here we just estimate the order of magnitude of possible effects.  One-loop 
corrections to the matrix elements of $f$ and $y$ can be schematically 
written as
\beq
(\delta f,\delta y) \sim \frac{(f^2,y^2)}{16\pi^2}
\log\frac{(M_i,M_{\Delta_L})}{v_R}\,,
\eeq
which, for judicious choices of parameters, should be at or below the percent 
level. For comparison, the present precision of most of the input parameters 
in $m_\nu$ is at about $10\%$. 
Moreover, the loops induced by $N_{Ri}$ exchanges are typically proportional 
to $y^2$. The lighter $N_{Ri}$, the smaller the corresponding Yukawa coupling, 
since $m_\nu\sim y^2/M_ {Ri}$. Therefore, for lighter RH neutrinos, a RG 
evolution over a wider range may be partly compensated by smaller couplings. 
Similarly, loops induced by $\Delta_L$ exchanges are proportional to $f^2$ 
and, since $m_\nu\sim v_L f \sim v^2 f/M_{\Delta_L}$, the lighter the 
$\Delta_L$, the smaller the expected coupling $f$ and the corresponding 
radiative correction. As a simplified numerical test, we chose certain 
structures of $f$ and $y$ at $v_R$ and modified their elements in the seesaw 
formula by order percent corrections. 
Next, we implemented the usual bottom-up procedure to 
reconstruct $f$ and found that, among the 8 dual solutions, one reproduces the 
original structure of $f$ within percent errors. However, this is not the 
case when strong hierarchies among the entries of $f$ and $y$ are present 
{\it and} small matrix elements receive corrections proportional to the large 
ones. We expect that only in these special regions of the parameter space 
instabilities may occur and large departures of the reconstructed structures 
of $f$ from the true ones may arise.

To complete the discussion of RG effects, one should consider the evolution 
of $m_\nu$ from $M_s$ down to the to electroweak scale, where it can be 
compared with experimental data. If only the standard model (or MSSM) 
particles contribute to the RG evolution, the running of the mass squared 
differences and mixing angles is rather small and negligible for the purposes 
of this paper, except perhaps in the case of quasi-degenerate light neutrinos 
of same CP parity. This could affect the reconstruction of $f$ in example 
{\bf 5} of section \ref{numex}.

\subsection{Leptogenesis}
\label{lept}

As we demonstrated above, in the realistic case of three lepton generations 
there are eight different matrices $f_i$ which, for a given $y$, result in  
exactly the same mass matrix of light neutrinos $m_\nu$. A natural question 
is then how one can discriminate between these eight possible solutions. The 
seesaw formula (\ref{master}) cannot tell us more than it already did, and a 
new independent source of information is necessary.  Such a source could be 
provided by the ability of $f_i$ to reproduce the observed baryon asymmetry 
of the universe.

It is well known that the seesaw mechanism not only explains nicely the 
smallness of neutrino mass, but also has a built-in mechanism for generating 
the baryon asymmetry of the universe through leptogenesis \cite{FY}. 
In this mechanism, 
first a lepton asymmetry is produced in out-of-equilibrium CP-violating 
decays of heavy RH neutrinos and/or Higgs triplets, which is then reprocessed 
into baryon asymmetry by electroweak sphalerons. The amount of the produced 
lepton asymmetry depends on (i) mass matrix of RH neutrinos $M_R$ (masses, 
mixing angles and CP phases); (ii) Majorana-type Yukawa coupling of 
leptons to Higgs triplets; (iii) Dirac-type Yukawa coupling of leptons to 
Higgs doublets. In this paper, we consider theories with $f_L=f_R\equiv f$, 
and so the parameters involved in (i) and (ii) 
essentially coincide. This renders the computation of the lepton asymmetry 
more predictive than in generic type I+II seesaw scenarios.

The impact of multiple solutions for $f$ on leptogenesis has been already
analyzed in a class of $SO(10)$ models in \cite{HLS} and very non-trivial 
results were found. Here we will not undertake a quantitative analysis 
of leptogenesis, but rather will make some general remarks, which may be 
of guidance for the development of specific models:

\bei

\item It is recognized by now \cite{JPR,hamsen,antkin2} 
that the presence of both RH neutrinos and 
Higgs triplets may lead to leptogenesis scenarios that are qualitatively 
different from those in pure type I seesaw models. We would like to 
stress, as a minimal possibility,  
that leptogenesis can work in models with  $\Delta_L$ and just one 
RH neutrino species $N_R$ (this effective pattern occurs, e.g., when the 
other two RH neutrinos are super-heavy and/or very weakly coupled).
Since $N_R$ couples to a unique linear combination $L$ of the flavor 
eigenstates of lepton doublets, this scenario may be called ``flavorless
leptogenesis''. This is a viable possibility since the relevant 
interactions of $N_R$ and $\Delta_L$ contain an 
unremovable CP-violating phase 
(which manifests itself in the one-generation seesaw formula, see section 
\ref{1g}). 
A recent study of this scenario can be found in \cite{GZZ}.

\item In many realistic cases an extra freedom gained by the interplay of 
type I and type II seesaw terms may turn out to be insufficient to cure
the shortcomings of thermal leptogenesis in 
the pure type I scenario. In particular, it remains true 
 \cite{hamsen,HRS} 
that the decaying particle should generically be very heavy (above $\gtrsim 
10^8$ GeV) to produce a sufficient asymmetry, thus requiring large reheating 
temperature with the associated problems. Also, the hierarchical structure 
of Yukawa couplings, which is natural in unified scenarios, may strongly 
suppress the asymmetry. The existence of various solutions for $f$, in 
particular of those with a non-hierarchical structure, may alleviate this 
problem. Further details can be found in \cite{HLS}.

\item  Given the rich structure of the Higgs sector at the seesaw scale in
LR symmetric and $SO(10)$ models, sources of lepton asymmetry other than 
the decays of RH neutrinos and/or $\Delta_L$ into left-handed leptons 
should not be overlooked. As an example \cite{FHM}, a successful leptogenesis 
may be due to decays of RH neutrinos into a RH charged lepton and an 
$SU(2)_L$ singlet charged scalar (contained, e.g., in the ${\bf 120}_H$ 
multiplet of $SO(10)$). The coupling $f_R$ between the RH leptons and 
$\Delta_R$ (see eq.~(\ref{Lyuk})) 
may also affect the evolution of the asymmetry.

\item 
In minimal models with ``natural'' values of the seesaw scale and of 
Yukawa couplings,
in most regions of parameter space the produced lepton asymmetry tends to be 
too small.  A way out may be provided by the resonant enhancement of the 
lepton asymmetry which occurs when two RH neutrinos are quasi-degenerate
in mass \cite{lise}. 
As we demonstrated in several examples presented in section \ref{numex}, the 
crossing of the RH neutrino mass levels indeed occurs in the LR symmetric 
seesaw for certain choices of input parameters and for some of the 
solutions $f_i$.  
The requirement of level crossing may be a powerful constraint for model 
building.

\eei

\section{Discussion and summary}

In models with more than one source of neutrino mass, to disentangle 
different contributions using only low-energy data is not in general 
possible. The situation is different when these contributions have a common 
origin, as it is the case in left-right symmetric seesaw models. 
In a wide class of such models and their partially unified or Grand Unified 
extensions, the mass matrix of light neutrinos $m_\nu$ contains type I and 
type II seesaw contributions which depend on the same Majorana-like triplet 
Yukawa coupling matrix $f$. In this paper we undertook for the first time a 
thorough phenomenological bottom-up analysis of such a scenario. We have 
shown that the interplay of type I and type II seesaw terms in $m_\nu$ may 
change our interpretation of neutrino data and provide an insight into the 
underlying theory at the seesaw scale.

We have adopted an approach in which the mass matrix of light neutrinos 
$m_\nu$ and the matrix of Dirac-type Yukawa couplings $y$ are considered  
known, and the seesaw relation is solved for the matrix of the Majorana-like 
Yukawa couplings $f$, which coincides (up to a constant factor) with the 
mass matrix of RH neutrinos and deeply characterizes the structure of the 
underlying theory. To this end, we have developed a linearization procedure 
which allowed us, for symmetric or antisymmetric $y$, to solve the seesaw 
non-linear matrix equation and obtain exact analytic expressions for the 
matrix $f$ in a compact form. 

For symmetric $y$, the overall number of solutions was shown to be $2^n$ for 
$n$ lepton generations. Thus, in the realistic case of three generations, 
there are eight different matrices $f$ which, for a given $y$, result in  
exactly the same mass matrix of light neutrinos $m_\nu$. 

We have studied implications of an intriguing duality property \cite{AF} of 
the LR symmetric seesaw mechanism with symmetric or antisymmetric $y$, which 
relates pairwise different solutions for the matrix $f$. The eight solutions 
of the seesaw equation for $f$ form four dual pairs. 
We have demonstrated that, if one of the solutions $f$ of the seesaw 
equation corresponds to type I dominance, then the dual solution corresponds 
to type II seesaw dominance and vice versa. The other solutions then in 
general correspond to a hybrid seesaw with type I and type II contributions 
to $m_\nu$ being of the same order. An important consequence of this result 
is that, 
{\it knowing only the values of the input parameters, one can determine if 
there are solutions with one seesaw type dominance, but cannot decide which 
particular seesaw type dominates}.

We have explored the behaviour of the analytic solutions for $f$ in a number 
of interesting limiting cases: 
(a) one seesaw type dominates; (b) the two seesaw contributions almost cancel 
each other; (c) the eigenvalues of $y$ are strongly hierarchical; (d) one 
light neutrino mass vanishes. By analyzing several numerical examples, we 
found that the masses of RH neutrinos exhibit the following generic 
features:  
(i) differ strongly for different dual solutions for $f$; 
(ii) depend crucially on the light neutrino mass spectrum;
(iii) depend weakly on variations of $\theta_{13}$ within its allowed range 
and on variations of the low-energy leptonic CP-violating phases;
(iv) depend strongly on the hierarchy among the eigenvalues of $y$ 
only in the cases with predominant type I seesaw contribution to $m_\nu$; 
(v) may become quasi-degenerate in several regions of the parameter 
space which are close to the points where the level crossings occur;
(vi) can be at TeV scale or lighter even when $v_R$ is much larger and/or some
entries of the neutrino Yukawa coupling matrix $y$ are of order one.

We have also studied (in Appendix \ref{antiY}) the case of antisymmetric 
Dirac-type Yukawa coupling matrices $y$ and showed that our linearization 
procedure works in that case as well, leading to simple analytic expressions 
for $f$. The multiplicity of solutions in that case was found to be 1, 2 and 
2 for one, two and three lepton generations. 

In addition, we considered (in Appendix \ref{star}) an alternative 
realization of the discrete LR symmetry, in which it acts as parity 
transformation rather than charge conjugation. This leads to type II and 
type I contributions to $m_\nu$ depending on $f$ and $f^*$ rather than on 
the same matrix $f$. In that case 
exactly the same duality of solutions holds, provided that the matrix $y$ 
is Hermitian or anti-Hermitian and invertible. In the present paper we gave 
explicit formulas for the solutions for one and two lepton generations, and 
briefly outlined the approach to the three-generation case. The multiplicity 
of solutions for $n$ lepton generations was shown to be $2^n$ in the case 
of Hermitian $y$, whereas for anti-Hermitian $y$ there are either $2^n$ or 
no solutions. The latter possibility is a consequence of the non-analytic 
dependence of the seesaw relation on $f$ in this realization of the discrete 
LR symmetry. 
Our results on the numbers of solutions for $f$ in various 
cases are summarized in Table 1.

\TABLE[t]{
\begin{tabular}{|c|c|c|c|c|c|}
\hline
seesaw formula & neutrino Yukawa y & section &
one gen. & two gen. & three gen. \\
\hline \hline
eq.~(\ref{master}) & generic (no duality) & 3.1 and 3.2 &  2 & 4 & -- \\
\hline
eq.~(\ref{master}) & symmetric & 2.3 & 2 & 4 & 8 \\
\hline
eq.~(\ref{master}) & antisymmetric & App. A & 1& 2 & 2 \\
\hline \hline
eq.~(\ref{master2}) & generic (no duality) & App. C & 2 or 0 & -- & -- \\
\hline
eq.~(\ref{master2}) & Hermitian & App. C & 2 & 4 & 8 \\
\hline
eq.~(\ref{master2}) & anti-Hermitian & App. C & 2 or 0 & 4 or 0 & 8 or 0 \\
\hline
\end{tabular}
\caption{Multiplicity of solutions for the Majorana-type Yukawa 
coupling matrix $f$. The first column indicates the seesaw equation under 
consideration, the second column specifies the assumption made on the 
structure of the Dirac-type Yukawa coupling matrix $y$, the third column 
indicates the section where the multiplicity was derived. Fourth, fifth and 
sixth columns give the number of solutions in the case of one, two and 
three lepton generations, respectively. The dash indicates the cases for 
which the number of solutions was not derived.}
}

Since one of our main goals was to provide a guidance for building specific
models incorporating the LR-symmetric seesaw, we have discussed the structure 
of the seesaw formula in the minimal LR model, Pati-Salam model and in 
several $SO(10)$ models with renormalizable Yukawa couplings. Our analysis 
shows that in most cases the phenomenon of seesaw duality is realized: one 
can identify a matrix of couplings $f$ which enters in the seesaw  formula 
both directly and through its inverse. Such matrix can be reconstructed 
by making use of the methods developed in this paper.

We have discussed briefly the stability of our results with respect to the 
renormalization group running effects, which below the LR symmetry breaking 
scale  may result in modifications of the relations $f_L=f_R$ and 
$y^T=\pm y$ that were at the basis of our analysis. Conditions were found 
under which these renormalization effects will not spoil our reconstruction 
of the matrix $f$ from the input data. 
A comprehensive study of this issue, 
however, will require a dedicated effort.

Finally, we pointed out that the usual leptogenesis mechanism of the 
generation of baryon asymmetry of the universe produces an asymmetry 
that strongly depends on the adopted solution for $f$ and therefore may in 
principle help discriminate between the eight allowed solutions. Moreover, 
the level crossing between the masses of the two lightest RH neutrinos, which 
occurs for a number of solutions, may provide a resonant enhancement of the 
produced asymmetry.

\section*{Acknowledgments}
EA was supported by the Wenner-Gren Foundation as an Axel Wenner-Gren 
visiting professor at the Royal Institute of Technology. MF was partially 
supported by the RTN European Program MRTN-CT-2004-503369. The authors are 
grateful to Alexei Smirnov for very helpful discussions and comments. 
MF thanks the authors of ref. \cite{HLS} for many fruitful 
discussions and the comparison of respective results.

\appendix

\section{Reconstruction of $f$ in the case of antisymmetric $y$
 \label{antiY}}

Consider the case of antisymmetric Dirac-type neutrino Yukawa coupling, 
$y=-y^T$. A crucial difference with respect to the case of symmetric $y$ is 
the fact that $(2n+1)$-dimensional antisymmetric matrices are not invertible, 
so that the usual duality among the solutions for $f$ does not hold for an odd 
number of lepton generations. In the case of one generation, one trivially has 
$y=0$, and the seesaw is purely of type II.

The case of two generations is more interesting. An antisymmetric $2\times 2$ 
matrix is defined by a single parameter. Consider the system (\ref{system}) 
for $y_{\mu 2} = y_{\tau 3}=0$ and $y_{\mu 3} = - y_{\tau 2} \equiv \bar{y}$.  
Using the same linearization procedure as in section \ref{2g}, one 
easily finds the general solution:  
\beq 
f = \frac{x\lambda}{x\lambda-\bar{y}^2} m ~~~~~~~~{\rm with~~~~~~~}
(x\lambda)^2 - \left(2\bar{y}^2 + x \det m \right)x\lambda + \bar{y}^4 =0\,.
\label{antis}\eeq 
The solutions $\lambda_{\pm}$ of the quadratic equation correspond to a pair 
of dual solutions $f_{\pm}$. The duality relation (\ref{detdu}) is replaced by 
$x^2 \lambda_+ \lambda_{-}= \bar{y}^4$. Notice that both solutions are 
proportional to the matrix $m$.

While in general the system (\ref{system}) has four solutions, two of them 
are singular when $y$ is exactly antisymmetric, as assumed above. This can 
be understood as follows: consider the case where the diagonal entries of 
$y$ are equal to zero and the off-diagonal ones satisfy $y_{\mu 3} 
\approx \bar{y}$ and $y_{\tau 2} \approx -\bar{y}$, so that $y$ is nearly 
antisymmetric, Then $f$ is still (approximately) given by eq. (\ref{antis}), 
while the quartic equation for $\lambda$ reads 
\beq
\left[(x\lambda)^2 - \left(2\bar{y}^2 + x \det m \right)x\lambda +
\bar{y}^4\right](x\lambda -\bar{y}^2)^2 \approx 0 \,,
\eeq 
to be compared with eq.~(\ref{antis}). The pair of dual solutions 
corresponding to $x\lambda \rightarrow \bar{y}^2$ is not physical in the limit,
since all the matrix elements of $f$ diverge.

Consider now the relative size of type I and II seesaw contributions to $m_\nu$. 
The two 
solutions of the characteristic equation for $\lambda$ in eq.~(\ref{antis}) 
can be written as
\beq
x\lambda_\pm = \frac{x\det m}{2}\left[1+\frac d2 \pm (1+d)^{1/2}\right]\,, 
~~~~~d\equiv \frac{4\bar{y}^2}{x\det m} \,.
\label{lampm}
\eeq
A straightforward calculation shows that both $m_\nu^{II}$ and
$m_\nu^I$ are proportional to $m_\nu$ and their ratio is 
\beq
r^{II/I}_\pm\equiv \left.\frac{(m_\nu^{II})_{ij}}{(m_\nu^I)_{ij}}\right|_\pm 
= - \frac 2d \left[1+\frac d2 \pm (1+d)^{1/2}\right] \,. 
\eeq 
For $|d|\ll1$, $r_\pm^{II/I} \approx -(4/d)^{\pm1}$, so that one seesaw type 
dominates. For $|d|\sim 1$, $r_\pm^{II/I} \sim 1$, corresponding to hybrid 
seesaw. For $|d|\gg 1$, $r_\pm^{II/I} \approx -1$, which implies a 
considerable cancellation of the two seesaw contributions. These three cases 
resemble closely the corresponding limits in the one lepton generation 
scenario (section \ref{1g}).

Let us now turn to the realistic case of three lepton generations. As was 
pointed out above, the usual duality does not apply since $y$ has one vanishing 
eigenvalue. However, a simple analytic reconstruction of the matrix $f$ 
is still possible, as we show below.

Any antisymmetric $3\times 3$ matrix can be written as
\beq
y=\left(\bea{ccc}
0 & y_3 & y_2 \\ -y_3 & 0 & y_1 \\ -y_2 & -y_1 & 0 \\
\eea\right)= U\left(\bea{ccc}
0 & 0 & 0 \\ 0 & 0 & -y_{123} \\ 0 & y_{123} & 0 \\
\eea\right)U^T \equiv U y' U^T\,,
\eeq
where $y_{123}\equiv\sqrt{|y_1|^2+|y_2|^2+|y_3|^2}$ and the unitary 
matrix $U$ is given by
\beq
U=\frac{1}{y_{123}y_{23}}\left(\bea{ccc}
y_1^*y_{23} & - y_{23}^2 & 0 \\
-y_2^* y_{23} & -y_2^*y_1 & y_3 y_{123} \\
y_3^* y_{23} & y_3^*y_1 & y_2 y_{123}
\eea\right)\,,
\label{Ua}\eeq
with $y_{23}\equiv\sqrt{|y_2|^2+|y_3|^2}$. Defining $m'\equiv U^\dag m U^*$ 
and $f'\equiv U^\dag f U^*$, one can write the seesaw equation as 
\beq
m' = f' + \frac 1x y' f'^{-1} y' \,.
\eeq
The entries of the first row of $f'$ are uniquely determined as  
$f'_{1i}=m'_{1i}$, whereas to find the 2-3 block one needs to implement 
the usual linearization procedure. This yields 
\beq
f'_{22}=\frac{x\lambda m'_{22}-y_{123}^2{m'}^2_{12}}
{x\lambda-y_{123}^2m'_{11}} \,,~~~~
f'_{23}=\frac{x\lambda m'_{23}-y_{123}^2m'_{12}m'_{13}}
{x\lambda-y_{123}^2m'_{11}} \,,~~~~
f'_{33}=\frac{x\lambda m'_{33}-y_{123}^2{m'}_{13}^2}
{x\lambda-y_{123}^2m'_{11}} \,,
\eeq
where $\lambda$ is the solution of the quadratic equation
\beq
(x\lambda)^2 - (2y_{123}^2 m'_{11} + x \det m')x\lambda 
+ y_{123}^4 {m'}_{11}^2 = 0\,.
\eeq
The two solutions $\lambda_{\pm}$ have the same form as in 
eq. (\ref{lampm}), but with $d \equiv 4 y_{123}^2 m'_{11}/(x\det m')$.
Even though the usual duality property does not hold in this case, a 
{\it different duality} between the two solutions is present: one finds
$x\lambda_+ \lambda_- = y_{123}^4{m'}_{11}^2$
and $f'_+ + f'_- = \tilde{m}$, where $\tilde{m}_{ij}\equiv m'_{ij}+
m'_{1i}m'_{1j}/m'_{11}$. The two solutions for $f$ in the original basis
are obtained as $f_{\pm} = U f'_{\pm}U^T$, where $U$ is given in
eq. (\ref{Ua}).

Thus, for an antisymmetric $y$ the realistic case of three lepton generations 
can be reduced to that of an effective two-generation system, so that there 
are only two solutions of the LR seesaw formula for $f$, which are related by 
a modified duality.

Note that the method of counting the solutions developed in 
section \ref{1n} does not apply to the case of antisymmetric $y$: 
for one and three generations, because $y$ is not invertible, and for two 
generations because this is a degenerate case.
Indeed, for $n=2$ 
one finds $\tilde{f}^{-1}=[-\det y / (x\det f)]\cdot \tilde{f}$, so that  
the equation $\tilde{m}=\tilde{f}+\tilde{f}^{-1}$, which replaces 
eq.~(\ref{tildess}) in the case $y^T=-y$, is linear rather than quadratic 
in $\tilde{f}$. 
This equation results in a quadratic equation for $\det f$, leading to two 
solutions for $f$.

\section{Relative size of type I and II seesaw in the presence of flavor mixing
 \label{dominance}}

In the simple case of one or more unmixed generations the relative size of 
type I and type II seesaw contribution to $m_\nu$ was discussed in section 
\ref{1g}. Here we analyze this issue 
in the case of 2 and 3 generations with flavor mixing.

Due to the mixing, each entry of $m_\nu$ receives contributions from type
I and II seesaw in different proportions. 
For simplicity, we will present only the conditions for the dominance of 
one seesaw type in all the matrix elements of $m$ 
(extensions to more general cases can be easily obtained). In what follows, 
the relations involving $m_{\alpha\beta}$ and $y_i$ will therefore be 
assumed to hold for each $\alpha,\beta$ and $i$, if not otherwise stated.

Consider first the two-generation case. From eq.~(\ref{sol2}) one can 
see that, when a solution $\lambda_1$ satisfies $|x\lambda_1| \gg |y_i 
y_j|$ ($i,j=2,3$), one obtains $f_1 \simeq m$ (assuming that $m_{\mu\mu}$ and 
$m_{\tau\tau}$ are of the same order). This corresponds to the dominant 
type II seesaw. Then the dual solution $x\lambda_2=x\hat{\lambda} _1=y_2^2 
y_3^2/(x\lambda_1)$ has modulus $\ll |y_i y_j|$ and the corresponding
matrix $f_2=\hat{f}_1$ takes the form obtained in type I seesaw case. 
In analogy with eq. (\ref{domina}), let us define the set of parameters
\beq
d^{ij}_{\alpha\beta\gamma\delta} \equiv \frac{v_L}{v_R} \frac{4 v^2 y_i y_j}
{(m_\nu)_{\alpha\beta} (m_\nu)_{\gamma\delta}} =
\frac{4 y_i y_j}
{x m_{\alpha\beta} m_{\gamma\delta}} \,,~~~~~ 
i,j =2,3\,,~~~\alpha,\beta,\gamma,\delta = \mu,\tau \,.
\label{manyd}
\eeq
Notice that eqs. (\ref{solu}) and (\ref{KR}) imply that a necessary condition 
to have $|x\lambda_1| \gg |y_i y_j|$ is
$|m_{\alpha\beta} m_{\gamma\delta}| \gg 4 |y_i y_j / x|$, 
that is, 
\beq
|d^{ij}_{\alpha\beta\gamma\delta}| \ll 1\,.
\label{domps}\eeq
Therefore, this limit ensures the existence of a pair of dual solutions
$f_{1,2}$ with the dominance of one seesaw type, in analogy with the
limit $|d|\ll 1$ in the one-generation case.  The solutions
$\lambda_{1,2}$ are defined by choosing $r_+$ in eq. (\ref{solu}) and
can be expanded in terms of the small parameters as 
\beq
\bea{c}
x\lambda_1 = x\det m \left[1+\dfrac{k}{x(\det m)^2} +\dfrac{3(\det m)^2
y_2^2 y_3^2 - k^2}{x^2(\det m)^4} + \dots \right] \,,\\ \\
x\lambda_2\equiv x\hat{\lambda}_1 = \dfrac{y_2^2 y_3^2}{x\det m}
\left[1-\dfrac{k}{x(\det m)^2} + \dots \right] \,.\\ \\
\eea\label{la12}
\eeq 
The solutions for $f$ are related by $f_1 = m -f_2$ with 
\beq
f_2 \approx \dfrac{1}{x\det m}\left(\bea{cc}
-y_2^2 \left[ m_{\tau\tau}+ \dfrac{y_3^2 m_{\mu\mu} \det m  
-  m_{\tau\tau} k }{x(\det m)^2}
\right] & 
y_2y_3 m_{\mu\tau} \left[1 -\dfrac{y_2 y_3 \det m +k}{x(\det m)^2} \right] \\
\dots & 
-y_3^2  \left[ m_{\mu\mu} + \dfrac{y_2^2 m_{\tau\tau} \det m  
- m_{\mu\mu} k }
{x(\det m)^2}\right]
\eea\right) \,.
\eeq
The second terms in the square brackets represent the leading order correction 
to the pure type I seesaw.

In general, condition (\ref{domps}) 
does not guarantee that the other 
dual pair of solutions $f_{3,4}$ corresponds to one seesaw type 
dominance.  If no special cancellations occur, eq. (\ref{domps}) implies, 
in particular, $|\det m| \gg 4|y_i y_j /x|$. If this condition is 
satisfied, then $f_3$ and $f_4$ are of hybrid type.
This novel feature with respect to the one-generation case is a genuine
effect of flavor mixing. In fact, the solutions corresponding to the choice 
of $r_-$ in eq.~(\ref{solu}) are given, up to higher orders in small 
parameters, by 
\beq
x\lambda_{3,4} = -\dfrac{k\pm s}{2\det m} \left[ 1 \mp \dfrac{s}{x(\det
m)^2} +\dots \right] \,,~~~~~~ s\equiv\sqrt{k^2 - (2 y_2 y_3 \det m)^2}
\,. 
\label{la34}
\eeq 
Notice that $\lambda_{3,4}$ are dual to each other. The corresponding 
solutions for $f$ (to leading order) are 
\beq
f_{3,4}\approx\dfrac{1}{2s}\left(\bea{cc} m_{\mu\mu}(s \pm k) \mp 2
m_{\tau\tau} y_2^2 \det m & m_{\mu\tau}(s \pm k \pm 2 y_2 y_3 \det m)\\
\dots & m_{\tau\tau}(s \pm k) \mp 2 m_{\mu\mu} y_3^2 \det m \eea\right)\,. 
\label{f34}
\eeq 
Roughly, $(f_{3,4})_{ij}\sim m_{\alpha\beta}$ and $F_{3,4} = \lambda_{3,4} 
\sim y_i y_j /x $. Therefore, type II and type I contributions to $m_\nu$ 
are of the same order: $m_\nu^{II} \sim v_L m_{\alpha\beta}$ and $m_\nu^I 
\sim v^2 y_i y_j f_{kl} / (v_R F) \sim v_L m_{\alpha\beta} $; this 
proves that the solutions $f_{3,4}$ are of hybrid type.

If a special cancellation in $\det m$ occurs, 
also the solutions $f_{3,4}$ are dominated by one 
type of seesaw. Indeed, in the limit $\det m \rightarrow 0$ 
eq. (\ref{solu}) becomes 
\beq
x\lambda_{1,3} \approx \pm \frac 12 \left(\sqrt{kx+4y_2^2y_3^2} +
\sqrt{kx} \right) \,,~~~~~ x\lambda_{2,4} \equiv x\hat{\lambda}_{1,3}
\approx \pm \frac 12 \left(\sqrt{kx +4y_2^2y_3^2} - \sqrt{kx} \right)
\,,~~~~~ 
\eeq 
where $k \approx (m_{\mu\mu}y_3 + m_{\tau\tau} y_2)^2$. When eq. (\ref{domps}) 
holds, one has $|kx|\gg 4y^2_2y^2_3$. As a consequence, $|x\lambda_1| 
= |x\lambda_3|\gg | y_i y_j|$, so that $f_{1,3}$ are solutions with dominant 
type II seesaw. By duality, $f_{2,4} $ correspond to the dominant 
type I seesaw.

Note that a suppression in $\det m_\nu$ is phenomenologically motivated if the 
neutrino spectrum has normal mass hierarchy. In this case the full $3\times 3$ 
mass matrix $m_\nu$ is dominated by large entries in the $2\times 2$ 
$\mu\tau$-block. This block has to incorporate the maximal mixing observed in 
the atmospheric neutrino oscillations as well as the hierarchy 
$\Delta m^2_{sol} \ll \Delta m^2_{atm}$. These two requirements imply that the 
determinant of the $\mu\tau$-block is suppressed.

When one abandons the condition (\ref{domps}), that is, when at least
one $|d^{ij}_ {\alpha\beta\gamma\delta}| \gtrsim 1$, all four solutions 
are of hybrid type. In other words, {\em some sets of input 
parameters necessarily imply a hybrid seesaw scenario.}

In the limit when all $|d^{ij}_{\alpha\beta\gamma\delta}| \gg 1$, one has
$x\lambda_i\approx r_\pm /4 \approx \pm y_2 y_3$, and
eq. (\ref{sol2}) implies that (at least the diagonal) matrix elements 
of $f$ grow as $1/(|x\lambda|-|y_2y_3|)$, thus ending up to be much larger 
than $m_{\alpha\beta}$. This means that one enters the cancellation region,
where type I and II contributions are almost equal in absolute value and 
opposite in sign. In this limit, the matrix elements of $f$ will violate 
perturbative unitarity. This means that for some values of the input 
parameters, there are no physically acceptable solutions for $f$. In other 
words, {\em the LR symmetric seesaw mechanism cannot reproduce certain sets 
of the input parameters and thus, in principle, can be ruled out.}

Consider now the three-generation case. In analogy with eq. (\ref{manyd}), one 
finds that a necessary condition for the dominance of one seesaw type is
\beq
|d^{ij}_{\alpha\beta\gamma\delta}| 
\equiv \frac{v_L}{v_R} \left|\frac{4 v^2 y_i y_j}
{(m_\nu)_{\alpha\beta} (m_\nu)_{\gamma\delta}}\right| =
\left|\frac{4 y_i y_j}
{x m_{\alpha\beta} m_{\gamma\delta}}\right|
\ll 1\,,~~~~
i,j =1,2,3\,,~~~\alpha,\beta,\gamma,\delta = e,\mu,\tau\,.
\label{domps3}\eeq
In this case, the solutions $\lambda_i$ of eq. (\ref{final}) can be
expanded in the small parameters $d^{ij}_{\alpha\beta\gamma\delta}$,
which we denote collectively by $d$. The largest solution turns out to be
\beq
\lambda_1 = \det m \left[1+\frac{A}{(\det m)^2}+
\frac{3Y^2S(\det m)^2-A^2}{(\det m)^4}+{\cal O}(d^3)\right]\,.
\label{lala}\eeq
Noting that $Y^2/\lambda_1^2\sim d^3$, it is straightforward to verify 
with eq. (\ref{solone}) that $f_1$ corresponds to the dominant type II 
seesaw. As a consequence, the three RH neutrino masses satisfy 
$M_{1,2,3}\gg y_i y_j M_s$, where $M_s$ is the seesaw scale defined as
$M_s \equiv v^2/m_3$ (normal mass ordering) of $v^2/m_2$ (inverted mass 
ordering).

Three more solutions are given by $\lambda_{2,3,4}=\bar{\lambda}_{2,3,4}[1+
{\cal O}(d)]$, where $\bar{\lambda}_{2,3,4}$ are the roots of the cubic 
equation
\beq
\bar\lambda^3+\frac{A}{\det m}\bar\lambda^2+Y^2S\bar\lambda+Y^2
\det m = 0\,.
\eeq
It is easy to see that $\lambda_{2,3,4}/\lambda_1 \sim d$. From 
eq. (\ref{solone}) one finds
\beq
(f_k)_{ij} \approx \frac{
(\bar\lambda_k^3-Y^2\det m)m_{ij}+\bar\lambda_k^2 A_{ij} -Y^2 \bar
\lambda_k S_{ij}
}{
\bar\lambda_k^3-Y^2\bar\lambda_k S - 2 Y^2 \det m
} \,,~~~~k=2,3,4 \,.
\eeq
One can verify that for these three solutions type I and II contributions to 
$m_\nu$ are of the same order (hybrid seesaw). Since $\lambda_k=\det f_k$ are 
suppressed with respect to $\lambda_1$, only two RH neutrinos are have masses 
that are larger than the seesaw scale, $M_{2,3}\gg y_i y_j M_s$, while 
$M_1\sim y_i y_j M_s$.

The dual solutions, $\hat{\lambda}_{2,3,4}=-Y^2/\lambda_{2,3,4}$, are
suppressed by a factor $\sim d^2$ with respect to $\lambda_1$. The
corresponding structures for $f$ are  
\beq
(\hat{f}_k)_{ij} \approx \frac{-Y^2(\bar\lambda_k S + \det m)m_{ij}-
\bar\lambda_k^2 A_{ij} + Y^2 \bar\lambda_k S_{ij}}{
\bar\lambda_k^3-Y^2\bar\lambda_k S - 2 Y^2 \det m
} \,,~~~~k=2,3,4 \,.
\eeq
Note that $f_k +\hat f_k = m$, as required by duality. The solutions 
$\hat{f}_k$ also correspond to type I and II contributions to $m_\nu$ 
being of the same order (hybrid  seesaw). The masses of the RH neutrinos 
satisfy $M_{3}\gg y_i y_j M_s$, $M_{1,2}\sim y_i y_j M_s$.

Finally, the solution dual to $\lambda_1$ is given by
\beq
\hat{\lambda}_1 = -\frac{Y^2}{\det m}
\left[1-\frac{A}{(\det m)^2}+{\cal O}(d^2)\right]\,.
\label{lasm}\eeq
This is the smallest solution, since $\hat{\lambda}_1/\lambda_1 \sim d^3$. By 
duality, it corresponds to the dominant type I seesaw, as one can check 
explicitly expanding eq. (\ref{solone}). All three RH neutrino masses are 
approximately at the seesaw scale, $M_{1,2,3}\sim y_i y_j M_s$.

As already mentioned, the parameters $d^{ij}_{\alpha\beta\gamma\delta}$ may 
not all be small at the same time. In that case, for any of the eight 
solutions $f$, only some of the elements of the matrix $m_\nu$ may be 
dominated by one seesaw type, while one or more other elements receive 
significant contributions from both seesaw types.

A remark is in order on the special case when $\det m$ is suppressed. 
For example, one may have $|\det m| \ll |Y|$ even if eq. (\ref{domps3}) holds.
Clearly, in the limit $\det m \rightarrow 0$ the expansions 
(\ref{lala})-(\ref{lasm}) do not apply. We showed in section \ref{det0} that, 
when $\det m=0$, two pairs of dual solutions for $\lambda$ are equal in 
absolute value and opposite in sign to the other two dual pairs. Moreover,  
the equation for $\lambda$ has the same form as in the two-generation case
(with the redefinitions (\ref{ridef})).  In particular, the relative size of 
type I and II seesaw contributions to $m_\nu$ may be evaluated as we did above 
for the two-generation case: when eq.~(\ref{domps3}) holds, there are two 
opposite solutions $\pm\lambda_1$ which lead to the dominant type II seesaw, 
while their duals correspond to the dominant type I seesaw. The other two 
opposite pairs of dual solutions correspond in general to hybrid seesaw, 
unless a special suppression in $\sqrt{A}$ occurs.

\section{Alternative realization of the discrete left-right symmetry 
\label{star}}

Consider a different realization of the discrete LR symmetry in the minimal LR 
model, in which it acts as parity transformation. Eq. (\ref{Dp}) is then 
replaced by   
\beq 
L\leftrightarrow (L^{c})^c\equiv \left(\begin{array}{c} N_R \\
l_R\end{array}\right) \,,~~~~~ \Phi
\leftrightarrow \Phi^\dag \,,~~~~~ \Delta_L\leftrightarrow \Delta_R^* \,.
\label{Dp2}\eeq 
This possibility is almost as popular in the literature 
(see, e.g., \cite{DGKO,kiers}) as the one we adopted before, even though it does 
not allow an $SO(10)$ embedding of the minimal LR model. Eqs. (\ref{Dp2}) 
and (\ref{Lyuk}) imply
\beq 
f\equiv f_L = f_R^* \,,~~~~~ g=g^\dag \,,~~~~~
h= h^\dag \,.
\label{symm2}
\eeq 
In this case the seesaw relation is  
\beq 
m_\nu \simeq v_L f - \frac{v^2}{v_R} y
(f^*)^{-1} y^T \,,
\label{master2} 
\eeq 
where $y$ is defined in eq. (\ref{dirac}). The difference in 
complex phases between the seesaw formulas (\ref{master}) and (\ref{master2}) 
can, in principle, lead to a different phenomenology if CP is violated in the 
leptonic sector.

It can be easily checked by direct substitution into eq.~(\ref{master2}) that, 
if the matrix $y$ is Hermitian or anti-Hermitian and invertible, this equation 
possesses the same duality property as eq.~(\ref{master}): namely if a matrix 
$f$ solves eq.~(\ref{master2}), so does its dual $\hat{f}\equiv m_\nu/v_L-f$. 
The above conditions on $y$ are both necessary and sufficient for the duality 
to hold.

Eqs. (\ref{symm2}) and (\ref{dirac}) imply that $y$ is actually Hermitian 
if $h=0$ (which holds, e.g., in the SUSY case) or if the two VEVs of the 
bidoublet $v_1$ and $v_2$ are both real (this is not the case in general, 
since both VEVs can be non-zero, and their relative phase cannot be 
rotated away \cite{DGKO}). If $y$ is Hermitian, one can diagonalize it 
according to $y=U y_d U^\dag$ with $U$ unitary, and go to the basis where 
$y$ is diagonal and real by absorbing $U$ in the definition of $m_\nu$ 
and $f$ (analogously to eq. (\ref{madia})).

Consider first the one-generation case with an arbitrary complex $y$. 
Defining $f\equiv |f| e^{i\phi}$, one 
can obtain from eq.~(\ref{master2})  
\beq e^{i\phi} = \dfrac{m|f|}{|f|^2 - y^2/x} \,. 
\label{argu}
\eeq 
The value of $|f|$ is determined by the requirement that that numerator and 
denominator on the right hand side have the same moduli (recall that in our 
conventions all the VEVs are real and positive, and so is $x$): 
\beq
|f|^4 - \left[|m|^2 + \frac{2\cos\chi |y|^2}{x}\right] |f|^2 + \frac{|y|^4}
{x^2} = 0\,.
\label{qum}
\eeq
Here we used $y^2\equiv|y|^2 e^{i\chi}$. Eq.~(\ref{qum}) has (two) real and 
positive solutions for $|f|^2$ if and only if 
\beq
|m|^2 > \frac{2|y|^2}{x} (1-\cos\chi) \,.
\label{cond}
\eeq
This means that, if the input parameters do not obey this inequality, 
no value of $f$ satisfies the seesaw formula (\ref{master2}).
In other words, such a set of input parameters cannot be explained by the LR 
symmetric seesaw mechanism. This problem never occurs in the case of 
eq. (\ref{master}), which is analytic in the elements of $f$ and thus 
always has solutions.

The requirement that $y$ be Hermitian or anti-Hermitian reduces in the one 
generation case to the condition that $y$ be real ($\chi = 0$) or purely 
imaginary ($\chi=\pi$). Let us focus on the $\chi=0$ case. Eq. (\ref{cond}) 
is then always satisfied and the two solutions of eq. (\ref{qum}) are 
\beq
|f_\pm| = \frac{|m|}{2} \left[\left(1+ |d|\right)^{1/2} \pm 1 \right]\,,
\label{solm}\eeq
where $d$ was defined in eq. (\ref{domina}). The discussion of the dependence 
of $|f_\pm| $ on $|d|$ in different limiting cases is analogous to that in 
section \ref{1g}. Finally, the phases $\phi_\pm \equiv \arg f_\pm$ are 
determined uniquely by plugging eq. (\ref{solm}) into eq. (\ref{argu}). Thus, 
if $y$ is real, eq. (\ref{master2}) has two solutions.

Consider now two lepton generations, limiting ourselves to the case $y=y^\dag$ 
and working in the basis where $y$ is diagonal and real. The linearization 
procedure is the same as in section \ref{2g}. However, to close the system of 
equations one has to consider both eq.~(\ref{master2}) and its complex 
conjugate, so that the linearized system contains 6 equations. Defining
$\lambda=| \lambda|e^{i\rho}$, one arrives at the solution which is 
analogous to that in eq. (\ref{sol2}): 
\beq
f=\dfrac{x|\lambda|}{(x|\lambda|)^2-y_2^2 y_3^2}
\left(\bea{cc}
x|\lambda| m_{\mu\mu} + y_2^2 m^*_{\tau\tau} e^{i\rho} & 
x|\lambda| m_{\mu\tau} - y_2 y_3 m^*_{\mu\tau} e^{i\rho} \\
\dots & x|\lambda| m_{\tau\tau} + y_3^2 m^*_{\mu\mu} e^{i\rho} 
\eea\right) \,.
\label{sol2*}\eeq
The parameter $\lambda$ is determined, as usual, from the equation
$\lambda=\det f (\lambda)$, which reads:
\beq\bea{c}
\left(x^2|\lambda|^2 - y_2^2 y_3^2\right)^2 - 
x^3|\lambda|^2
(|m_{\mu\mu}|^2 y_3^2 + 2|m_{\mu\tau}|^2 y_2 y_3+|m_{\tau\tau}|^2 
y_2^2)  \\
-x^2 |\lambda| \left[\det m e^{-i\rho} x^2|\lambda|^2 + (\det m)^* e^{i\rho} 
y_2^2 y_3^2  \right] = 0 \,.
\eea
\label{lam2*}\eeq
The imaginary part of this equation implies $\rho\equiv\arg\lambda=\arg\det m 
+ l\pi$ with $l=0, 1$ (barring the special cases $\det m =0$ and $x|\lambda| 
= y_2 y_3$). Then $|\lambda|$ satisfies a quartic equation, which is completely 
analogous to that in eq. (\ref{lam2}). Therefore, its four solutions are 
easily found analytically and one can verify that for each $l=0$ and $l=1$ 
there is only one pair of dual solutions which yield real and positive 
$|\lambda|$ and therefore are acceptable. As a consequence, also in this 
scenario there are four 
solutions for the matrix $f$.

The three-generation case can be considered quite analogously. The 
linearization procedure yields a closed set of 12 linear equations for 
the elements of the matrices $f$ and $f^*$, and solving the characteristic 
equation one arrives at eight possible solutions for $f$.

It is actually not difficult to show that the number of solutions for $n$ 
lepton generations is always $2^n$, provided that the matrix $y$ is Hermitian 
and invertible. Indeed, going into the basis where $y$ is diagonal an real 
and performing the same transformations as in section \ref{1n}, one arrives at 
the equation
\beq
\tilde{m}=\tilde{f}-(\tilde{f}^*)^{-1}\,,
\label{tildess2}
\eeq
where $\tilde{m}$ and $\tilde{f}$ are symmetric. The matrix $\tilde{f}$ 
can be diagonalized as $\tilde{f}=V \tilde{f}_d V^T$, where $V$ is unitary 
and $\tilde{f}_d$ diagonal and real; it is easy to see that then 
$(\tilde{f}^*)^{-1}$ is diagonalized by the same transformation. Therefore, 
multiplying eq.~(\ref{tildess2}) by $V^\dag$ on the left and by $V^*$ on the 
right, one diagonalizes its right hand side, and so also the left hand side.  
Thus, $V$ is a unitary matrix that diagonalizes the matrix $\tilde{m}$ and 
therefore is determined by the known quantities $m$ and $y$. In the diagonal 
basis eq.~(\ref{tildess2}) yields
\beq
\tilde{f}_{d i}^2-\tilde{m}_{d i} \tilde{f}_{d i}-1=0\,,\qquad i=1,\dots\,n\,.
\label{tildes3}
\eeq
This gives two values of $\tilde{f}_{di}$ for each $i$:
\beq
\tilde{f}_{d i}=\frac{\tilde{m}_{d i}}{2}\pm\sqrt{\frac{\tilde{m}_{d i}^2}{4}+1}\,.
\label{tildess3}
\eeq
A full $n$-generation solution is obtained by picking one of these two values 
for each $i$, which yields $2^n$ solutions for the matrix $\tilde{f}$ and 
therefore for $f$.

The same method  can also be applied to the case of anti-Hermitian $y$ 
(making use of the fact that $iy$ is Hermitian). Eq. (\ref{tildess3}) is then 
replaced by 
\beq
\tilde{f}_{d i}=\frac{\tilde{m}_{d i}}{2}\pm 
\sqrt{\frac{\tilde{m}_{d i}^2}{4}-1}\,,
\label{tildess4}
\eeq
This leads to $2^n$ solutions for 
$f$, provided that 
all $\tilde{m}_{di}$ satisfy $\tilde{m}_{di}^2>4$; 
otherwise the seesaw equation (\ref{master2}) 
has no solutions.
 
A procedure similar to that described below eq.~(\ref{tildess2}) 
(but making use of a complex orthogonal transformation rather than of a 
unitary one) was first used in \cite{HLS} in order to solve eq.~(\ref{master})
with symmetric $y$.  
In both cases, this approach allows not only to count the 
solutions of the seesaw equation, but also to obtain them explicitly.



\providecommand{\href}[2]{#2}\begingroup\raggedright\endgroup

\end{document}